\documentclass[11pt]{article}

\usepackage[top=1in,bottom=1in,left=1in,right=1in]{geometry}
\usepackage{natbib}
\usepackage[unicode=true,bookmarks=true,bookmarksnumbered=false,bookmarksopen=false,breaklinks=false,pdfborder={0 0 1},backref=false,colorlinks=true,hypertexnames=false]{hyperref}
\hypersetup{citecolor=blue}
\usepackage{url}

\renewcommand{\baselinestretch}{1.5}

\usepackage{amsmath}
\usepackage{amssymb}
\usepackage{amsthm}
\usepackage{mathtools}
\usepackage{bm}

\newtheorem{theorem}{Theorem}
\newtheorem{lemma}{Lemma}
\newtheorem{corollary}{Corollary}
\newtheorem{proposition}{Proposition}

\newtheorem{assumption}{Assumption}
\newtheorem{definition}{Definition}
\newtheorem{remark}{Remark}

\usepackage{float}

\usepackage{algpseudocode,algorithm}
\algnewcommand\algorithmicinput{\textbf{Input}:}
\algnewcommand\algorithmicoutput{\textbf{Output}:}
\algnewcommand\INPUT{\item[\algorithmicinput]}
\algnewcommand\OUTPUT{\item[\algorithmicoutput]}

\usepackage{graphicx}
\usepackage[caption=false,labelformat=simple]{subfig}

\graphicspath{}

\usepackage{booktabs}
\usepackage{multirow}

\newsavebox{\simtablebox}
\newcommand{\simtablefit}[1]{%
  \sbox{\simtablebox}{#1}%
  \typeout{SIMTABLEWIDTH natural=\the\wd\simtablebox; target=\the\textwidth}%
  \ifdim\wd\simtablebox>\textwidth
    \resizebox{\textwidth}{!}{\usebox{\simtablebox}}%
  \else
    \usebox{\simtablebox}%
  \fi
}

\usepackage{enumerate}
\usepackage{paralist}

\newcommand*{\affaddr}[1]{#1}
\newcommand*{\affmark}[1][*]{\textsuperscript{#1}}

\newcommand{\E}{\mathbb{E}}
\newcommand{\Prob}{\mathbb{P}}
\newcommand{\R}{\mathbb{R}}

\newcommand{\Pn}{\mathbb{P}_n}

\newcommand{\ind}{\mathbf{1}}
\newcommand{\norm}[1]{\left\lVert #1 \right\rVert}

\newcommand{\indep}{\perp\!\!\!\perp}
\newcommand{\NDE}{\mathrm{NDE}}
\newcommand{\NIE}{\mathrm{NIE}}
\newcommand{\TE}{\mathrm{TE}}
\newcommand{\diag}{\mathrm{diag}}
\newcommand{\sign}{\mathrm{sign}}
\newcommand{\op}{o_p}

\usepackage{xcolor}

\global\long\def\bZ{\mathbf{Z}}
\global\long\def\bz{\mathbf{z}}
\global\long\def\ba{\mathbf{a}}
\global\long\def\bb{\mathbf{b}}
\global\long\def\bD{\mathbf{D}}
\global\long\def\bA{\mathbf{A}}
\global\long\def\bB{\mathbf{B}}

\global\long\def\bG{\mathbf{G}}
\global\long\def\bH{\mathbf{H}}

\global\long\def\bW{\mathbf{W}}

\global\long\def\bU{\mathbf{U}}
\global\long\def\bu{\mathbf{u}}
\global\long\def\bV{\mathbf{V}}
\global\long\def\bv{\mathbf{v}}

\global\long\def\bR{\mathbf{R}}
\global\long\def\bM{\mathbf{M}}
\global\long\def\bfm{\mathbf{m}}
\global\long\def\bS{\mathbf{S}}
\global\long\def\bs{\mathbf{s}}
\global\long\def\bT{\mathbf{T}}
\global\long\def\bt{\mathbf{t}}
\global\long\def\bc{\mathbf{c}}

\global\long\def\be{\mathbf{e}}
\global\long\def\bg{\mathbf{g}}
\global\long\def\matzero{\boldsymbol{\mathrm{0}}}
\global\long\def\matI{\boldsymbol{\mathrm{I}}}
\global\long\def\matone{\boldsymbol{\mathrm{1}}}

\global\long\def\balpha{\boldsymbol{\alpha}}
\global\long\def\bepsilon{\boldsymbol{\epsilon}}

\global\long\def\bSigma{\boldsymbol{\Sigma}}
\global\long\def\bOmega{\boldsymbol{\Omega}}

\global\long\def\bdelta{\boldsymbol{\delta}}
\global\long\def\bdeta{\boldsymbol{\eta}}
\global\long\def\btheta{\boldsymbol{\theta}}

\global\long\def\bpsi{\boldsymbol{\psi}}
\global\long\def\bchi{\boldsymbol{\chi}}

\global\long\def\bmu{\boldsymbol{\mu}}

\title{Semiparametric causal mediation analysis with high-dimensional mediators: A sparse single-index model approach}

\author{%
    Yi Zhao\affmark[1], Siyuan Ping\affmark[1], Wanzhu Tu\affmark[1], and Fan Li\affmark[2]
    \\
    \affaddr{\affmark[1]Department of Biostatistics and Health Data Science, \\ Indiana University School of Medicine} \\
    \affaddr{\affmark[2]Department of Biostatistics, Yale University} \\
}

\date{}

\begin{document}

\maketitle

\thispagestyle{empty}

\begin{abstract}
Causal mediation analysis has been widely used in biomedical research to understand underlying mechanisms. With multiple candidate mediators, identification of the causal effect and related mediators becomes challenging. In this study, a semiparametric causal mediation framework is proposed for studies with multiple candidate mediators. A single-index model with a smooth link is considered in the outcome model to study the nonlinear joint mediation effect. Causal identification assumptions are discussed under the proposed framework. Multiply robust estimators are introduced for both full-ratio and fixed-index mediation scores. Sparsity in the mediator index is imposed that enables mediator selection and better interpretation. An index-score augmented procedure is introduced for inference. Asymptotic properties of the proposed estimators are discussed. Simulation studies demonstrate the performance of the proposed framework. We apply the method to the Anti-Amyloid Treatment in Asymptomatic Alzheimer's Disease (A4) study, a randomized trial in preclinical Alzheimer's disease. Because the hypothesized therapeutic pathway runs through neurodegeneration as a whole rather than through any single brain region, we target the joint mediation effect through regional MRI volume changes. The estimated mediation effect of brain atrophy on cognitive decline is not statistically significant, but the brain regions identified are consistent with existing knowledge.
\end{abstract}


\newpage

\section{Introduction}
\label{sec:intro}

Mediation analysis seeks to understand the mechanisms through which an exposure affects an outcome. In biomedical studies with many candidate mediators, a central question is how much of the exposure effect is transmitted through the mediator collection as a whole. This question is especially relevant when the mediating mechanism is a process that no single variable fully represents. Several measurements are then needed to capture it, and their individual mediation signals are often too weak to discern. We therefore focus on the natural indirect effect (NIE) transmitted through the mediators jointly, rather than through individual components, and on the corresponding natural direct effect (NDE). Mediator-specific natural effects generally require assumptions about the causal structure among the mediators~\citep{vanderweele2014mediation}, and such assumptions are rarely defensible when the mediators measure the same process. In such settings, the joint effect is not merely a convenient summary but the effect that the design can identify. It addresses a question complementary to, and often logically prior to, the search for mediator-specific or path-specific mechanisms. Since the classical coefficient-product framework of \citet{baron1986moderator}, mediation analysis has been extensively developed within the potential outcomes framework \citep{robins1992identifiability, pearl2001direct, imai2010identification, vanderweele2015explanation}. Identification of natural effects requires assumptions about mediator--outcome confounding and cross-world counterfactual relationships beyond exposure exchangeability \citep{andrews2021insights}.

Our study is motivated by the Anti-Amyloid Treatment in Asymptomatic Alzheimer's Disease (A4) study, a randomized, placebo-controlled trial of solanezumab in cognitively unimpaired older adults with elevated brain amyloid~\citep{sperling2014a4, sperling2023trial}. The promise of anti-amyloid therapy rests on a mechanistic chain: lowering amyloid is expected to slow neurodegeneration, which in turn should slow cognitive decline. Within this biological framework, neurodegeneration is treated as a single construct, measured in practice by atrophy distributed across medial temporal, lateral temporal and parietal regions~\citep{jack2018nia}. The clinically relevant question is therefore whether treatment affects cognition through this process as a whole, not through any particular brain region. Its primary outcome was change in the Preclinical Alzheimer Cognitive Composite (PACC), which combines measures of memory, executive function and global cognition~\citep{donohue2014preclinical}. The outcome is itself a composite, adopted because no single test captures early cognitive decline reliably. Neuroimaging research has followed the same logic on the structural side, summarizing regional measurements in composite atrophy signatures~\citep{dickerson2009cortical, davatzikos2009longitudinal, jack2015different}.

Although the A4 trial did not demonstrate slowing of cognitive decline, a null total effect does not preclude asking whether treatment altered the neurodegenerative pathway and how that pathway relates to cognition. Imaging studies have reported associations of distributed atrophy and joint structural changes with cognitive decline~\citep{lee2026atrophy, menze2026joint}. We consider 43 regional MRI volume-change rates as candidate mediators of the treatment--PACC relationship. Our primary question concerns the joint natural indirect effect through this collection, rather than the causal contribution of any individual region. Existing atrophy signatures, however, were constructed to distinguish patients from controls, and their fixed weights need not reflect the structural changes most relevant to cognition. The distinction matters for anti-amyloid agents, which have been associated with accelerated brain volume loss that may reflect amyloid removal rather than neuronal injury~\citep{alves2023accelerated}. A representation weighted by its relationship with the outcome offers a possible alternative: it aggregates information across correlated regions, may discount volume change unrelated to cognition, and suggests structural patterns for subsequent investigation. Analyses are conducted separately among APOE-$\varepsilon4$ carriers and non-carriers, where APOE-$\varepsilon4$ carriage was associated with elevated amyloid in A4 screening data~\citep{sperling2020association}.

Estimating mediation through a collection of correlated measurements presents several challenges. Separate analyses of individual mediators generally do not recover their joint natural indirect effect \citep{imai2013identification, vanderweele2014mediation}. High-dimensional mediation methods have addressed mediator selection and coefficient-product estimation \citep{zhang2016estimating, song2020bayesian, zhao2022pathway}, as well as inference on aggregate indirect effects \citep{zhou2020estimation, jones2025causal}. These developments are particularly relevant to genomic, epigenomic, and neuroimaging studies, where many correlated measurements may contribute to a common process. Under linear structural models, aggregate mediation can be represented by a sum of coefficient products. When the joint mediator-outcome relationship is nonlinear, however, constant coefficient products may not adequately characterize the indirect effect. This motivates methods that combine a parsimonious representation of the mediator collection with flexible outcome modeling and inference on joint natural effects.

Dimension reduction offers another approach to correlated mediator measurements. Spectral transformations and sparse principal components have been used to construct mediation measures and interpretable mediator features~\citep{huang2016hypothesis, zhao2020sparse}. Latent mediating variables have also been introduced, with biomarkers serving as their indicators~\citep{derkach2019high}. For multiple exposures and mediators, principal component mediation analysis jointly estimates exposure and mediator projections under linear structural equation models~\citep{zhao2024mediation}. For covariance-based transformations, however, a direction explaining mediator variation need not characterize its relationship with the outcome. This consideration motivates dimension reduction that directly incorporates the joint mediator--outcome relationship.

Nonlinear modeling and data-adaptive nuisance estimation have also been considered in mediation analysis. \citet{loh2022nonlinear} developed nonlinear analysis with high-dimensional mediators using mediator-distribution interventions. These interventional effects differ from the natural effects considered here~\citep{vansteelandt2017interventional}. For natural effects, efficient influence functions and multiply robust estimators have been derived~\citep{tchetgen2012semiparametric}. Consistency is obtained under specified combinations of nuisance models, with additional conditions required for efficiency. Orthogonal scores and cross-fitting allow data-adaptive nuisance estimation~\citep{chernozhukov2018double}, with applications to mediation with high-dimensional covariates~\citep{farbmacher2022causal} and mediation across principal strata in the presence of a post-treatment event~\citep{cheng2026identification}. \citet{liu2024general} proposed general cross-fitted one-step estimators for several mediation estimands with multivariate mediators using sequential regressions and Riesz learning. The present study focuses on sparse single-index modeling and scalar reweighting under index sufficiency.


A single-index outcome model represents the joint mediator contribution through a linear projection, $\btheta^\top\bfm$, and an unknown smooth function. The target estimands remain the natural direct and indirect effects defined using the full potential mediator vector \(\bM(x)\). The index enters the outcome model and estimation procedure, providing a lower-dimensional representation of the mediator-outcome relationship. Under conditional-mean index sufficiency and the causal identification assumptions, the full-vector mediation formula admits an equivalent representation involving the scalar index. This structure can reduce the dimensional burden of nuisance estimation and may help aggregate information across correlated measurements whose individual contributions are difficult to detect.  In the A4 application, the index can be viewed as an atrophy signature whose weights are learned from the relationship between regional change and cognition, after adjustment for treatment and baseline covariates. Sparsity further yields a parsimonious set of variables contributing to the outcome index. Semiparametric and sparse single-index methods provide tools for estimating this representation \citep{ichimura1993semiparametric, hardle1993optimalbandwidth, alquier2013sparse, radchenko2015high, neykov2016l1}. Incorporating these tools into natural effect estimation additionally requires accounting for estimation of the index and its relationship with the mediation score.

In this study, we develop a semiparametric framework for joint natural direct and indirect effects with multiple, potentially high-dimensional mediators, combining a sparse single-index outcome model with influence-function-based estimation. It makes three methodological contributions. First, the unknown smooth link accommodates a nonlinear joint mediator--outcome relationship, while sparsity identifies the variables contributing to the index. Second, we develop full-ratio and index-ratio formulations for multiply robust estimation. The former uses the joint mediator distribution and the latter a scalar index distribution under conditional-mean index sufficiency, with both targeting the same natural effects. Third, we develop a cross-fitted index-score augmentation that accounts for the first-order contribution of index estimation and its covariance with the mediation score. We establish selection and inference results under stated regularity conditions and evaluate estimation, coverage, and selection in simulations. An application to the A4 study illustrates joint mediation through regional MRI changes and exploratory characterization of the associated structural pattern.

The rest of the manuscript is organized as follows. Section~\ref{sec:setup} presents the causal definitions, assumptions, identification and proposed model. The influence functions and multiply robust estimators are discussed in Section~\ref{sec:semiparam}. The estimation procedure and asymptotic properties are presented in Sections~\ref{sec:estimation} and~\ref{sec:asymptotic}. Via simulation studies in Section~\ref{sec:simulation}, the performance of the proposed framework is evaluated and compared with competing approaches. Section~\ref{sec:a4} applies the framework to the A4 study. Section~\ref{sec:discussion} summarizes the proposed framework and discusses limitations and future directions.


\section{Causal definitions, assumptions, and identification}
\label{sec:setup}

Following the potential outcomes framework~\citep{rubin1974estimating}, let $\bM(x)\in\mathcal{M}\subseteq\mathbb{R}^{p}$ denote the potential outcome of mediator vector under treatment $X = x$, for $x\in\mathcal{X}=\{0,1\}$. Let $Y(x, \bfm)\in\mathcal{Y}\subseteq\mathbb{R}$ denote the outcome under treatment $X = x$ with the mediators set to $\bM = \bfm$. For treatment $x$ and the mediator value under $x^*$, the composite potential outcome is $Y(x, \bM(x^*))$. 
For a single unit, let $\mathcal{O} = (\bZ, X, \bM, Y)$ denote the observed data, where $\bZ = (Z_1, \ldots, Z_q)^\top \in \R^q$ denotes pre-treatment covariates (confounders) measured before exposure, $X \in \{0, 1\}$ denotes the binary treatment/exposure, $\bM = (M_1, \ldots, M_p)^\top \in \mathcal{M}$ is the vector of mediators measured post-treatment, and $Y \in \mathcal{Y}$ denotes continuous outcome. Considering a study of $n$ units, let $\mathcal{O}_i = (\bZ_i, X_i, \bM_i, Y_i)$ denote the observation of unit $i$, $i = 1, \ldots, n$.

Following \citet{pearl2001direct} and \citet{robins2003semantics}, the \emph{natural direct effect} (NDE) and \emph{natural indirect effect} (NIE) for treatment $x = 1$ relative to $x^* = 0$ are defined as
\begin{align}
  \NDE &= \E\!\left\{Y\!\left(1, \bM(0)\right) - Y\!\left(0, \bM(0)\right)\right\},
  \label{eq:NDE}\\
  \NIE &= \E\!\left\{Y\!\left(1, \bM(1)\right) - Y\!\left(1, \bM(0)\right)\right\}.
  \label{eq:NIE}
\end{align}
The total effect decomposes as $\TE = \NDE + \NIE$, where $\TE = \E\{Y(1, \bM(1)) - Y(0, \bM(0))\}$.
Let $\zeta_0=\E\{Y(1,\bM(0))\}$ denote the cross-world mean, and let $\delta_x=\E\{Y(x,\bM(x))\}$ for $x\in\{0,1\}$. Then $\NDE=\zeta_0-\delta_0$ and $\NIE=\delta_1-\zeta_0$. 

We impose the following assumptions for identification.
\begin{assumption}[Stable unit treatment value assumption (SUTVA)]
  \label{ass:sutva}
  Potential mediators and outcomes are well defined under the specified exposure and joint mediator interventions, with no interference between units and no relevant hidden version of these interventions.
\end{assumption}
\begin{assumption}[Consistency]
  \label{ass:consistency}
  For each unit $i$, $\bM_i = \bM(X_i)$ and $Y_i = Y(X_i, \bM_i)$ almost surely.
\end{assumption}

\begin{assumption}[No Unmeasured Exposure-Outcome Confounding]
  \label{ass:conf1}
  The joint collection of potential outcomes and potential mediators is independent of exposure given baseline covariates,
  \[
    \{Y(x,\bfm),\bM(x^*):x,x^*\in\{0,1\},\bfm\in\mathcal{M}\}
       \indep X\mid\bZ.
  \]
  In particular, $Y(x,\bfm)\indep X\mid\bZ$ and $Y(x,\bM(x))\indep X\mid\bZ$.
\end{assumption}

\begin{assumption}[No Unmeasured Exposure-Mediator Confounding]
  \label{ass:conf_m}
  $\bM(x) \indep X \mid \bZ$ for all $x \in \{0,1\}$.
\end{assumption}

\begin{assumption}[No Unmeasured Mediator-Outcome Confounding]
  \label{ass:conf2}
  $Y(x, \bfm) \indep \bM \mid X=x, \bZ$ for all $x \in \{0,1\}$ and $\bfm \in \mathcal{M}$.
\end{assumption}

\begin{assumption}[Cross-world Independence]
  \label{ass:cross}
  $Y(x, \bfm) \indep \bM(x^*) \mid \bZ$ for all $x, x^* \in \{0,1\}$ and $\bfm \in \mathcal{M}$.
\end{assumption}

\begin{assumption}[Positivity]
  \label{ass:positivity}
  There exists $\delta > 0$ such that almost surely,
  \begin{enumerate}
    \item[\textup{(i)}] $\delta \le \pi(\bZ) \le 1 - \delta$;
    \item[\textup{(ii)}] $f_{\bM|X,\bZ}(\bfm \mid 1, \bZ) > 0$ whenever $f_{\bM|X,\bZ}(\bfm \mid 0, \bZ) > 0$,
  \end{enumerate}
  where $\pi(\bZ)=\mathbb{P}(X=1\mid \bZ)$ is the propensity score.
\end{assumption}

The imposed assumptions are natural extensions of those under the scenario of a single mediator~\citep{imai2010identification} by treating the $p$ mediators as a joint mediator. Assumptions~\ref{ass:consistency} and~\ref{ass:positivity} are standard in causal inference. Assumptions~\ref{ass:conf1}--\ref{ass:conf2} are the sequential ignorability assumptions in causal mediation analysis. Together with the cross-world independence assumption, these are not testable. It should be noted that the assumptions are imposed on the full vector $\bM$. Marginal ignorability for each $M_{j}$ does not imply joint ignorability for $\bM$.

The sequential ignorability assumptions, together with consistency, identify $\E\{Y(1,\bfm)\mid\bZ=\bz\}$ and $f_{\bM(0)\mid\bZ=\bz}(\bfm)$. Assumption~\ref{ass:cross} gives a sufficient cross-world condition for the identification of $\E\{Y(1,\bM(0))\mid \bZ=\bz\}$~\citep{pearl2001direct, robins2003semantics}. As discussed by \citet{andrews2021insights}, observed data cannot verify independence between quantities from distinct counterfactual worlds. Cross-world violations may occur even without intermediate or exposure-induced mediator-outcome confounding, depending on the causal model. When full cross-world independence is of concern, other approaches include additional model assumptions, bounds, sensitivity analysis, and alternative estimands~\citep{vanderweele2009conceptual, tchetgen2012semiparametric, vansteelandt2017interventional, andrews2021insights}.

\subsection{Nonparametric Identification}
\label{sub:nonparametric_id}

For nonparametric identification of the causal effects, the following nuisance functions are defined.
\begin{align}
  \pi(\bz) &= \Prob(X = 1 \mid \bZ = \bz),
  \label{eq:ps} \\
  \mu(x, \bfm, \bz) &= \E\!\left(Y \mid X = x, \bM=\bfm, \bZ = \bz\right),
  \label{eq:or_full} \\
  r_{\bM}(\bfm, \bz) &=
  \frac{f_{\bM|X,\bZ}(\bfm \mid 0, \bz)}
       {f_{\bM|X,\bZ}(\bfm \mid 1, \bz)},
  \label{eq:full_dr} \\
  \phi(x^*, \bz) &= \E\!\left\{\mu(1, \bM, \bz) \mid X = x^*, \bZ = \bz\right\}.
  \label{eq:blended_full}
\end{align}
The propensity score $\pi(\bz)$ describes the exposure mechanism. The outcome regression $\mu(x,\bfm,\bz)$ is conditional on the full mediator vector. The ratio $r_{\bM}(\bfm,\bz)$ reweights the treated mediator distribution to the control distribution. The function $\phi(x^*,\bz)$ averages the treated outcome regression over the mediator distribution under exposure $x^*$.

The cross-world expectation is identified by the mediation g-formula \citep{pearl2001direct, robins2003semantics, tchetgen2012semiparametric}.
\begin{equation}
  \zeta_0 = \E\!\left\{\phi(0,\bZ)\right\}
           = \E\!\left\{\int \mu(1,\bfm,\bZ)\,
             f_{\bM|X,\bZ}(\bfm\mid 0, \bZ)\,\mathrm{d}\bfm\right\}.
  \label{eq:id_formula}
\end{equation}
Consistency and mediator-outcome ignorability connect $\mu(1,\bfm,\bz)$ to the expectation of treated potential outcome. Exposure ignorability allows averaging this mean over the population conditional on $\bZ$. Exposure-mediator ignorability and consistency identify the law of $\bM(0)$ from the control arm. Cross-world independence allows integrating the treated potential-outcome mean against this law. Positivity ensures that the treated outcome regression is defined over the required control mediator support. The outer expectation averages over the baseline covariates.

The single-treatment means satisfy $\delta_x=\E\{\E(Y\mid X=x,\bZ)\}$ for $x\in\{0,1\}$ under consistency and joint exposure ignorability. Identification does not require the mediator-outcome ignorability. Consequently, $\NDE=\E\{\phi(0,\bZ)\}-\delta_0$ and $\NIE=\delta_1-\E\{\phi(0,\bZ)\}$. The full-ratio influence functions in Section~\ref{sec:semiparam} are derived from this unrestricted observed-data representation.

When Assumption~\ref{ass:cross} is not valid, an interventional indirect effect provides an alternative estimand~\citep{vansteelandt2017interventional, andrews2021insights}. Let $\bG_x$ be drawn from $f_{\bM\mid X=x,\bZ}$ independently of individual potential outcomes given $\bZ$. The effect is $\E\{Y(1,\bG_1)-Y(1,\bG_0)\}$. The same g-formula identifies this population intervention contrast without the natural-effect cross-world condition. 

\subsection{The Proposed Semiparametric Model}
\label{sub:proposed_model}

A \emph{single-index} model is proposed for the potential outcome, together with an additive model for the potential mediators.
\begin{align}
 Y(x,\bfm)&=\gamma_0+\gamma x+f(\btheta^\top\bfm)+g_2(\bZ)+\eta,
 \nonumber\\
 \bM(x)&=\balpha_0+\balpha x+\bg_1(\bZ)+\bepsilon,
 \qquad x\in\{0,1\},\quad \bfm\in\mathcal{M}.
 \label{eq:potential_structural}
\end{align}
Here, $f:\R\to\R$, $g_2:\R^q\to\R$, and the components $g_{1j}:\R^q\to\R$ of $\bg_1$ are unknown smooth functions, and $\eta$ and $\bepsilon=(\epsilon_1,\ldots,\epsilon_p)^\top$ are model errors with $\E(\eta\mid\bZ)=0$ and $\E(\bepsilon\mid\bZ)=\matzero$.
Additionally, it is assumed that
\begin{equation}
 X\indep(\bepsilon,\eta)\mid\bZ,
 \label{eq:joint_error_exposure}
\end{equation}
and the mediator and outcome errors also satisfy
\begin{equation}
 \bepsilon\indep\eta\mid\bZ.
 \label{eq:error_indep}
\end{equation}
Together with \eqref{eq:potential_structural}, these conditions imply the sequentional ignorability and cross-world independence. They provide a sufficient structural formulation of the identification assumptions, while allowing dependence among the components of $\bepsilon$. Joint independence in~\eqref{eq:joint_error_exposure} is stronger than separate independence of $X$ from each error. A weaker condition is discussed in Section~\ref{sub:model_id}.

Under consistency (Assumption~\ref{ass:consistency}), substituting $x=X$ and $\bfm=\bM$ in \eqref{eq:potential_structural} gives the observed outcome model
\begin{equation}
  Y = \gamma_0 + \gamma X + f\!\left(\btheta^\top \bM\right) + g_2(\bZ) + \eta.
  \label{eq:outcome}
\end{equation}
The centered errors and conditions~\eqref{eq:joint_error_exposure}--\eqref{eq:error_indep} additionally give $\E(\eta\mid X,\bM,\bZ)=0$. Hence, the conditional outcome mean has the proposed single-index form.
Through the unknown $f$, this index allows nonlinear mediated effects. Sparsity in $\btheta$ identifies the subset of variables entering the joint outcome index. A variable may have a nonzero index weight without its distribution changing under the exposure, so index selection is distinguished from mediator-specific causal selection. 

Consistency also gives the observed mediator models
\begin{align}
  M_j &= \alpha_{0j} + \alpha_j X + g_{1j}(\bZ) + \epsilon_j, \qquad
  j = 1, \ldots, p.
  \label{eq:mediator}
\end{align}
The centered mediator errors and~\eqref{eq:joint_error_exposure} give $\E(\bepsilon\mid X,\bZ)=\matzero$. The errors may be correlated across mediators. The corresponding model for the observed index is
\begin{equation}
  \btheta^\top \bM = \alpha_0 + \alpha X + g_1(\bZ) + \epsilon,
  \label{eq:index_mediator}
\end{equation}
where $\alpha_0 = \btheta^\top \balpha_0$, $\alpha = \btheta^\top \balpha$, $g_1 = \btheta^\top \bg_1$, and $\epsilon = \btheta^\top \bepsilon$.
To remove scale and sign ambiguity in the index, the following normalizations are imposed,
\begin{equation}
  \norm{\btheta}_2 = 1, \qquad \theta_{j_*}>0,\quad j_*=\min\{j:\theta_j\ne0\},
  \label{eq:normalization}
\end{equation}
Let $\Theta$ denote the set of oriented unit vectors. Under sparsity, let $\mathcal{S} = \{j : \theta_j \neq 0\}$ denote the active set, with $|\mathcal{S}| = s_0 \ll p$, and the true index direction is denoted by $\btheta_0$.

The functions $g_{1j}(\bZ)$ and $g_2(\bZ)$ adjust for pre-treatment covariates in the counterfactual and observed models. The functions enter nuisance estimation rather than the NDE and NIE influence functions directly. In Section~\ref{sec:semiparam}, multiply robust estimation uses either the full mediator functions $\{\mu(x,\bfm,\bz), r_{\bM}(\bfm,\bz), \phi(x^*,\bz)\}$ or the reduced index functions $\{\mu_{\btheta}(x,v,\bz), r_{\btheta}(v,\bz), \phi_{\btheta}(x^*,\bz)\}$. The models provide an interpretable index vector $\btheta$ and a basis for choosing between full-ratio and index-ratio nuisance estimation.

\paragraph{Mediator dependence.}

The mediators can be correlated and the error vector $\bepsilon$ may have an arbitrary covariance $\bSigma_\epsilon = \mathrm{Cov}(\bepsilon)$. The covariance working model additionally assumes $\mathrm{Cov}(\bepsilon\mid\bZ)=\bSigma_\epsilon$, which implies $\mathrm{Cov}(\bepsilon\mid X,\bZ)=\bSigma_\epsilon$ under \eqref{eq:joint_error_exposure}. Its estimation is described in Section~\ref{sub:stage2}. The full-ratio estimator uses the $p$-dimensional ratio $r_{\bM}(\bfm,\bz)$ and is most appropriate when $p$ is small or moderate relative to $n$. Under the single-index outcome model, the mediator contribution to the cross-world expectation depends on $V_{\btheta}=\btheta^\top\bM$. The index-ratio estimator therefore replaces $r_{\bM}$ with the scalar ratio $r_{\btheta}(v,\bz)$ in~\eqref{eq:index_dr}. The Gaussian working model in Section~\ref{sec:semiparam} further assumes $\bepsilon\mid\bZ\sim\mathcal{N}(\matzero,\bSigma_\epsilon)$. Together with \eqref{eq:joint_error_exposure}, this gives $\btheta^\top\bM \mid X,\bZ \sim \mathcal{N}\!\bigl(\btheta^\top(\balpha_0 + \balpha X + \bg_1(\bZ)),\; \btheta^\top\bSigma_\epsilon\btheta\bigr)$. Thus, Gaussian integration and the index ratio depend on the error covariance only through the projected variance $\btheta^\top\bSigma_\epsilon\btheta$. Gaussianity is a working distribution assumption and is not required for the mediation g-formula. Section~\ref{sec:simulation} compares the performance of these two ratios under various scenarios.

When mediators are highly correlated, individual index weights can be difficult to recover. Assumption~\ref{ass:irrep} in Appendix~\ref{sub:app_index_conditions} gives a sufficient local support condition for an ideal profile solution. Nonzero $\theta_j$ identify membership in the joint outcome index, rather than mediator-specific indirect effects. A selected variable may have $\alpha_j=0$, and nonlinear averaging may cancel an aggregate mediated effect. For comparison, the simplified HIMA implementation applies marginal screening followed by joint linear regression with a minimax concave penalty~\cite[MCP][]{zhang2010nearly}.

\paragraph{Reduced nuisance functions.}

Let $V_{\btheta}=\btheta^\top\bM$ and $\mu_{\btheta}(x,v,\bz)=\E(Y\mid X=x,V_{\btheta}=v,\bZ=\bz)$.

\begin{definition}[Conditional-mean index sufficiency]
\label{def:index_sufficiency}
Suppose $\E|Y|<\infty$. For a fixed $\btheta\in\Theta$, the index $V_{\btheta}$ is conditionally mean-sufficient for $Y$ with respect to $\bM$ given $(X,\bZ)$ if
\[
  \E(Y\mid X,\bM,\bZ)
  =\E(Y\mid X,V_{\btheta},\bZ)
  \quad\text{almost surely}.
\]
Equivalently, there exists a measurable function $h_{\btheta}$ such that $\mu(x,\bfm,\bz)=h_{\btheta}(x,\btheta^\top\bfm,\bz)$ for almost every $(x,\bfm,\bz)$ under the joint law of $(X,\bM,\bZ)$. 
\end{definition}

In the A4 application, index sufficiency means that, given treatment and baseline covariates, expected cognitive change depends on regional atrophy only through a single weighted pattern, so that regions matter to the extent that they contribute to the pattern. The condition would fail if distinct structural patterns drove different cognitive domains,  in which case a composite outcome such as the PACC would depend on more than one index.

Throughout, index sufficiency refers to this conditional mean restriction. Under the single-index outcome model, $\mu(x,\bfm,\bz)=\mu_{\btheta}(x,\btheta^\top\bfm,\bz)$. The index-ratio estimator uses the reduced nuisance functions,
\begin{align}
  \mu_{\btheta}(x, v, \bz) &= \E\!\left\{Y \mid X = x, V_{\btheta} = v, \bZ = \bz\right\},
  \label{eq:or_index} \\
  r_{\btheta}(v, \bz) &=
  \frac{f_{V_{\btheta}|X,\bZ}(v \mid 0, \bz)}
       {f_{V_{\btheta}|X,\bZ}(v \mid 1, \bz)},
  \label{eq:index_dr} \\
  \phi_{\btheta}(x^*, \bz) &= \E\!\left\{\mu_{\btheta}(1, V_{\btheta}, \bz) \mid X = x^*, \bZ = \bz\right\}
                = \int \mu_{\btheta}(1, v, \bz)\,
                f_{V_{\btheta}|X,\bZ}(v \mid x^*, \bz)\,\mathrm{d}v.
  \label{eq:blended}
\end{align}
The index ratio $r_{\btheta}(v,\bz)$ reweights the scalar index distribution. The conditional cross-world mean $\phi_{\btheta}(x^*,\bz)$ averages the treated index regression over the index law under exposure $x^*$.

\subsection{Identification under the Proposed Model}
\label{sub:model_id}

Under Assumptions~\ref{ass:sutva}--\ref{ass:positivity} and the single-index outcome regression induced by the counterfactual model, the mediation g-formula depends on $\bM$ only through $V_{\btheta}=\btheta^\top\bM$, giving the equivalent representation
\[
  \zeta_0 = \E\!\left\{\int \mu_{\btheta}(1,v,\bZ)\, f_{V_{\btheta}|X,\bZ}(v\mid 0,\bZ)\,\mathrm{d}v\right\},
\]
with the one-dimensional density-ratio relation
\[
  f_{V_{\btheta}|X=0,\bZ=\bz}(v) = r_{\btheta}(v,\bz)\cdot f_{V_{\btheta}|X=1,\bZ=\bz}(v).
\]
The full and reduced representations give \textsc{SIME-Full} and \textsc{SIME-Index}, respectively. 
Under the reduced representation, the resulting identification formulas for NDE and NIE are
\begin{align}
  \NDE &= \E\{\phi_{\btheta}(0, \bZ)\} - \E\{Y(0, \bM(0))\}, \label{eq:NDE_id}\\
  \NIE &= \E\{\phi_{\btheta}(1, \bZ)\} - \E\{\phi_{\btheta}(0, \bZ)\}.
  \label{eq:NIE_id}
\end{align}
In~\eqref{eq:potential_structural}, the common outcome error cancels from the natural-effect contrasts. The direct contrast reduces to the exposure coefficient, and the indirect contrast is a difference of the link evaluated at the two potential mediator indices. The identification assumptions connect these counterfactual contrasts to the observed data. Under model~\eqref{eq:outcome}, $\mu(1,\bfm,\bz) = \gamma_0 + \gamma + f(\btheta^\top\bfm) + g_2(\bz)$. Substitution into \eqref{eq:NDE_id}--\eqref{eq:NIE_id} gives
\begin{align}
  \NDE &= \gamma,
  \label{eq:NDE_param}\\
  \NIE &= \E\!\left\{\phi_{\btheta}(1,\bZ) - \phi_{\btheta}(0,\bZ)\right\}
        = \E\!\left\{\E\!\left\{f(\btheta^\top\bM)\mid X=1,\bZ\right\}
          - \E\!\left\{f(\btheta^\top\bM)\mid X=0,\bZ\right\}\right\}.
  \label{eq:NIE_semi}
\end{align}
Section~\ref{sub:app_identifySI} provides a detailed discussion of the identification of the causal effects.

Section~\ref{sec:semiparam} distinguishes efficiency under two models. The full-ratio influence function is efficient under the nonparametric law of $(\bZ,X,\bM,Y)$. For a fixed index satisfying the single-index assumption, the index-ratio influence function is efficient under the reduced law of $(\bZ,X,V_{\btheta},Y)$. Models~\eqref{eq:outcome}--\eqref{eq:mediator} provide a low-dimensional way to estimate $\mu_{\btheta}$, $r_{\btheta}$, and $\phi_{\btheta}$. The full-ratio approach remains available when the full mediator density ratio can be estimated stably.

\paragraph{Connection to multiply robust estimation.}

The counterfactual models induce nuisance functions for estimation. The influence-function estimators do not require all working nuisance models to be correct. In the full model, $\phi(0,\bz)$ is determined by $\mu(1,\bfm,\bz)$ and the mediator distribution under control. It can also be estimated by regressing the treated-mediator outcome regression on $\bZ$ in the control arm. Similarly, $\phi_{\btheta}(0,\bz)$ is determined by $\mu_{\btheta}$ and the index distribution under control, or can be estimated by regressing the treated-index outcome regression on $\bZ$ in the control arm. These relationships support the \emph{multiply robust} estimators under the nuisance-model combinations given in Section~\ref{sec:semiparam}~\citep{tchetgen2012semiparametric}. The robustness concerns nuisance estimation under valid causal identification assumptions. The index-ratio formulation additionally requires a sufficient index. The NDE and NIE estimators also require consistent estimation of the corresponding single-treatment means. Cross-fitting allows data-adaptive nuisance estimation without Donsker restrictions under the stated convergence and moment conditions~\citep{cheng2026identification}. For a learned index, additional score and remainder conditions in Appendix~\ref{sub:app_causal_asym} are required.

\paragraph{Observed regression simplification.}

Under model~\eqref{eq:outcome}, $\mu(x,\bfm,\bz) = \gamma_0 + \gamma x + f(\btheta^\top\bfm) + g_2(\bz)$ depends on $\bfm$ through $\btheta^\top\bfm$. Instead of a $p$-dimensional regression, nuisance estimation uses the univariate link $f$ and parameters $(\gamma_0,\gamma,\btheta)$. The g-formula in~\eqref{eq:id_formula} then requires integration only over $\btheta^\top\bM$, as described in Section~\ref{sec:estimation}. This avoids nonparametric regression over high-dimensional $\bM$, while imposing the robustness target to the reduced model. If the outcome depends on features of $\bM$ beyond $\btheta^\top\bM$, the index-ratio estimator generally loses the unconditioned nonparametric efficiency interpretation. The full-ratio estimator retains unrestricted robustness but requires a $p$-dimensional density ratio.

\paragraph{A weaker error condition for natural effects.}
Within the counterfactual model~\eqref{eq:potential_structural}, evaluated at the true direction $\btheta_0$, a weaker structural model replaces \eqref{eq:error_indep} by
\begin{equation}
 \E(\eta\mid\btheta_0^\top\bepsilon,\bZ)=0.
 \label{eq:index_error_mean_indep}
\end{equation}
Under \eqref{eq:potential_structural} and~\eqref{eq:joint_error_exposure}, this condition gives both
\begin{equation}
 \E\{Y(x,\bfm)\mid\btheta_0^\top\bM(x^*)=v,\bZ=\bz\}
 =\E\{Y(x,\bfm)\mid\bZ=\bz\}
 \label{eq:index_mean_indep}
\end{equation}
for all $x,x^*\in\{0,1\}$, $\bfm\in\mathcal{M}$, and index values on the conditional support, and the index regression
\[
 \E(Y\mid X=x,V_{\btheta_0}=v,\bZ=\bz)
 =\gamma_0+\gamma x+f(v)+g_2(\bz).
\]
$V_{\btheta_0}(x^*)=\btheta_0^\top\{\balpha_0+\balpha x^*+\bg_1(\bZ)\}+\btheta_0^\top\bepsilon$, and conditional averaging of the invariant outcome equation removes $\eta$. Thus the reduced mediation g-formula identifies the mean natural effects under these additional structural assumptions, consistency, and positivity, even if dependence remains between $\eta$ and components of $\bepsilon$ outside the index. This approach does not imply full mediator-outcome ignorability or the full-data regression condition $\E(Y\mid X,\bM,\bZ)=\mu_{\btheta_0}(X,V_{\btheta_0},\bZ)$. The influence function analysis below retains the stronger identification assumptions and index sufficiency. The observed single-index regression alone, or \eqref{eq:index_mean_indep} without the invariant counterfactual condition, does not justify this weaker approach.

The sensitivity analysis in Appendix~\ref{sub:app_sensitivity_model} assumes~\eqref{eq:potential_structural} and~\eqref{eq:joint_error_exposure} while relaxing~\eqref{eq:index_error_mean_indep} through aligned dependence between $\eta$ and $\btheta_0^\top\bepsilon$. Its discrepancy formulas concern that specific structural construction. Residual based calibration requires the variance adjustment given there.
For the interventional contrast defined in Section~\ref{sub:nonparametric_id}, the single-index outcome model gives $\E\{Y(1,\bG_1)-Y(1,\bG_0)\}=\E\{f(\btheta^\top\bG_1)-f(\btheta^\top\bG_0)\}$. This representation requires the sequential ignorability and positivity for the intervention contrast.


\section{Semiparametric Efficiency Theory}
\label{sec:semiparam}

\subsection{Full-Ratio and Index-Ratio Influence Functions}

Let $\mathcal{M}_{\mathrm{np}}$ denote the nonparametric law of $(\bZ,X,\bM,Y)$.

\begin{theorem}[Full-Ratio Influence Function]
  \label{thm:eif_full}
  Under Assumptions~\ref{ass:sutva}--\ref{ass:positivity}, with the score assumed to be square integrable, the efficient influence function for $\zeta_0 = \E\{Y(1, \bM(0))\}$ in $\mathcal{M}_{\mathrm{np}}$ is given by
  \begin{equation}
    \varphi_{\zeta_0}^{\mathrm{F}}(\mathcal{O}; P_0)
    = \frac{X}{\pi(\bZ)} r_{\bM}(\bM,\bZ)
      \left\{Y - \mu(1, \bM, \bZ)\right\}
    + \frac{1-X}{1-\pi(\bZ)}
      \left\{\mu(1, \bM, \bZ) - \phi(0, \bZ)\right\}
    + \phi(0, \bZ) - \zeta_0.
    \label{eq:eif_full}
  \end{equation}
  The semiparametric efficiency bound for $\sqrt{n}$-regular estimators of $\zeta_0$ in $\mathcal{M}_{\mathrm{np}}$ is $\mathrm{Var}\{\varphi_{\zeta_0}^{\mathrm{F}}(\mathcal{O}; P_0)\}$.
\end{theorem}

The estimator \textsc{SIME-Full} is constructed from the full-ratio influence function. The semiparametric robustness and efficiency properties are retained. This approach requires estimation of the $p$-dimensional mediator density ratio $r_{\bM}(\bfm,\bz)$.
For a fixed sufficient direction $\btheta=\btheta_0$, let $\mathcal M_{\mathrm{red}}(\btheta_0)$ denote the nonparametric model for the reduced observations $(\bZ,X,V_{\btheta_0},Y)$. Under index sufficiency, $\mu(1,\bM,\bZ)=\mu_{\btheta_0}(1,V_{\btheta_0},\bZ)$. The reduced mediation functional equals the full cross-world mean. The additive and Gaussian assumptions impose additional structural conditions. 

\begin{theorem}[Index-Ratio Influence Function]
  \label{thm:eif}
  Under Assumptions~\ref{ass:sutva}--\ref{ass:positivity} and index sufficiency, with the score assumed to be square integrable, the efficient influence function for the reduced mediation functional in $\mathcal M_{\mathrm{red}}(\btheta_0)$ with fixed and known $\btheta_0$ is given by
  \begin{equation}
    \varphi_{\zeta_0}(\mathcal{O}; P_0)
    = \frac{X}{\pi(\bZ)} r_{\btheta}(V_{\btheta},\bZ)
      \left\{Y - \mu_{\btheta}(1, V_{\btheta}, \bZ)\right\}
    + \frac{1-X}{1-\pi(\bZ)}
      \left\{\mu_{\btheta}(1, V_{\btheta}, \bZ) - \phi_{\btheta}(0, \bZ)\right\}
    + \phi_{\btheta}(0, \bZ) - \zeta_0.
    \label{eq:eif}
  \end{equation}
  The semiparametric efficiency bound for $\sqrt{n}$-regular estimators of $\zeta_0$ in this reduced model is $\mathrm{Var}(\varphi_{\zeta_0}(\mathcal{O}; P_0))$.
\end{theorem}

The estimator \textsc{SIME-Index} is constructed from the index-ratio influence function, which is efficient under the reduced law with a fixed and known index. It is generally not efficient in $\mathcal{M}_{\mathrm{np}}$. Estimation of the index direction introduces additional uncertainty, as discussed in Appendix~\ref{sub:app_causal_asym}, and the fixed-index efficiency bound does not apply. In high dimensions, the full-ratio estimator achieves nonparametric robustness but can be unstable for large $p$. The index-ratio estimator reduces the dimension under the single-index structure, with efficiency defined under the reduced law.

The efficient influence functions for the NDE and NIE are obtained by subtractions:
\[
  \varphi_{\NDE} = \varphi_{\zeta_0} - \varphi_{\delta_0}, \qquad \varphi_{\NIE} = \varphi_{\delta_1} - \varphi_{\zeta_0},
\]
where $\delta_0 = \E\{Y(0, \bM(0))\}$ and $\delta_1 = \E\{Y(1, \bM(1))\}$ are the standard doubly robust functionals with their corresponding efficient influence functions~\citep{bang2005doubly}.

In the following, write $\varphi_{\zeta_0,\mathrm F}=\varphi_{\zeta_0}^{\mathrm F}$ from \eqref{eq:eif_full} and $\varphi_{\zeta_0,\mathrm I}=\varphi_{\zeta_0}$ from \eqref{eq:eif}. Define $\varphi_{\NDE,a}=\varphi_{\zeta_0,a}-\varphi_{\delta_0}$ and $\varphi_{\NIE,a}=\varphi_{\delta_1}-\varphi_{\zeta_0,a}$ for $a\in\{\mathrm F,\mathrm I\}$. The subscripted index scores here treat $\btheta_0$ as fixed. The complete learned-index scores are denoted by $\Phi$ in \eqref{eq:joint_effect_if}.

\subsection{Multiply Robust Estimation}

The full-ratio influence function in \eqref{eq:eif_full} includes four nuisance components: the propensity score $\pi(\bz)$, the full outcome regression $\mu(1,\bfm,\bz)$, the full mediator density ratio $r_{\bM}(\bfm,\bz)$, and the conditional cross-world mean $\phi(0,\bz)$. A \emph{multiply robust} estimator is proposed, which is consistent when any of the following three full-model combinations is correctly specified:
\begin{itemize}
  \item[(F-a)] the full outcome regression $\mu(1,\bfm,\bz)$ and the control mediator density $f_{\bM\mid X,\bZ}(\bfm\mid0,\bz)$;
  \item[(F-b)] the full outcome regression $\mu(1,\bfm,\bz)$ and the propensity score $\pi(\bz)$;
  \item[(F-c)] the full mediator density ratio $r_{\bM}(\bfm,\bz)$ and the propensity score $\pi(\bz)$.
\end{itemize}
Under consistent estimation of all full nuisance functions and the rate, moment, and sample-reuse conditions stated below, the full-ratio estimator achieves the efficiency bound in $\mathcal M_{\mathrm{np}}$.

For the index-ratio influence function in \eqref{eq:eif}, the reduced nuisance components are given by:
\begin{enumerate}
  \item[\textup{(1)}] Propensity score: $\pi(\bz) = \Prob(X=1 \mid \bZ=\bz)$;
  \item[\textup{(2)}] Reduced outcome regression: $\mu_{\btheta}(1, v, \bz) = \E\{Y \mid X=1, V_{\btheta}=v, \bZ=\bz\}$;
  \item[\textup{(3)}] Index density ratio: $r_{\btheta}(v, \bz) = f_{V_{\btheta}|X,\bZ}(v \mid 0, \bz) / f_{V_{\btheta}|X,\bZ}(v \mid 1, \bz)$;
  \item[\textup{(4)}] Conditional cross-world mean: $\phi_{\btheta}(0, \bz) = \E\{\mu_{\btheta}(1, V_{\btheta}, \bz) \mid X=0, \bZ=\bz\}$.
\end{enumerate}
The corresponding \emph{multiply robust} estimator is consistent when any of the following three reduced-model combinations is correctly specified:
\begin{itemize}
  \item[(I-a)] the reduced outcome regression $\mu_{\btheta}(1, v, \bz)$ and the control index density $f_{V_{\btheta} \mid X,\bZ}(v \mid 0, \bz)$;
  \item[(I-b)] the reduced outcome regression $\mu_{\btheta}(1, v, \bz)$ and the propensity score $\pi(\bz)$;
  \item[(I-c)] the index density ratio $r_{\btheta}(v,\bz)$ and the propensity score $\pi(\bz)$.
\end{itemize}

\begin{theorem}[Efficiency at a Correct Nuisance Law]
  \label{thm:localeff}
  Under consistent estimation of all corresponding nuisance functions and the score, rate, and sample-reuse conditions in Appendices~\ref{sub:app_cov_asym}--\ref{sub:app_causal_asym}, \textsc{SIME-Full} constructed from \eqref{eq:eif_full} is locally efficient in $\mathcal{M}_{\mathrm{np}}$. For fixed and known $\btheta_0$, the corresponding estimator, \textsc{SIME-Index}, is locally efficient under the reduced observed-data law of $(\bZ,X,V_{\btheta_0},Y)$.
\end{theorem}

\subsection{Machine Learning Nuisance Estimation via Cross-Fitting}

\citet{cheng2026identification} showed that cross-fitting allows data-adaptive nuisance estimation in mediation influence functions and inference without Donsker-class restrictions. Following this, cross-fitting is applied to the full nuisance components $\nu_{\mathrm{F}}=(\pi,\mu,r_{\bM},\phi)$ or the reduced nuisance components $\nu_{\mathrm{I}}=(\pi,\mu_{\btheta},r_{\btheta},\phi_{\btheta})$. Let $\nu=(\nu_1,\ldots,\nu_4)$ denote either ordered tuple, and let $\hat\nu_j$ estimate its $j$th component, $j=1,\ldots,4$. It is assumed that these estimators satisfy
\begin{equation}
  \norm{\hat{\nu}_a - \nu_a}_2 \cdot \norm{\hat{\nu}_b - \nu_b}_2 = \op(n^{-1/2}),
  \label{eq:rate_condition}
\end{equation}
for component indices $(a,b)\in\{(1,2),(1,4),(2,3)\}$. Together with individual nuisance and score convergence and appropriate moment conditions, the full-ratio and fixed-index estimators have an asymptotically linear expansion. Contributions to the errors from covariance estimation are discussed in Appendix~\ref{sub:app_cov_asym}. For an estimated index, additional conditions in Appendix~\ref{sub:app_causal_asym} are required. Under the single-index models in \eqref{eq:outcome}--\eqref{eq:mediator}, the reduced nuisance components can be estimated through low-dimensional models, as described in Section~\ref{sec:estimation}. The full-ratio approach is used when the full mediator law can be stably estimated.

\subsection{Gaussian Working Models for the Nuisance Functions}

Under models~\eqref{eq:outcome}--\eqref{eq:mediator} with $\norm{\btheta}_2 = 1$, and assuming Gaussian mediators, $\bM \mid X, \bZ \sim \mathcal{N}(\balpha_0 + \balpha X + \bg_1(\bZ), \bSigma_\epsilon)$, the full mediator density ratio is given by
\begin{equation}
  r_{\bM}(\bfm,\bz) =
  \exp\!\left[
  -\frac{1}{2}\{\bfm-\bmu_0(\bz)\}^\top\bSigma_\epsilon^{-1}
  \{\bfm-\bmu_0(\bz)\}
  +\frac{1}{2}\{\bfm-\bmu_1(\bz)\}^\top\bSigma_\epsilon^{-1}
  \{\bfm-\bmu_1(\bz)\}
  \right],
  \label{eq:lr_full}
\end{equation}
where $\bmu_x(\bz)=\balpha_0+\balpha x+\bg_1(\bz)$. Estimation of this ratio requires a high-dimensional covariance matrix. Thus, the approach may be effective in small or moderate dimensions but unstable when $p$ is large relative to $n$. For the projected index, the density ratio is given by
\begin{equation}
  r_{\btheta}(v, \bz) =
  \exp\!\left\{-\frac{(v - m_0(\bz))^2}{2\sigma_{\btheta}^2}
    + \frac{(v - m_1(\bz))^2}{2\sigma_{\btheta}^2}\right\},
  \label{eq:lr}
\end{equation}
where $m_x(\bz) = \btheta^\top\{\balpha_0 + \balpha x + \bg_1(\bz)\}$ and $\sigma_{\btheta}^2=\btheta^\top\bSigma_\epsilon\btheta$. The conditional cross-world mean is obtained from the one-dimensional integral
\begin{equation}
  \phi_{\btheta}(0, \bz) = \gamma_0 + \gamma +
  \int f(v) \,
  \mathcal{N}(v; m_0(\bz), \sigma_{\btheta}^2) \, \mathrm{d}v
  + g_2(\bz).
  \label{eq:phi_si}
\end{equation}
These Gaussian expressions are used as working models for nuisance estimation. Deriving the efficient influence function requires a separate analysis.

\section{Estimation}
\label{sec:estimation}

\begin{sloppypar}
An estimation procedure with four stages is proposed:
\begin{inparaenum}[(1)]
  \item training-fold index estimation,
  \item training-fold nuisance estimation,
  \item evaluation-fold effect estimation, and
  \item causal effect inference.
\end{inparaenum}
\end{sloppypar}

\subsection{Stage 1: Sparse Single-Index Estimation}
\label{sub:stage1}

The index is estimated on the training complement of each evaluation fold. Let $\{I_k\}_{k=1}^K$ denote a random partition of the samples, with training sample size $N_k=|I_k^c|$. Below, $n$ denotes the sample size of the current training complement.

Represent the functionals by a smooth spline, $f(v)=\bb(v)^\top\ba$ and $g_{2}(\bv)=\bb(\bv)^\top\bdelta$, where $\bb(v)$ is a spline basis and $\ba$ and $\bdelta$ are coefficients. For an oriented unit direction $\btheta\in\Theta$, define
\begin{equation}
 Q_n(\btheta)=\min_{\gamma_0,\gamma,\bdelta,\ba}
 \left[\frac1n\sum_{i=1}^n
 \{Y_i-\gamma_0-\gamma X_i-\bb(\btheta^\top\bM_i)^\top\ba
                         -\bb(\bZ_i)^\top\bdelta\}^2
       +\kappa_n\ba^\top\bR_n\ba\right],
 \label{eq:profile}
\end{equation}
where $\kappa_n\geq0$ is a smoothing parameter, $\bR_n$ is a specified nonnegative roughness-penalty matrix. A sparse index is estimated by
\begin{equation}
 \hat{\btheta}\in\underset{\btheta\in\Theta}{\arg\min} \left\{Q_n(\btheta)+\lambda_n\|\btheta\|_1\right\},
 \label{eq:penalized}
\end{equation}
where $\lambda_{n}\geq0$ is the tuning parameter.
Estimation alternates between fitting the spline and linear coefficients at a fixed direction and updating the direction under the unit-norm and orientation constraints. The smoothing and sparsity parameters are selected by cross-validation within the training complement. The resulting estimate for fold $k$ is denoted by $\hat{\btheta}^{(k)}$.

\subsection{Stage 2: Cross-Fitted Nuisance Estimation}
\label{sub:stage2}

For fold $k$, Stage~1 yields $\hat{\btheta}^{(k)}$ and training-fold link tuning. For $i\in I_k$, set $\hat V_i^{(k)}=\hat{\btheta}^{(k)\top}\bM_i$. All nuisance fits, including the profile-score projections used in Stage~3, use $I_k^c$ only.

\noindent\textbf{Propensity score.} Estimate $\hat\pi^{(k)}(\bz)$ on $I_k^c$, such as by logistic regression. Clipping may be implemented to avoid extreme values.

\noindent\textbf{Outcome regression.} With estimated index and coefficients from Stage~1, estimate
 \[
 \hat\mu_{\hat{\btheta}^{(k)}}^{(k)}(x,v,\bz)
   =\hat\gamma_0^{(k)}+\hat\gamma^{(k)}x+\hat f^{(k)}(v)
                                      +\hat g_2^{(k)}(\bz).
 \]

 \noindent\textbf{Mediator means and density ratio.} Estimate $\hat{\bmu}_x^{(k)}(\bz)=\hat{\balpha}_0^{(k)}+\hat{\balpha}^{(k)}x+\hat{\bg}_1^{(k)}(\bz)$ and set $\hat m_x^{(k)}(\bz)=\hat{\btheta}^{(k)\top}\hat{\bmu}_x^{(k)}(\bz)$. For the principal Gaussian index estimator, use the projected residual variance in~\eqref{eq:projected_residual_variance} and
 \[
 \hat r_{\hat{\btheta}^{(k)}}^{(k)}(v,\bz)
  =\exp\!\left\{-\frac{(v-\hat m_0^{(k)}(\bz))^2}{2\hat\sigma_{V,k}^2}
              +\frac{(v-\hat m_1^{(k)}(\bz))^2}{2\hat\sigma_{V,k}^2}\right\}.
 \]
 A non-Gaussian conditional index-law estimator can be substituted if the required nuisance and derivative conditions hold. For SIME-F, estimate the full mediator ratio using~\eqref{eq:lr_full} and a suitable training-fold covariance estimate.

 \noindent\textbf{Conditional cross-world mean and arm means.} Compute
 \[
 \hat\phi_{\hat{\btheta}^{(k)}}^{(k)}(0,\bz)
 =\hat\gamma_0^{(k)}+\hat\gamma^{(k)}
  +\int\hat f^{(k)}(v)\mathcal N(v;\hat m_0^{(k)}(\bz),\hat\sigma_{V,k}^2)\,\mathrm dv
  +\hat g_2^{(k)}(\bz).
 \]
Gaussian integration is performed analytically for a piecewise-linear approximation of the fitted link, also used for prediction; the formula and numerical-error conditions are given in Appendix~\ref{sub:app_gaussian_integration}. For the AIPW arm means, estimate $\xi_x(\bz)=\E(Y\mid X=x,\bZ=\bz)$ directly, or integrate the same outcome regression under the fitted arm-$x$ mediator law.

\paragraph{Projected residual variance for the principal index estimator.}
For a linear mediator mean design of rank $d_k$, define $\hat e_i^{(k)}=\hat{\btheta}^{(k)\top}\{\bM_i-\hat{\bmu}_{X_i}^{(k)}(\bZ_i)\}$, $i\in I_k^c$, and use
\begin{equation}
 \hat\sigma_{V,k}^2=\frac1{N_k-d_k}
                   \sum_{i\in I_k^c}(\hat e_i^{(k)})^2.
 \label{eq:projected_residual_variance}
\end{equation}
This degrees-of-freedom-adjusted form incorporates mediator correlation through the projected residuals and avoids estimating or inverting the full covariance matrix. The correction is unbiased for a fixed direction under the correctly specified Gaussian linear mediator model. 

\paragraph{Full covariance estimation.}
Full covariance estimates are constructed from training fold under the common conditional covariance assumption. Options include a working identity matrix, the sample covariance, and a shrinkage estimator, such as the Ledoit--Wolf (LW) estimator~\citep{ledoit2004well}. Estimation details are given in Appendix~\ref{sub:app_covariance_estimation}. Simulations in Section~\ref{sec:simulation} compare different options.
For SIME-F, the estimated Gaussian log ratio is
\begin{equation}
 \log\hat r_{\bM}^{(k)}(\bfm,\bz)
 =-\left\{\bfm-\frac{\hat{\bmu}_0^{(k)}(\bz)+\hat{\bmu}_1^{(k)}(\bz)}{2}\right\}^{\!\top}
       (\hat{\bSigma}_\epsilon^{(k)})^{-1}\hat{\balpha}^{(k)},
 \label{eq:lr_full_est}
\end{equation}
where the covariance estimate must be invertible. For a covariance-based SIME-I, replace the principal scalar estimate by $\hat{\btheta}^{(k)\top}\hat{\bSigma}_\epsilon^{(k)}\hat{\btheta}^{(k)}$ and positivity is required.

\subsection{Stage 3: Index-Augmented Causal Effect Estimators}
\label{sub:stage3}

The cross-fitted full-ratio benchmark for $\zeta_0=\E\{Y(1,\bM(0))\}$ is
\begin{equation}
 \begin{split}
 \hat\zeta_{0,\mathrm F}^{\mathrm{mr}}
 =\frac1n\sum_{k=1}^K\sum_{i\in I_k}
 \Big[&\hat\phi^{(k)}(0,\bZ_i)
 +\frac{X_i}{\hat\pi^{(k)}(\bZ_i)}\hat r_{\bM}^{(k)}(\bM_i,\bZ_i)
                  \{Y_i-\hat\mu^{(k)}(1,\bM_i,\bZ_i)\}\\
 &+\frac{1-X_i}{1-\hat\pi^{(k)}(\bZ_i)}
                  \{\hat\mu^{(k)}(1,\bM_i,\bZ_i)-\hat\phi^{(k)}(0,\bZ_i)\}\Big].
 \end{split}
 \label{eq:mr_estimator_full}
\end{equation}
Define the index score of unit $i$ by
\begin{align}
 \hat h_{\mathrm I}^{(k)}(\mathcal O_i)
 &=\hat\phi_{\hat{\btheta}^{(k)}}^{(k)}(0,\bZ_i)
 +\frac{X_i}{\hat\pi^{(k)}(\bZ_i)}
       \hat r_{\hat{\btheta}^{(k)}}^{(k)}(\hat V_i^{(k)},\bZ_i)
       \{Y_i-\hat\mu_{\hat{\btheta}^{(k)}}^{(k)}(1,\hat V_i^{(k)},\bZ_i)\}
 \nonumber\\
 &\quad
 +\frac{1-X_i}{1-\hat\pi^{(k)}(\bZ_i)}
 \{\hat\mu_{\hat{\btheta}^{(k)}}^{(k)}(1,\hat V_i^{(k)},\bZ_i)
                   -\hat\phi_{\hat{\btheta}^{(k)}}^{(k)}(0,\bZ_i)\}.
\end{align}
The index-ratio based estimates of $\zeta_{0}$ is then
\begin{equation}
  \hat\zeta_{0,\mathrm I}^{\mathrm{mr}}
 =\frac1n\sum_{k=1}^K\sum_{i\in I_k}\hat h_{\mathrm I}^{(k)}(\mathcal O_i).
 \label{eq:mr_estimator}
\end{equation}
The unaugmented NDE and NIE benchmarks for $a\in\{\mathrm F,\mathrm I\}$ are
\begin{equation}\label{eq:NDENIE_est}
  \widehat{\NDE}_a=\hat\zeta_{0,a}^{\mathrm{mr}}-\hat\delta_0^{\mathrm{dr}}, \qquad \widehat{\NIE}_a=\hat\delta_1^{\mathrm{dr}}-\hat\zeta_{0,a}^{\mathrm{mr}}. 
\end{equation}

Let $\xi_x(\bz)=\E(Y\mid X=x,\bZ=\bz)$ and $p_x(\bz)=x\pi(\bz)+(1-x)\{1-\pi(\bz)\}$. The cross-fitted AIPW arm mean is
\begin{equation}
 \hat\delta_x^{\mathrm{dr}}=\frac1n\sum_{k=1}^K\sum_{i\in I_k}
 \left[\hat\xi_x^{(k)}(\bZ_i)
 +\frac{\ind(X_i=x)}{\hat p_x^{(k)}(\bZ_i)}
                         \{Y_i-\hat\xi_x^{(k)}(\bZ_i)\}\right],
 \label{eq:aipw_delta}
\end{equation}
where $\hat p_x^{(k)}=x\hat\pi^{(k)}+(1-x)(1-\hat\pi^{(k)})$. In addition to a correct nuisance combination for the mediation benchmark, NDE consistency requires a correct propensity or correct $\xi_0$, and NIE consistency requires a correct propensity or correct $\xi_1$. Under SIME-I, the estimate of $\xi_{x}$ depends on the accuracy of estimating the index parameter $\btheta$.

Consider a local perturbation of $\btheta$ within its active support along the tangent direction. Let $\hat\bs_T^{(k)}$ denote the change in the outcome least-squares loss along that direction, $\hat\bH_T^{(k)}$ its tangent derivative, and $\hat\bG_T^{(k)}$ the tangent derivative of the index-score mean, as specified in Appendix~\ref{sub:app_causal_asym}. Define
\begin{equation}
 \hat h_{\mathrm I,\mathrm{aug}}^{(k)}(\mathcal O)
 =\hat h_{\mathrm I}^{(k)}(\mathcal O)
  -\hat\bG_T^{(k)\top}(\hat\bH_T^{(k)})^{-1}\hat\bs_T^{(k)}(\mathcal O),
 \qquad 
 \hat\zeta_{0,\mathrm I}^{\mathrm{aug}}
 =\frac1n\sum_{k=1}^K\sum_{i\in I_k}
                       \hat h_{\mathrm I,\mathrm{aug}}^{(k)}(\mathcal O_i).
 \label{eq:mr_estimator_aug}
\end{equation}
The revised augmented estimation of the effects are
\begin{equation}
 \widehat{\NDE}_{\mathrm I}^{\mathrm{aug}}
     =\hat\zeta_{0,\mathrm I}^{\mathrm{aug}}-\hat\delta_0^{\mathrm{dr}},
 \qquad
 \widehat{\NIE}_{\mathrm I}^{\mathrm{aug}}
     =\hat\delta_1^{\mathrm{dr}}-\hat\zeta_{0,\mathrm I}^{\mathrm{aug}}.
 \label{eq:effects_aug}
\end{equation}
The index augmentation cancels for $\widehat{\TE}=\hat\delta_1^{\mathrm{dr}}-\hat\delta_0^{\mathrm{dr}}$. The theory in Appendix~\ref{sub:app_causal_asym} requires first-order accuracy of the profile-score nuisance projection and derivative estimates. This active-space construction is the benchmark for the combined extension in Appendix~\ref{sub:app_combined_strategy}.

\subsection{Stage 4: Inference on Causal Effects}
\label{sub:stage4}

\paragraph{Joint score-based inference.}
For $i\in I_k$, let $\hat a_{xi}$ be the AIPW arm-score contribution in~\eqref{eq:aipw_delta}, and write $\hat h_i=\hat h_{\mathrm I,\mathrm{aug}}^{(k)}(\mathcal O_i)$. The complete held-out contrast vector is
\begin{equation}
 \hat{\mathbf q}_i=
 \begin{pmatrix}
  \hat h_i-\hat a_{0i}\\
  \hat a_{1i}-\hat h_i\\
  \hat a_{1i}-\hat a_{0i}
 \end{pmatrix},
 \qquad
 \bar{\mathbf q}=\frac1n\sum_{i=1}^n\hat{\mathbf q}_i
 =\begin{pmatrix}
   \widehat{\NDE}_{\mathrm I}^{\mathrm{aug}}\\
   \widehat{\NIE}_{\mathrm I}^{\mathrm{aug}}\\
   \widehat{\TE}
  \end{pmatrix}.
 \label{eq:complete_effect_scores}
\end{equation}
Estimate the joint influence-function covariance and the covariance of the effect estimates by
\begin{equation}
 \hat{\bV}=\frac1{n-1}\sum_{i=1}^n
 (\hat{\mathbf q}_i-\bar{\mathbf q})
 (\hat{\mathbf q}_i-\bar{\mathbf q})^\top,
 \qquad
 \widehat{\operatorname{Cov}}(\bar{\mathbf q})=\hat{\bV}/n.
 \label{eq:effect_covariance}
\end{equation}
Under the conditions of Corollary~\ref{cor:ci}, a marginal $(1-\alpha)$ Wald interval for component $j$ is
\begin{equation}
 \bar q_j\ \pm\ z_{1-\alpha/2}\sqrt{\hat V_{jj}/n}.
 \label{eq:stage4_ci}
\end{equation}
The validity of this inference procedure depends on successful screening, an accurate local index fit, and negligible joint score remainders. Oracle inference for partially linear single-index models can likewise require minimum-signal conditions for the index support \citep{cui2025estimation}. Details of the combined strategy with link refitting and broader correction are given in Appendix~\ref{sub:app_combined_strategy}.

\section{Asymptotic Theory}
\label{sec:asymptotic}

This section discusses the asymptotic properties of the proposed estimators.  Detailed regularity conditions and proofs are given in Appendix~\ref{app:proofs}.
Let $\mathcal S_0$ denote the support of the true index $\btheta_0$, with $s_0=|\mathcal S_0|$.

\begin{theorem} 
 \label{thm:selection}
 Under Assumptions~\ref{ass:index_ident}--\ref{ass:irrep}, with probability tending to one, there exists a local solution $\tilde\btheta$ of \eqref{eq:penalized}, constructed in Appendix~\ref{sub:app_index_conditions}, with support $\mathcal S_0$ and error $\|\tilde\btheta-\btheta_0\|_2=O_p(\sqrt{s_0}\lambda_n)$. If the stated conditions hold with $\lambda_n\asymp\sqrt{\log p/n}$, this error is $O_p(\sqrt{s_0\log p/n})$.
\end{theorem}

The selected support describes the outcome index. For the fixed-dimensional oracle result, let $\bU$ be an orthonormal basis of the tangent space at $\btheta_{0,\mathcal S_0}$. Let $\bs_T$ be the profile least-squares score in these coordinates, $\bH_T$ the derivative of its population mean, and $\bOmega_T=\operatorname{Var}(\bs_T)$. The local chart and score definitions are provided in Appendix~\ref{sub:app_scad_asym}.

\begin{theorem} 
 \label{thm:oracle}
 Under consistent screening and Assumption~\ref{ass:oracle_chart}, the SCAD local solution $\btheta^{\mathrm{SCAD},*}$ and unpenalized oracle local solution agree with probability tending to one. In particular,
 \begin{equation}
  \sqrt n\,\bB(\btheta^{\mathrm{SCAD},*}_{\mathcal S_0}
                         -\btheta_{0,\mathcal S_0})
  \ \xrightarrow{d}\ \mathcal N(\matzero,
                         \bB\bSigma_\theta\bB^\top),
  \label{eq:clt_theta}
 \end{equation}
 for any fixed matrix $\bB$ of appropriate dimension, where
 \begin{equation}
  \bSigma_\theta=\bU\bH_T^{-1}\bOmega_T\bH_T^{-1}\bU^\top.
  \label{eq:avar_theta}
 \end{equation}
 The same expansion holds for a computed $\hat{\btheta}$ only if it is $o_p(n^{-1/2})$ from this local solution and has the same support with probability tending to one.
\end{theorem}

For causal-effect inference, define $\bpsi_T=-\bH_T^{-1}\bs_T$ and let $\bG_T$ be the derivative of the population mean of the index mediation score in the same tangent coordinates. The joint score and nuisance conditions are stated in Appendix~\ref{sub:app_causal_asym}.

\begin{theorem} 
 \label{thm:causal_an}
 Suppose the identification assumptions, Assumption~\ref{ass:cov_rates}, and Assumption~\ref{ass:smooth_theta} hold, and the sufficient-index outcome and additive conditional mean are correctly specified. Then
 \begin{align}
 \sqrt n(\hat\zeta_{0,\mathrm I}^{\mathrm{aug}}-\zeta_0)
  &=\frac1{\sqrt n}\sum_{i=1}^n
    \{\varphi_{\zeta_0,\mathrm I}(\mathcal O_i)
                  -\bG_T^\top\bH_T^{-1}\bs_T(\mathcal O_i)\}
       +o_p(1),\label{eq:theta0_clt}\\
 \Phi_{\NDE,\mathrm I}
  &=\varphi_{\NDE,\mathrm I}
                   -\bG_T^\top\bH_T^{-1}\bs_T
    =\varphi_{\NDE,\mathrm I}+\bG_T^\top\bpsi_T,\nonumber\\
 \Phi_{\NIE,\mathrm I}
  &=\varphi_{\NIE,\mathrm I}
                   +\bG_T^\top\bH_T^{-1}\bs_T
    =\varphi_{\NIE,\mathrm I}-\bG_T^\top\bpsi_T.
 \label{eq:joint_effect_if}
 \end{align}
 Consequently, $\sqrt n(\widehat{\NDE}_{\mathrm I}^{\mathrm{aug}}-\NDE)$ and $\sqrt n(\widehat{\NIE}_{\mathrm I}^{\mathrm{aug}}-\NIE)$ converge jointly to a centered Gaussian distribution with covariance equal to that of $(\Phi_{\NDE,\mathrm I},\Phi_{\NIE,\mathrm I})$.

\end{theorem}

\begin{corollary} 
 \label{cor:ci}
 Under Theorem~\ref{thm:causal_an} and conditions of Lemma~\ref{lem:score_covariance}, the sample variances of augmented NDE and NIE scores consistently estimate their asymptotic variances. The confidence intervals are constructed accordingly. Analogous intervals apply to SIME-F under its stated conditions.
\end{corollary}

\section{Simulation Studies}
\label{sec:simulation}

Linear (Setting~1) and nonlinear (Setting~2) outcome links are considered. The data-generating settings and implementation details are provided in Appendices~\ref{sub:app_simulation_settings} and~\ref{sub:app_simulation_implementation}.

The compared methods include SIME-I, SIME-F, their support-Oracle counterparts, LinMed, HIMA and JonesPRD. SIME-I uses the index density ratio and SIME-F uses the full mediator density ratio. The principal SIME-I combines link refitting with a broader index-score correction, as described in Appendix~\ref{sub:app_combined_strategy}. Support-Oracle methods use the true candidate support and re-estimate the index and link. LinMed uses LASSO-based linear mediation. The simplified HIMA comparator uses marginal screening and MCP \citep{zhang2016estimating}. JonesPRD uses product-weight adaptive LASSO \citep{jones2025causal}. Further implementation details for these methods are provided in Appendix~\ref{sub:app_simulation_implementation}.

Performance is evaluated using bias and root mean squared error (RMSE) for NDE and NIE, and coverage probability (CP) of nominal 95\% intervals. True-positive and false-positive rates (TPR and FPR) assess index selection. The summary tables in Appendix~\ref{sub:app_simulation_results} report descriptive full-sample selection, while training-fold selection is reported in Appendix~\ref{sub:app_covariance}. Definitions and details are given in Appendix~\ref{sub:app_simulation_evaluation}.

Tables~\ref{tab:sim_linear_p50}--\ref{tab:sim_nonlinear_p500_n5000} and Figure~\ref{fig:inference_comparison} in Appendix~\ref{sub:app_simulation_results} summarize effect estimation, selection and interval coverage.
Overall, the proposed SIME-I performs well across scenarios and the performance improves as the sample size increases.
Tables~\ref{tab:aug_linear} and~\ref{tab:aug_nonlinear} demonstrate the improvement of the proposed index-augmented estimators.
HIMA and JonesPRD are competitive under the linear link with substantial bias under the nonlinear link. Full-ratio reweighting can be unstable in high dimensions, and index selection is less reliable in smaller samples. Detailed effect, covariance and selection results, sensitivity analyses and component diagnostics are provided in Appendix~\ref{app:simulation}.

\section{Application to the A4 Study}
\label{sec:a4}

We apply the proposed approach to the Anti-Amyloid Treatment in Asymptomatic Alzheimer's Disease (A4) study, a randomized, placebo-controlled trial of solanezumab in cognitively unimpaired older adults with elevated brain amyloid. The analysis examines whether MRI volume changes mediate the treatment effect on change in the Preclinical Alzheimer Cognitive Composite (PACC) over 240 weeks. A higher PACC score indicates better cognitive performance \citep{sperling2023trial}. APOE-$\varepsilon4$ carriers and non-carriers are analyzed separately.

Treatment assignment is the exposure ($X$), volume-change rates from 43 regional brain and ventricular MRI measures are considered as candidate mediators ($\bM$), and change in PACC score is the outcome ($Y$). Age, sex, education and body mass index are included as covariates ($\bZ$). The analysis includes 487 carriers and 304 non-carriers. Mediators, outcome and covariates are standardized within each subgroup before five-fold cross-fitting. The principal SIME-I combines link refitting and broader index-score correction and is compared with LinMed, HIMA and JonesPRD. Effects are measured in subgroup-specific PACC change. Preprocessing and fitting details are provided in Appendix~\ref{sub:app_a4_analysis}.

Table~\ref{tab:a4_effects} presents the effect estimates and 95\% confidence intervals. For SIME-I, the estimated NDE and NIE are $-0.033$ and $-0.053$ among carriers, and $0.068$ and $0.045$ among non-carriers. All available intervals contain zero. The linear mediation methods give estimates of similar direction and magnitude.

\begin{table}
\caption{\label{tab:a4_effects}Estimated effects and 95\% confidence intervals (CIs) by APOE-$\varepsilon4$ status in the A4 study. Selected: number of active mediators in the estimated index.}
\begin{center}
  \begin{tabular}{l l r r c r r}
  \hline
  & & \multicolumn{2}{c}{Carrier ($n=487$)} & & \multicolumn{2}{c}{Non-carrier ($n=304$)} \\
  \cline{3-4}\cline{6-7}
  \multicolumn{1}{c}{\multirow{-2}{*}{Method}} & \multicolumn{1}{c}{\multirow{-2}{*}{Quantity}} & \multicolumn{1}{c}{Estimate} & \multicolumn{1}{c}{95\% CI} &   & \multicolumn{1}{c}{Estimate} & \multicolumn{1}{c}{95\% CI} \\
  \hline
  & NDE & $-0.033$ & $[-0.173,\,0.107]$ &  & $0.068$ & $[-0.156,\,0.291]$ \\ 
  & NIE & $-0.053$ & $[-0.188,\,0.083]$ &  & $0.045$ & $[-0.146,\,0.236]$ \\ 
  & TE & $-0.085$ & $[-0.263,\,0.092]$ &  & $0.112$ & $[-0.116,\,0.340]$ \\ 
  \multirow{-4}{*}{\textbf{SIME-I}} & Selected & \multicolumn{2}{c}{10} &  & \multicolumn{2}{c}{7} \\ 
  \hline
  & NDE & $-0.030$ & $[-0.156,\,0.097]$ &  & $0.068$ & $[-0.111,\,0.248]$ \\ 
  & NIE & $-0.043$ & $[-0.115,\,0.030]$ &  & $0.043$ & $[-0.045,\,0.131]$ \\ 
  & TE & $-0.072$ & -- &  & $0.112$ & -- \\ 
  \multirow{-4}{*}{LinMed} & Selected & \multicolumn{2}{c}{18} &  & \multicolumn{2}{c}{15} \\ 
  \hline
  & NDE & $-0.039$ & -- &  & $0.060$ & -- \\ 
  & NIE & $-0.033$ & -- &  & $0.052$ & -- \\ 
  & TE & $-0.072$ & -- &  & $0.112$ & -- \\ 
  \multirow{-4}{*}{HIMA} & Selected & \multicolumn{2}{c}{8} &  & \multicolumn{2}{c}{8} \\ 
  \hline
  & NDE & $-0.042$ & $[-0.168,\,0.066]$ &  & $0.056$ & $[-0.101,\,0.228]$ \\ 
  & NIE & $-0.042$ & $[-0.136,\,0.092]$ &  & $0.037$ & $[-0.118,\,0.176]$ \\ 
  & TE & $-0.083$ & -- &  & $0.093$ & -- \\ 
  \multirow{-4}{*}{JonesPRD} & Selected & \multicolumn{2}{c}{19} &  & \multicolumn{2}{c}{15} \\ 
  \hline
  \end{tabular}
\end{center}
\end{table}

\begin{figure}
  \begin{center}
    \subfloat[APOE-$\varepsilon4$ carrier ($n=487$)]{\includegraphics[width=\textwidth]{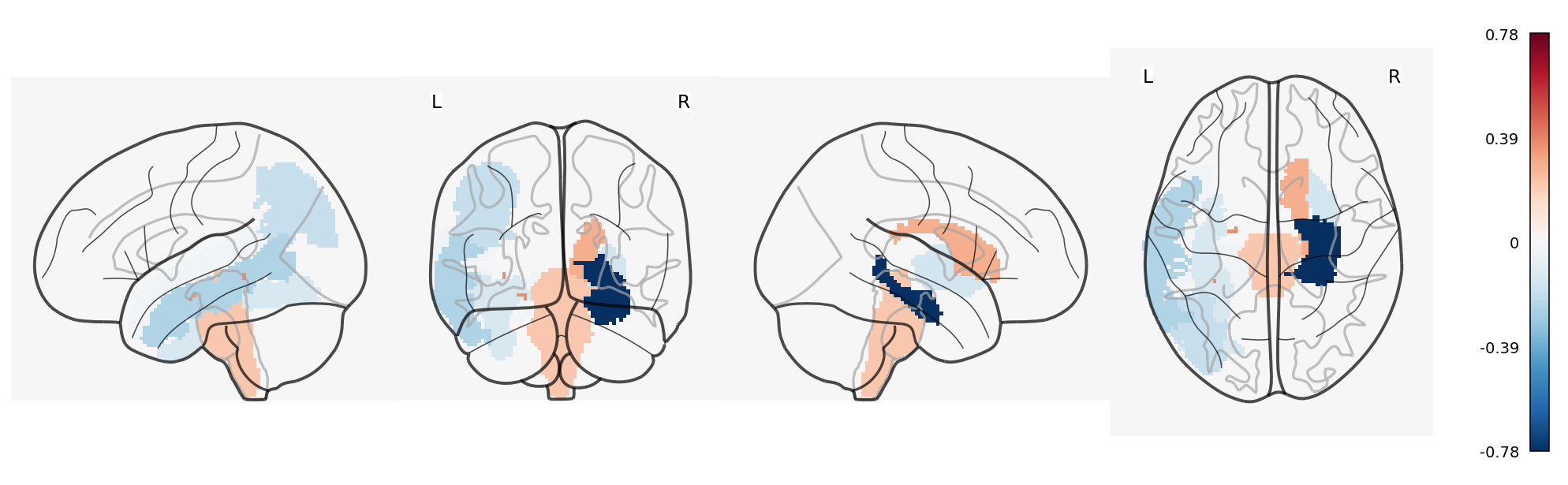}}\par\smallskip
    \subfloat[APOE-$\varepsilon4$ non-carrier ($n=304$)]{\includegraphics[width=\textwidth]{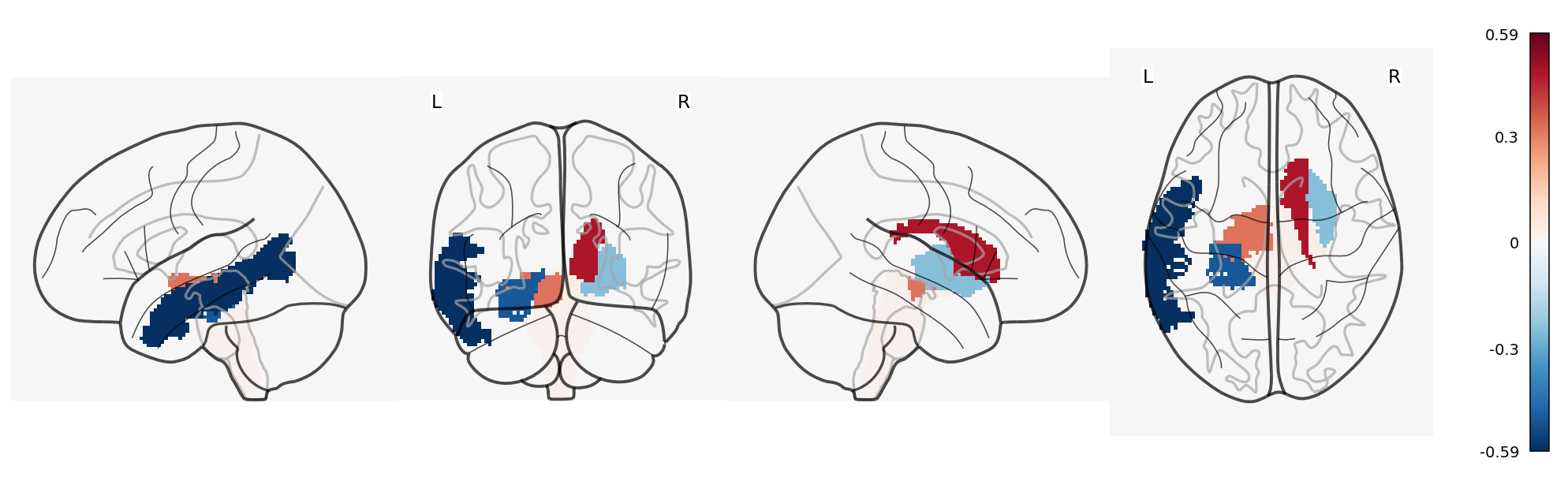}}
  \end{center}
\caption{\label{fig:a4_glass}Identified brain regions from SIME-I by APOE-$\varepsilon4$ status in the A4 study.. Red: positive weights; blue: negative weights. L and R: left and right brain hemisphere.}
\end{figure}

Comparing APOE-$\varepsilon4$ carriers and non-carriers, the effects are in the opposite directions, suggesting the potential moderating effect of APOE genotype.
In addition, the proportion of mediation effect in carriers ($62.3\%$) is higher than in non-carriers ($40.2\%$).
Table~\ref{tab:a4_weights} presents the index weights of brain regions from SIME-I and Figure~\ref{fig:a4_glass} shows the corresponding brain maps. 
For the carriers, regions with high weight magnitude include the right hippocampus, left inferior lateral ventricle, right caudate, and left middle temporal cortex. For non-carriers, high magnitude regions include the left middle temporal cortex, left parahippocampal cortex, right caudate, left ventral diencephalon, and right putamen.
Five regions appear in both groups, including the right caudate, right putamen, left middle temporal cortex, left parahippocampal cortex and brainstem.  
These coefficients summarize the joint MRI-change pattern.

The opposing hippocampal and ventricular weights in carriers define a structural pattern in which hippocampal volume loss and ventricular enlargement both increase the index. This configuration suggests that the carrier index captures complementary aspects of medial temporal tissue loss and expansion of cerebrospinal-fluid spaces. Longitudinal medial temporal atrophy has been associated with Alzheimer's disease pathology and cognitive decline in cognitively unimpaired adults~\citep{xie2025multi}. Similarly, declining medial temporal tissue-to-inferior-lateral-ventricle ratios were associated with less favorable cognitive trajectories in the longitudinal cognitive impairment and dementia cohort~\cite[DELCODE,][]{menze2026joint}.

The shared middle temporal and parahippocampal contributions suggest that temporal structural change is a common feature of the two subgroups. The additional inferior parietal, fusiform and superior temporal contributions in carriers extend this pattern across temporal and parietal association cortex. This distribution is consistent with the atrophy network identified in A4/LEARN, which included middle temporal and lateral parietal regions and was associated with subsequent cognitive decline~\citep{lee2026atrophy}.

The caudate, putamen and brainstem contributions indicate that both patterns extend beyond cortical and medial temporal structures. Cortex-to-caudate and cortex-to-putamen volume ratios have been associated with cognitive change in patients with Alzheimer's disease or amnestic mild cognitive impairment~\citep{na2024cortex}. Reduced whole-brainstem volume has also been associated with mild cognitive impairment and elevated genetic risk for Alzheimer's disease~\citep{lancaster2025reductions}. Locus coeruleus degeneration has been related to tau pathology and memory decline~\citep{wearn2026locus}, although the whole-brainstem measure used here cannot localize that nucleus-specific process. The additional bilateral ventral diencephalon contributions in non-carriers further broaden the subcortical pattern. 
It shall be noted that these joint index weights do not identify regional causal effects or establish APOE-specific mechanisms, but rather to suggest potential contributing factors.
Additional comparisons and diagnostics are provided in Appendix~\ref{app:a4}.

\begin{table}
  \caption{\label{tab:a4_weights}Index weights of brain regions estimated from SIME-I by APOE-$\varepsilon4$ status in the A4 study.}
  \begin{center}
    \begin{tabular}{l r r}
    \hline
    \multicolumn{1}{c}{Region} & \multicolumn{1}{c}{Carrier ($n=487$)} & \multicolumn{1}{c}{Non-carrier ($n=304$)} \\
    \hline
    Left inferior lateral ventricle & $0.3673$ & -- \\ 
    Right caudate & $0.2851$ & $0.4799$ \\ 
    Right putamen & $-0.1411$ & $-0.2513$ \\ 
    Right hippocampus & $-0.7836$ & -- \\ 
    Left ventral diencephalon & -- & $0.3217$ \\ 
    Right ventral diencephalon & -- & $0.0350$ \\ 
    Left fusiform cortex & $-0.1283$ & -- \\ 
    Left inferior parietal cortex & $-0.1793$ & -- \\ 
    Left middle temporal cortex & $-0.2378$ & $-0.5907$ \\ 
    Left parahippocampal cortex & $-0.0216$ & $-0.5023$ \\ 
    Left superior temporal cortex & $-0.0191$ & -- \\ 
    Brainstem & $0.2093$ & $0.0237$ \\ 
    \hline
    \end{tabular}
  \end{center}
\end{table}

\section{Discussion}
\label{sec:discussion}



We developed a semiparametric framework for estimating natural direct and indirect effects through a collection of potentially high-dimensional mediators. The central scientific question concerns the amount of mediation through the collection as a whole, which can be informative before attributing effects to individual mediators or specific pathways. A sparse single-index outcome model provides a parsimonious representation of the joint mediator–outcome relationship while allowing nonlinear responses through an unknown smooth link. Combined with multiply robust estimation and index-score augmentation, this representation supports inference on joint natural effects under the stated identification and regularity conditions.

A key distinction is between the causal estimand and the representation used for estimation. The natural effects remain defined with respect to the full potential mediator vector \(\bM(x)\). Under conditional-mean index sufficiency, the mediation formula can be evaluated through the distribution of \(\btheta^\top \bM\), yielding a lower-dimensional representation of the same target. This distinction is particularly useful when candidate mediators characterize a common biological process or measurement modality. In such settings, the index summarizes an outcome-relevant pattern across measurements and may aggregate information that is difficult to extract from separate mediator analyses. 


The single-index representation nonetheless raises a fair concern about interpretability. Index weights describe membership in an outcome-relevant pattern rather than mediator-specific causal effects: a variable can enter the index even when its distribution is unaffected by exposure, and an outcome-relevant pattern can coexist with a zero joint indirect effect. With correlated measurements, individual weights and selected supports can also vary across samples, as the fold diagnostics in Appendix~\ref{sub:app_a4_stability} show. We have argued that the pattern, rather than the individual measurement, is the appropriate level of interpretation for questions such as the one posed in the A4 study: the hypothesized pathway is neurodegeneration as a whole, region-specific natural effects are not identifiable without a causal ordering among regions, and composite measures are standard in both cognitive and imaging assessment of preclinical Alzheimer's disease. The selected variables are best viewed as candidates for subsequent investigation using substantive knowledge and methods designed for more specific causal questions. In the linear-link special case, the joint indirect effect reduces to the familiar sum of coefficient products; the nonlinear formulation extends this connection while generally precluding a simple additive attribution across mediators.

The full-ratio and index-ratio approaches offer complementary estimation strategies. SIME-F uses the joint mediator density ratio, with robustness characterized in the full observed-data model. SIME-I replaces this ratio with a scalar index ratio under conditional-mean index sufficiency. Its Gaussian implementation requires estimation of the projected residual variance rather than the full mediator covariance matrix, and the simulations indicate improved numerical stability in the settings examined. These computational advantages rely on the structural reduction. Index sufficiency is a substantive constraint: if distinct structural patterns drove different cognitive domains, a single index would not suffice. The disagreement between the full-ratio and index-ratio estimates in the A4 analysis is consistent with either instability of the full ratio or a departure from index sufficiency, and distinguishing the two is an important direction for future work. Multiple robustness then protects against specified combinations of nuisance-model misspecification.

Several limitations remain. The combined extension with link refitting and broader score correction improves small-sample bias and interval coverage. Independent evaluation and further assessment of inner-fold randomization are future directions. Learned-index inference requires support recovery, local profile curvature, adequate smoothing and nuisance convergence, and negligible score remainders. Additional numerical and theoretical evaluations are beyond the scope of the current study and are part of future directions. Multiply robust consistency under nuisance misspecification does not establish augmented interval validity. Appendix~\ref{sub:app_sensitivity_model} gives an aligned mediator--outcome error-dependence analysis with numerical results in Appendix~\ref{sub:app_sensitivity}.

Several extensions merit study. Future research may relax the Gaussian mediator working model through full or one-dimensional nonparametric density-ratio estimation under suitable nuisance and derivative conditions. Multiple-index outcome models could accommodate a low-dimensional mediator subspace, with additional identifiability and inference requirements~\citep{li1991sliced}. Post-treatment events, including truncation by death or noncompliance, may be studied by combining index models with principal-strata identification~\citep{cheng2026identification}.


\clearpage

\appendix
\counterwithin{figure}{section}
\counterwithin{table}{section}
\counterwithin{equation}{section}
\counterwithin{lemma}{section}
\counterwithin{theorem}{section}
\counterwithin{proposition}{section}
\counterwithin{assumption}{section}
\renewcommand{\theassumption}{\thesection\arabic{assumption}}

\section*{Appendix}

This appendix presents additional identification arguments, estimation details, theoretical results and proofs, and numerical results.

\section{Causal Identification}
\label{app:causal}

\subsection{Identification under the Single-Index Model}
\label{sub:app_identifySI}

Assumption~\ref{ass:sutva} allows $\bM(x)$ and $Y(x,\bfm)$ to be indexed by each unit's own specified interventions, with no interference between units or relevant hidden intervention versions. This makes the cross-world outcome $Y(1,\bM(0))$ well defined. The reduced identification formula under Assumptions~\ref{ass:sutva}--\ref{ass:positivity} and the single-index outcome model is
\begin{equation}
  \zeta_0 = \E\!\left\{\E\!\left\{\E\left\{Y \mid X=1, V_{\btheta}, \bZ\right\}
    \mid X=0, \bZ\right\}\right\} = \E\!\left\{\phi_{\btheta}(0, \bZ)\right\}.
  \label{eq:identification}
\end{equation}
The nested expectations in~\eqref{eq:identification} represent successive steps in identification.
The inner expectation $\E\{Y \mid X=1, V_{\btheta}=v, \bZ=\bz\} = \mu_{\btheta}(1,v,\bz)$ is the treated-arm outcome regression. Assumptions~\ref{ass:consistency} and~\ref{ass:conf2} connect this regression to potential outcomes. Consistency relates the observed outcome to $Y(1,\bfm)$. Assumption~\ref{ass:conf2} gives $Y(1,\bfm) \indep \bM \mid X=1,\bZ$. Assumption~\ref{ass:conf1} extends this relation from the treated arm to the population conditional on $\bZ$. The middle expectation averages $\mu_{\btheta}(1,v,\bz)$ over the \emph{control} law of $V_{\btheta}$ to obtain $\phi_{\btheta}(0,\bz)$. Under Assumption~\ref{ass:conf_m} and consistency, this law is the conditional law of $\btheta^\top\bM(0)$ given $\bZ=\bz$. Assumption~\ref{ass:cross} then gives
\[
  \E\{Y(1,\bM(0))\mid \bZ=\bz\} = \int \mu_{\btheta}(1,v,\bz)\, f_{V_{\btheta}|X=0,\bZ=\bz}(v)\,\mathrm{d}v,
\]
where Assumptions~\ref{ass:conf1}, \ref{ass:conf_m}, and~\ref{ass:conf2}, together with consistency, express the potential-outcome regression and mediator law in terms of observed-data quantities. The single-index formulation reduces integration to the scalar index. Positivity (Assumption~\ref{ass:positivity}) ensures that the control index law is absolutely continuous with respect to the treated law. The density ratio is therefore defined over the required control support. Finally, the outer expectation averages over $\bZ$.

Equations~\eqref{eq:NDE_id} and~\eqref{eq:NIE_id} give the corresponding reduced formulations of the natural effects.
For the NDE in~\eqref{eq:NDE_id}, $\E\{Y(0,\bM(0))\} = \E\!\left\{\E\!\left\{Y\mid X=0,\bZ\right\}\right\}$ is identified from the control arm under consistency and exposure ignorability. Identification of this single-treatment mean requires ignorability of $Y(x,\bM(x))$ given $\bZ$, as provided by randomized exposure or joint exposure--potential-outcome ignorability. Mediator--outcome exchangeability is not required for this step.
Under the semiparametric model assumptions, the NDE equals the coefficient of $X$ in \eqref{eq:outcome}. It is estimable at the $\sqrt{n}$ rate under regularity conditions for the partially linear single-index model. The multiply robust estimator $\hat\zeta_0^{\mathrm{mr}}-\hat\delta_0^{\mathrm{dr}}$ targets this coefficient when the observed outcome model and the baseline identification assumptions hold.
The NIE in \eqref{eq:NIE_semi} is obtained by averaging the unknown link $f$ over the mediator law. Under $\bM\mid X=x,\bZ=\bz\sim \mathcal{N}(\bmu_x(\bz),\bSigma_\epsilon)$, with $\bmu_x(\bz) = \balpha_0 + \balpha x + \bg_1(\bz)$, the index has distribution $\btheta^\top\bM\mid X=x,\bZ=\bz \sim \mathcal{N}(\btheta^\top\bmu_x(\bz),\,\btheta^\top\bSigma_\epsilon\btheta)$, giving
\begin{equation}
  \NIE = \E\!\left\{\int f(t)\,\mathcal{N}\!\left(t;\,
    \btheta^\top\bmu_1(\bZ),\,
    \btheta^\top\bSigma_\epsilon\btheta\right)\mathrm{d}t
    - \int f(t)\,\mathcal{N}\!\left(t;\,
    \btheta^\top\bmu_0(\bZ),\,
    \btheta^\top\bSigma_\epsilon\btheta\right)\mathrm{d}t\right\}.
  \label{eq:NIE_explicit}
\end{equation}
The Gaussian index distributions differ in their means by $\btheta^\top\balpha$. For the linear link $f$, with $f(t)=\beta t$, \eqref{eq:NIE_explicit} gives $\NIE=\beta\,\btheta^\top\balpha$, which is the usual product-of-coefficients formula. With nonlinear $f$, the mediated effect may be amplified or attenuated.

The NIE depends on $f$ through a smooth functional of $f$. Averaging $f$ over index distributions, rather than evaluating $f$ at a fixed point, reduces sensitivity to pointwise error in $\hat f$. With the one-step EIF correction and under the product-rate and other conditions of Theorem~\ref{thm:causal_an}, this formulation permits $\sqrt{n}$ inference for the final NIE estimator.

These results distinguish nonparametric identification from the structural model. The mediation g-formula is identified under the nonparametric model without imposing structural models~\eqref{eq:outcome}--\eqref{eq:mediator}. The full mediator representation is given in \eqref{eq:id_formula}. Under the single-index formulation, it reduces to \eqref{eq:identification} and the observed-data contrasts \eqref{eq:NDE_id}--\eqref{eq:NIE_id}. The expressions $\NDE=\gamma$ and \eqref{eq:NIE_semi} additionally require the observed outcome model and the baseline identification assumptions. If \eqref{eq:outcome} is misspecified, the natural effects remain defined and identifiable by the full g-formula. Consistency of the index-ratio estimator requires the reduced nuisance representation to remain correct.

\section{Additional Estimation Details}
\label{app:estimation}

This appendix describes full covariance estimation, Gaussian integration of the prediction link, and the combined strategy with link refitting and broader score correction.

\subsection{Full Covariance Matrix Estimation}
\label{sub:app_covariance_estimation}

The full covariance matrix is estimated on the training complement $I_k^c$ by pooling the exposure groups under the common conditional covariance assumption. Define the mediator residuals $\hat{\bepsilon}_i^{(k)}=\bM_i-\hat{\bmu}_{X_i}^{(k)}(\bZ_i)$ and their mean $\bar{\bepsilon}^{(k)}=N_k^{-1}\sum_{i\in I_k^c}\hat{\bepsilon}_i^{(k)}$. The residual sample covariance is
\begin{equation}
 \bS^{(k)}=\frac1{N_k}\sum_{i\in I_k^c}
 \{\hat{\bepsilon}_i^{(k)}-\bar{\bepsilon}^{(k)}\}
 \{\hat{\bepsilon}_i^{(k)}-\bar{\bepsilon}^{(k)}\}^{\!\top}.
 \label{eq:residual_cov}
\end{equation}
The sample covariance option uses $\bS^{(k)}$ without adjustment. For least-squares mediator means with a design of rank $d_k$ including an intercept, the degrees-of-freedom-adjusted covariance is $N_k\bS^{(k)}/(N_k-d_k)$. The identity covariance option uses $\matI_p$ as a working benchmark.

Linear shrinkage combines the sample covariance with a diagonal or scaled-identity target. Shrinkage toward $\diag(\bS^{(k)})$ preserves sample variances while reducing off-diagonal covariances. A positive weight on a positive definite target yields an invertible estimate even when the sample covariance is singular. Ledoit--Wolf (LW) shrinkage uses a scaled-identity target that matches the average residual variance, with a data-adaptive weight \citep{ledoit2004well}. Targets and weights are estimated within the training complement.

Under a correctly specified Gaussian linear mediator mean model fitted by least squares, let $\bU_k$ have orthonormal columns spanning the orthogonal complement of the training design. If $\bM_{\mathrm{tr}}^{(k)}$ contains the training mediator vectors as rows, the rows of $\bU_k^\top\bM_{\mathrm{tr}}^{(k)}$ are independent, mean-zero Gaussian residual contrasts conditional on the design. Their sample second moment with divisor $N_k-d_k$ equals the adjusted residual covariance above. The simulations and A4 analyses apply LW shrinkage to these contrasts using sample size $N_k-d_k$, without further centering. 

\subsection{Averaging of the Fitted Link}
\label{sub:app_gaussian_integration}

The fitted outcome link is averaged over the conditional Gaussian distribution of the mediator index using a piecewise-linear approximation. On the ordered grid $t_1<\cdots<t_J$, let $y_j=\hat f^{(k)}(t_j)$ and use linear interpolation with constant tails. On $(t_j,t_{j+1})$ the link is $c_j+s_jv$, where $s_j=(y_{j+1}-y_j)/(t_{j+1}-t_j)$ and $c_j=y_j-s_jt_j$. For $V\sim\mathcal N(m,\sigma^2)$, $\sigma>0$, define $z_j=(t_j-m)/\sigma$ and $P_j=\Phi(z_{j+1})-\Phi(z_j)$. Then
\begin{equation}
 \E\{\hat f^{(k)}(V)\}=y_1\Phi(z_1)
 +\sum_{j=1}^{J-1}\left[(c_j+s_jm)P_j
                  +s_j\sigma\{\varphi(z_j)-\varphi(z_{j+1})\}\right]
 +y_J\{1-\Phi(z_J)\},
 \label{eq:gaussian_piecewise_integral}
\end{equation}
where $\Phi$ and $\varphi$ denote the standard normal distribution and density. Stable tail-probability evaluation is used for extreme arguments. The formula exactly integrates the piecewise-linear link used for outcome prediction, eliminating Monte Carlo randomness.

As sample size increases, the grid is refined and its range enlarged using training information. Exact integration of the interpolant leaves approximation error relative to the smooth profile link. For a twice differentiable fitted link, the interior interpolation error is at most $\max_j(t_{j+1}-t_j)^2\sup|\hat f''|/8$. The constant-tail error must also be controlled under the evaluation laws. A strong sufficient numerical condition is $o_p(n^{-1/2})$ $L^2(P_0)$ error between the deployed and smooth-link mediation scores, uniformly over folds, with analogous control for both AIPW and complete augmented scores and consistent local score derivatives. 

\subsection{Combined Strategy: Link Refitting and Broader Correction}
\label{sub:app_combined_strategy}

The combined strategy refits the link at a fixed index direction and extends the correction to broader tangent directions. Within each training complement, the Stage~1 direction $\hat\btheta^{(k)}$ and its support $A_k$ are retained. The spline link and linear outcome coefficients are refitted at this fixed direction. The roughness penalty is selected as the largest value whose inner cross-validation prediction error is within one standard error of the minimum. The link derivative, profile nuisance projection and correction ingredients are then recomputed together. The smoother link also changes outcome predictions and Gaussian link integrals. Link refitting therefore changes effect estimates as well as their standard errors.

Let $B_k$ be a broader mediator coordinate set containing $A_k$, with all $p$ mediators used as candidates in the numerical implementation. A basis $\bW_k$ combines the active tangent basis with coordinate directions for $B_k\setminus A_k$, so $\bW_k^\top\hat\btheta_{B_k}^{(k)}=\matzero$. After residualization on the training index and the penalized link/covariate projection, let $\hat{\bD}_{B_k}(\mathcal O)$ be the residualized derivative of the fitted outcome mean in these directions, and let $\hat\eta_k$ be its outcome residual. The resulting score and curvature estimate are
\begin{equation}
 \hat\bs_{B_k}(\mathcal O)=-2\hat\eta_k(\mathcal O)\hat{\bD}_{B_k}(\mathcal O),
 \qquad
 \hat\bH_{B_k}=2P_{I_k^c}
       \{\hat{\bD}_{B_k}\hat{\bD}_{B_k}^{\top}\}.
 \label{eq:broad_profile_quantities}
\end{equation}
The Gaussian gradient formula in~\eqref{eq:index_gradient_gaussian} is estimated on $B_k$ and expressed in the same basis to obtain $\hat\bG_{B_k}$. 
The correction direction is estimated by minimizing the quadratic loss $\ba^\top\hat\bH_{B_k}\ba/2-\hat\bG_{B_k}^\top\ba$ with an $\ell_1$ penalty on the added coordinates. The active tangent block is unpenalized. The added coordinates are scaled to unit curvature after eliminating that block. The penalty is selected using inner training folds, and the selected correction coordinates are refitted without the lasso penalty. The added coordinates enter the correction score, while the fitted index support and direction remain fixed. For $i\in I_k$, the active-space augmentation is replaced by
\begin{equation}
 \hat h_{\mathrm I,B}^{(k)}(\mathcal O_i)
 =\hat h_{\mathrm I}^{(k)}(\mathcal O_i)
  -\hat\ba_{B_k}^{\top}\hat\bs_{B_k}(\mathcal O_i),
 \qquad
 \hat\zeta_{0,\mathrm I}^{B}
 =\frac1n\sum_{k=1}^K\sum_{i\in I_k}
                   \hat h_{\mathrm I,B}^{(k)}(\mathcal O_i).
 \label{eq:broad_augmented_score}
\end{equation}
Both the mediation score and the AIPW arm scores use the refitted link. These contributions are used in~\eqref{eq:complete_effect_scores}--\eqref{eq:stage4_ci} to form the combined NDE, NIE and TE estimates and complete-score standard errors. The broader correction cancels from TE, but link refitting can change the arm-score predictions and hence the TE estimate.

Sparse single-index debiasing provides related approaches for inference with imperfect selection \citep{eftekhari2021inference,wu2023model}, while regularized Riesz learning targets the functional directly \citep{chernozhukov2022automatic}. Adapting these ideas to causal mediation requires the additional score and representer conditions discussed in Appendix~\ref{sub:app_causal_asym}. The combined strategy addresses unstable fitted derivatives and omission of relevant tangent directions. The production comparisons and paired component diagnostics provide empirical evidence for the combined strategy. They do not establish the broader-space cancellation, representer-estimation rates, score convergence or negligible remainders needed for valid inference after imperfect index selection. Theorem~\ref{thm:causal_an} and Corollary~\ref{cor:ci} justify the active-space benchmark under their stated assumptions. A corresponding result for~\eqref{eq:broad_augmented_score} remains to be developed.

\section{Additional Theoretical Results and Proofs}
\label{app:proofs}

This section presents the regularity conditions, additional theoretical results and proofs for Section~\ref{sec:asymptotic}, together with the mediator--outcome error-dependence sensitivity analysis. The derivative and stochastic bounds are assumed. 

\subsection{Local Index Conditions and Support Recovery}
\label{sub:app_index_conditions}

The index profile is nonconvex under the sphere constraint. Conditional local results are considered for \eqref{eq:penalized} and the SCAD refit in Appendix~\ref{sub:app_scad_asym}. The simulation and A4 analyses use a constrained profile refit. Applying the results to a computed estimate requires selection-consistent screening, suitable tuning and smoothing, convergence to the specified local solution, and asymptotically negligible optimization error. The causal-effect results in Section~\ref{sec:asymptotic} require the same conditions.

\begin{assumption}[Identified oriented index]
 \label{ass:index_ident}
 The correctly specified conditional mean model has a unique oriented unit index $\btheta_0$ after adjustment for $(X,\bZ)$. Its support is $\mathcal S_0$, of size $s_0$, and an anchor coordinate $j_0\in\mathcal S_0$ satisfies $\theta_{0j_0}\geq c>0$. The local derivatives below exist.
\end{assumption}

Relabeling the anchor locally if necessary, write $\bu=\btheta_{-j_0}$ and
\[
 \btheta=\bpsi(\bu),\qquad
 \psi_{j_0}(\bu)=\sqrt{1-\|\bu\|_2^2},\qquad
 \bpsi_{-j_0}(\bu)=\bu.
\]
The chart is defined in a neighborhood of $\btheta_0$ and allows the first original mediator to be inactive. Let $A=\mathcal S_0\setminus\{j_0\}$, $\bu_0=\btheta_{0,-j_0}$, and $\ell_n(\bu)=Q_n\{\bpsi(\bu)\}$. In this chart, the penalized criterion is
\begin{equation}
 L_n(\bu)+\lambda_n\|\bu\|_1,\qquad
 L_n(\bu)=\ell_n(\bu)+\lambda_n\sqrt{1-\|\bu\|_2^2}.
 \label{eq:chart_loss}
\end{equation}
The anchor term contributes to the gradient and curvature of the smooth part of the penalized criterion. The sphere constraint therefore requires an adjustment to the ambient LASSO penalty and Hessian. The convex-loss results of \citet{lee2015model} motivate the local curvature and inactive-gradient conditions, but a change of coordinates does not justify applying those results directly to the nonconvex profile.

\begin{assumption}[Local curvature and score control]
 \label{ass:rsc}
 Let $r_n\to0$ satisfy $\sqrt{s_0}\lambda_n=o(r_n)$ and $r_n=o(c_{\min,n})$, where $c_{\min,n}=\min_{j\in\mathcal S_0}|\theta_{0j}|$. On the chart ball of radius $r_n$ about $\bu_0$, $L_n$ is twice continuously differentiable and its Hessian is locally bounded. With probability tending to one, its Hessian restricted to $A$ has minimum eigenvalue at least $m>0$. Moreover, $\|\nabla_A\ell_n(\bu_0)\|_2=O_p(\sqrt{s_0}\lambda_n)$ and $\lambda_n\to0$, with $\lambda_n>0$.
\end{assumption}

\begin{assumption}[Inactive-coordinate certificate]
 \label{ass:irrep}
 Let $\bar\bu$ minimize $L_n(\bu)+\lambda_n\|\bu_A\|_1$ over this ball with $\bu_{A^c}=\matzero$. For some fixed $\tau\in(0,1)$, with probability tending to one,
 \begin{equation}
  \|\nabla_{A^c} L_n(\bar\bu)\|_\infty
                 \leq(1-\tau)\lambda_n.
  \label{eq:irrep}
 \end{equation}
\end{assumption}

Condition~\eqref{eq:irrep} is a sample-level certificate that remains unverified for the simulation design. Under a population quadratic approximation with $\bH=\nabla^2\ell(\bu_0)$, the active penalty gradient is $\bs_A=\sign(\bu_{0A})-\bu_{0A}/\theta_{0j_0}$. The corresponding irrepresentability expression involves $\bH_{A^cA}\bH_{AA}^{-1}\bs_A$, together with control of the stochastic score and remainder, rather than the inverse of the ambient active Hessian \citep{zhao2006model}. These derivative and inactive-coordinate conditions require more than uniform consistency of a univariate smoother. For a twice differentiable link, the usual interior local linear level bias is of order $h^2$. Additional smoothness does not increase the polynomial order of the fit \citep{fan1996local}.

Under the conditions of Theorem~\ref{thm:selection}, with probability tending to one the restricted solution $\bar\bu$ is interior, has support $A$, and is a strict local minimizer of \eqref{eq:chart_loss}. Its image $\tilde\btheta=\bpsi(\bar\bu)$ is the local solution in that theorem.

Appendix~\ref{sub:app_proof_selection} establishes local optimality under the stated curvature, score and inactive-gradient bounds. Verification of these bounds for a specified profile fit and convergence from a numerical starting point to the local solution remain necessary. The radius conditions entail $\sqrt{s_0}\lambda_n=o(c_{\min,n})$. Unit normalization implies $c_{\min,n}\leq s_0^{-1/2}$, so $\lambda_n\asymp\sqrt{\log p/n}$ necessarily requires $s_0^2\log p/n\to0$, beyond $s_0\log p/n\to0$. Smaller signals require an even stronger rate. Assumption~\ref{ass:index_ident} also retains a fixed positive anchor. 

\subsection{Oracle Expansion and Profile Score}
\label{sub:app_scad_asym}

The fixed-dimensional result in Theorem~\ref{thm:oracle} assumes $\Prob(\hat{\mathcal S}=\mathcal S_0)\to1$. This condition may follow from Theorem~\ref{thm:selection} with a suitable optimizer or from another valid screening procedure. Let $\bU$ be an $s_0\times(s_0-1)$ orthonormal basis of the tangent space at $\btheta_{0,\mathcal S_0}$. Consider the smooth local chart
\[
 \bchi(\bt)=\frac{\btheta_{0,\mathcal S_0}+\bU\bt}
                         {\sqrt{1+\|\bt\|_2^2}},\qquad
 q_n(\bt)=Q_n\{\bchi(\bt),\matzero_{\mathcal S_0^c}\}.
\]

\begin{assumption}[Oracle chart expansion]
 \label{ass:oracle_chart}
 The active dimension $s_0$ and the law of the active variables and profile score are fixed, and the true active coefficients are bounded away from zero. On a neighborhood of $\bt=\matzero$, $\nabla^2q_n$ converges uniformly in probability to a continuous Hessian with positive definite value $\bH_T$ at zero, and
 \[
  \nabla q_n(\matzero)=\frac1n\sum_{i=1}^n\bs_T(\mathcal O_i)
                         +o_p(n^{-1/2}),\qquad
  \E\bs_T=\matzero,\qquad
  \operatorname{Var}(\bs_T)=\bOmega_T<\infty.
 \]
 The selected SCAD local solution and the unpenalized oracle local solution are consistent and lie in this neighborhood. The SCAD tuning satisfies $\mu_n\to0$ and $\min_{j\in\mathcal S_0}|\theta_{0j}|/\mu_n\to\infty$.
\end{assumption}

The covariance in \eqref{eq:avar_theta} is an $s_0\times s_0$ matrix with a zero direction along $\btheta_0$, as required by normalization. For $s_0=1$, selection fixes the oriented direction and the covariance is zero. The chart score incorporates estimation of the link and linear covariate coefficients. Its assumed asymptotic linearity requires more than pointwise kernel consistency. Covariance estimation requires the full sandwich, including $\bH_T$. The active-space simulation and A4 benchmarks use complete augmented effect scores involving the estimated profile Hessian and gradient. The principal combined estimates use the broader correction in Appendix~\ref{sub:app_combined_strategy}.

\paragraph{The profile score used for index-aware inference.}
The oracle expansion applies to the constrained profile refit rather than normalized working-regression slopes. Under the correctly specified additive conditional mean model with linear covariate adjustment, let $V_0=\btheta_0^\top\bM$ and assume $f_0'(V_0)\bU^\top\bM_{\mathcal S_0}$ is square integrable. Let $\mathcal H_0$ be the closed linear span in $L^2(P_0)$ of
\[
 a(V_0)+bX+\bc^\top\bZ,
\]
where $a$ ranges over square-integrable functions and the intercept is included in $a$. For each component, let $\Pi_{\mathcal H_0}$ denote its least-squares projection onto this span and define
\begin{align}
 \bD(\mathcal O)&=f_0'(V_0)\bU^\top\bM_{\mathcal S_0}
       -\Pi_{\mathcal H_0}\{f_0'(V_0)\bU^\top\bM_{\mathcal S_0}\},
       \nonumber\\
 \bs_T(\mathcal O)&=-2\eta\bD(\mathcal O),\qquad
 \bH_T=2\E\{\bD\bD^\top\},\qquad
 \bOmega_T=4\E\{\eta^2\bD\bD^\top\}.
 \label{eq:profile_tangent_score}
\end{align}
Here $\eta=Y-m_0(X,\bM,\bZ)$, where $m_0$ is the true conditional outcome mean. The profile projection removes the link and linear covariate directions. Residualizing $\bM$ against $V_0$ alone generally gives a different score. If $g_2$ is fitted more flexibly, its nuisance directions must be included in $\mathcal H_0$. The least-squares profile score in \eqref{eq:profile_tangent_score} corresponds to the index-estimation criterion. It is not asserted to be a likelihood efficient score under heteroscedastic errors.

The population profiled mean derivative in a tangent direction is $\bD$. Since $\E(\eta\mid X,\bM,\bZ)=0$, its population Hessian at the truth is $\bH_T$. Therefore the tangent index influence function is
\begin{equation}
 \bpsi_T=-\bH_T^{-1}\bs_T,\qquad
 \bpsi_\theta=\bU\bpsi_T
 \label{eq:profile_index_if}
\end{equation}
on the active coordinates, with zeros off the support. Consistent estimates of $f_0'$, the nuisance projection and the complete sandwich are required. The approximation and regularization bias from a finite spline representation or smoothing penalty must be negligible in this expansion. A fixed smoothing rule does not establish this requirement. Theorem~\ref{thm:oracle} is a fixed-law result. For a triangular array, additionally require $\bH_{T,n}\to\bH_T$, $\bOmega_{T,n}\to\bOmega_T$, and a multivariate Lindeberg condition for $\bs_{T,n}$, as well as the uniform remainder conditions. For $s_0=1$, the tangent score and all index corrections are empty and taken to be zero.

\subsection{Covariance Estimation and Nuisance Convergence Rates}
\label{sub:app_cov_asym}

Estimation of $\bSigma_\epsilon$ contributes to the errors of $\hat r$ and $\hat\phi$ through the fitted mediator law. It does not change the causal identification formula or the nuisance components of the mediation influence function. Stage~1 does not use this covariance estimate, so the Stage~2 covariance choice does not change the selection and oracle statements. Mediator correlation still affects the restricted curvature and irrepresentability conditions required for index recovery.

To state the convergence requirements, let $W=\bM$ for SIME-F and $W=V_{\btheta_0}$ for the fixed-index analysis. Denote the corresponding true nuisance functions by $\mu,r,\phi$, and let $\Delta\mu=\hat\mu-\mu$, with analogous notation for the remaining components. Define $\|h\|_x^2=\E_{\bZ}[\E\{h(W,\bZ)^2\mid X=x,\bZ\}]$. For each training fold, let
\[
  e_\pi=\|\hat\pi-\pi\|_{L^2(P_{\bZ})},\quad
  e_\mu=(\|\Delta\mu\|_0^2+\|\Delta\mu\|_1^2)^{1/2},\quad
  e_r=\|\hat r-r\|_1,\quad
  e_\phi=\|\hat\phi-\phi\|_{L^2(P_{\bZ})}.
\]

\begin{assumption}[Covariance-dependent nuisance rates]
  \label{ass:cov_rates}
  For a fixed number of balanced folds, the nuisance functions evaluated at the true index are estimated using training data only. The true and fitted propensity scores lie in $[c,1-c]$ for some $c>0$. Uniformly over folds, all four errors above are $\op(1)$ and
  \begin{equation}
    e_\pi(e_\mu+e_\phi)+e_r e_\mu=\op(n^{-1/2}).
    \label{eq:cov_product_rate}
  \end{equation}
  The estimated mediation score converges to its true counterpart in $L^2(P_0)$, and the latter has a finite, nonzero variance. The analogous score convergence and propensity--outcome product-rate conditions hold for the AIPW estimators of $\delta_x$, $x=0,1$. For a growing-dimensional sequence of laws, also assume a joint Lindeberg central limit theorem for the true mediation and AIPW scores, with covariance converging to a finite matrix and positive limiting NDE/NIE variances. Finite variance at each $n$ alone is insufficient for that sequence.
\end{assumption}

The score condition controls products of $Y$ and density-ratio weights. Heavy-tailed ratios may require more than unweighted nuisance convergence. Cross-fitting controls the centered empirical fluctuation, and \eqref{eq:cov_product_rate} controls the conditional mean drift. Covariance estimation contributes to $e_r$ and $e_\phi$. Under these conditions and a fully correct nuisance law, it introduces no additional first-order influence term. The root-$n$ result requires the stated product rates as well as consistency. A sufficient choice is $\op(n^{-1/4})$ for each nuisance error, although unequal rates are permitted \citep{chernozhukov2018double}. Numerical integration error is subject to the same requirements.

\paragraph{Covariance functionals required for consistency.}
For SIME-I with fixed $\btheta_0$, the Gaussian ratio and structural integral depend on the covariance only through $\tau_0=\btheta_0^\top\bSigma_\epsilon\btheta_0$. A sufficient condition is
\begin{equation}
  a_{\Sigma,\mathrm{I}}^{(k)}
  =\left|\btheta_0^\top
    (\hat{\bSigma}_\epsilon^{(k)}-\bSigma_\epsilon)\btheta_0\right|=\op(1),
  \qquad 0<c\leq\tau_0\leq C<\infty,
  \label{eq:cov_index_rate}
\end{equation}
with fitted projected variances also bounded away from zero. Under local smoothness and square-integrable derivative envelopes for the Gaussian ratio and fitted link, uniformly bounded over any changing sequence of laws, the covariance contributions to $e_r$ and $e_\phi$ are $O_p(a_{\Sigma,\mathrm{I}}^{(k)})$. Thus, a sufficient covariance contribution to \eqref{eq:cov_product_rate} is $a_{\Sigma,\mathrm{I}}^{(k)}(e_\mu+e_\pi)=\op(n^{-1/2})$, together with the required mean, link, and remaining nuisance rates. Accuracy is required for a single quadratic form rather than every entry of a $p\times p$ matrix. The sample projected variance may satisfy this condition even when the sample covariance matrix is singular. The principal SIME-I estimator uses the projected residual variance in \eqref{eq:projected_residual_variance}. For an estimated index, these rates and moment envelopes must hold uniformly on its local active chart, as specified below. The augmentation additionally requires consistency of the active covariance action and profile nuisance projection. Positivity of the scalar variance alone is insufficient.

For SIME-F, \eqref{eq:lr_full_est} identifies $\bSigma_\epsilon^{-1}\balpha$ and the fitted mean offset as the relevant precision and mean quantities. Accuracy of this precision action in the prediction norm induced by $\bM$ may suffice, without consistent estimation of every precision-matrix entry. In fixed dimension, operator-norm consistency of $\hat{\bSigma}_\epsilon$ is a stronger sufficient condition when its eigenvalues are bounded above and away from zero, the mediator means are consistently estimated, and the ratio has the required moment envelopes. In growing dimension, mean estimation, inversion, and exponentiation require dimension-dependent error control. 

\paragraph{Sample covariance and shrinkage bias.}
Under a correctly specified Gaussian linear mediator model with fixed $p$ and fixed design rank, the sample residual covariance estimates $\bSigma_\epsilon$ at rate $O_p(N_k^{-1/2})$. This rate is not automatic when $p$ is comparable to $N_k$. For a linear shrinkage estimator assigning weights $1-\hat\lambda_\Sigma$ and $\hat\lambda_\Sigma$ to a sample covariance $\bS$ and target $\bT$, respectively, with $\hat\lambda_\Sigma\in[0,1]$, the following operator-norm bound holds:
\begin{equation}
 \|\hat{\bSigma}_{\epsilon,\mathrm{sh}}-\bSigma_\epsilon\|_{\mathrm{op}}
 \leq(1-\hat\lambda_\Sigma)\|\bS-\bSigma_\epsilon\|_{\mathrm{op}}
      +\hat\lambda_\Sigma\|\bT-\bSigma_\epsilon\|_{\mathrm{op}}.
 \label{eq:cov_bias_bound}
\end{equation}
When $\|\bT-\bSigma_\epsilon\|_{\mathrm{op}}=O_p(1)$ and the sample covariance is consistent, a vanishing shrinkage weight removes target bias. Its rate must also satisfy \eqref{eq:cov_product_rate}. For SIME-I, the corresponding projected condition is sufficient. With fixed index support, a fixed diagonal-target weight $\lambda_\Sigma>0$, and consistent sample entries, the limiting projected bias is
\begin{equation}
 \lambda_\Sigma\btheta_0^\top
   \{\diag(\bSigma_\epsilon)-\bSigma_\epsilon\}\btheta_0,
 \label{eq:cov_projected_bias}
\end{equation}
which may be nonzero under correlated mediators. Under its assumptions, data-adaptive Ledoit--Wolf shrinkage estimates an oracle weight. 

\paragraph{Consistency under covariance misspecification.}
Accurate covariance functionals, correctly specified mediator means and the stated regularity conditions give consistent Gaussian mediator nuisance estimates. Some multiply robust combinations also permit an incorrect covariance limit. For example, consistency of the outcome regression and propensity score can preserve consistency under (F-b) or (I-b), provided the remaining fitted scores have suitable limits and moments. SIME-I also requires a consistent sufficient index. These conditions establish consistency but do not establish the efficient variance at the fully correct law. Root-$n$ inference under partial misspecification requires a separate expansion around the nuisance limits and may include first-order estimation terms. Fixed shrinkage remains a working \mbox{regularization} until the required functional accuracy or a suitable inference result under misspecification is established.

\paragraph{Numerical integration and clipping.}
The procedure evaluates the Gaussian integral of the prediction link, represented as a piecewise-linear function, using \eqref{eq:gaussian_piecewise_integral}. Interpolation and tail errors relative to the smooth profile fit must satisfy the stated score-drift and moment bounds as numerical accuracy increases. Exact integration of a fixed interpolant does not give exact integration of the population link. For the implementation using 500 common Monte Carlo draws, a fixed draw count and interpolation grid do not by themselves ensure negligible root-$n$ score error. With $\hat\xi_0=\hat\phi(0,\cdot)-\hat\gamma$, the integration error in the NDE score cancels algebraically. If $e_0(\bZ)$ and $e_1(\bZ)$ denote the control- and treated-arm integration errors, the NIE and TE score error is
\[
 -(X-\hat\pi)\left\{\frac{e_1}{\hat\pi}
                              +\frac{e_0}{1-\hat\pi}\right\}.
\]
At the true propensity, this error has conditional mean zero, so a fixed draw count does not necessarily cause inconsistency. It can still contribute nonvanishing root-$n$ randomness, and propensity estimation introduces a first-order product unless its rate is controlled. The effect-score standard errors from that implementation do not establish validity for the combined fitting and integration randomness. Fixed propensity clipping is consistent when the true probabilities lie in its interval or clipping becomes inactive asymptotically. The Gaussian-covariate logistic simulation has probabilities in $(0,1)$ without a uniform positive lower bound. 

\subsection{Local Conditions for Index-Aware Inference}
\label{sub:app_causal_asym}

The index and all nuisance components are estimated within the training complement of each evaluation fold. The index-ratio score is augmented by the tangent profile score. The combined evaluation score accounts for index estimation without requiring population stationarity. Theorem~\ref{thm:causal_an} and Corollary~\ref{cor:ci} concern the active-space estimator in~\eqref{eq:mr_estimator_aug}. Section~\ref{sec:simulation} evaluates its finite-sample performance, and the A4 application reports both active-space benchmarks and the adopted combined extension. 

Let $\bchi(\bt)$ be the active tangent chart from Appendix~\ref{sub:app_scad_asym}, and let $m_{\bt}$ be the uniquely identified population least-squares profiled conditional mean within the chosen additive single-index family at $\bchi(\bt)$. Assume a locally twice differentiable version with moment envelopes that justify differentiation under expectation. Define $\bs_{\bt}(\mathcal O)=\nabla_{\bt}\{Y-m_{\bt}(X,\bM,\bZ)\}^2$, whose mean is the derivative of the population profiled squared-error criterion. Write $\bs_{\bt}$ for this population additive least-squares profile score and $h_{\bt}$ for the uncentered index mediation score using the true reduced conditional outcome mean and other true reduced nuisance functions along the same direction. The additive profile mean $m_{\bt}$ need not equal the true reduced conditional mean away from $\bt=\matzero$. Correct specification is required at the true index. Local approximation error from the restricted outcome fit is included in the score-remainder conditions below. At $\bt=\matzero$,
\[
 P_0h_{\matzero}=\zeta_0,\quad P_0\bs_{\matzero}=\matzero,
 \quad\bs_{\matzero}=\bs_T,
 \quad\left.\nabla_{\bt}P_0\bs_{\bt}\right|_{\matzero}=\bH_T.
\]
Define
\begin{equation}
 \bG_T=\left.\nabla_{\bt}P_0h_{\bt}\right|_{\matzero},
 \qquad \bA_T=\bH_T^{-1}\bG_T,
 \qquad h_{\bt}^{\mathrm{aug}}=h_{\bt}-\bA_T^\top\bs_{\bt}.
 \label{eq:index_gradient}
\end{equation}
Equivalently, $\bG_T=\bU^\top\bG_{\mathcal S_0}$ for any ambient derivative $\bG$ at the truth. Changing the orthonormal tangent coordinates consistently across all three correction factors leaves the scalar correction unchanged.

For the common-covariance Gaussian mediator model,
\begin{equation}
 \bG=-\bar c\left\{\balpha-
  \frac{\btheta_0^\top\balpha}
       {\btheta_0^\top\bSigma_\epsilon\btheta_0}
       \bSigma_\epsilon\btheta_0\right\},\qquad \bar c=\E_{\bZ}[\E\{f_0'(V_0)\mid X=0,\bZ\}],
 \label{eq:index_gradient_gaussian}
\end{equation}
as derived in Appendix~\ref{sub:app_proof_causal}. For this Gaussian gradient calculation, assume that $f_0$ is twice continuously differentiable and that the conditional-regression derivatives and all control and ratio-weighted treated expectations below admit integrable local derivative envelopes. In particular, differentiation of the reduced regression at the degenerate conditional variance when $\btheta=\btheta_0$ must be justified. These local smoothness and moment bounds are additional conditions that Gaussianity alone does not establish. Estimating this gradient requires the active covariance action $(\bSigma_\epsilon\btheta_0)_{\mathcal S_0}$ as well as the projected variance. Active residual cross-products suffice and no inverse of the full covariance is required. Alternatively, a consistently estimated derivative of the population score may be used. 

The fully cross-fitted augmented cross-world mean is defined in \eqref{eq:mr_estimator_aug}. Its correction uses $\hat\bA_T^{(k)}=(\hat\bH_T^{(k)})^{-1}\hat\bG_T^{(k)}$, and all three fitted factors in the tangent score are evaluated using that fold's training complement. The NDE and NIE contrasts use this mean and the cross-fitted AIPW arm means. The two corrections have opposite signs, so they cancel from their sum, the TE estimator.

\begin{assumption}[Joint local score and nuisance conditions]
 \label{ass:smooth_theta}
 The number of balanced folds and active dimension are fixed, screening succeeds on all training complements with probability tending to one, and the oriented constrained profile estimates satisfy $\|\hat\bt_k\|=O_p(n^{-1/2})$ uniformly over folds. The population means $P_0h_{\bt}$ and $P_0\bs_{\bt}$ are twice differentiable locally, with bounded second derivatives, and their derivatives at zero are $\bG_T^\top$ and $\bH_T$, respectively. For $s_0>1$, the matrix $\bH_T$ is positive definite, and $\hat\bA_T^{(k)}=\bA_T+o_p(1)$ uniformly over folds.

 Let $\hat h_k,\hat\bs_k$ denote the fitted functions evaluated at $\hat\bt_k$. Conditional on their training data, their population reference functions satisfy, uniformly over folds,
 \begin{equation}
 |P_0(\hat h_k-h_{\hat\bt_k})|
   +\|P_0(\hat\bs_k-\bs_{\hat\bt_k})\|
       =o_p(n^{-1/2}),\qquad
 \|\hat h_k-h_{\matzero}\|_{L^2(P_0)}
   +\|\hat\bs_k-\bs_T\|_{L^2(P_0)}=o_p(1).
 \label{eq:joint_score_remainders}
 \end{equation}
 The first mediation bound is ensured by the fixed-index drift/product-rate conditions of Assumption~\ref{ass:cov_rates} when they hold uniformly on the local chart. The profile-score bound is a separate nuisance-profiling requirement. It includes estimation of the link derivative and the full nuisance projection, spline approximation, smoothing bias, and optimization error. It does not follow merely from consistency of the fitted link. Preprocessing, tuning, and all numerical approximations are chosen using training data and satisfy these score bounds. The analogous conditional-drift and $L^2$ conditions hold for both AIPW arm scores.

 The true joint vector of the mediation, profile, and AIPW scores has finite second moments and a multivariate central limit theorem, with positive NDE and NIE limiting variances. For triangular arrays this includes convergence of its covariance, a joint Lindeberg condition, and uniform versions of all bounds above.
\end{assumption}

A constrained profile expansion in every training fold can provide the sufficient root-$n$ pilot rate. A weaker pilot rate requires explicit quadratic and coefficient-estimation product bounds and is not asserted here. The nuisance estimation bounds in \eqref{eq:joint_score_remainders} remain statistical assumptions requiring verification for the implementation. Specifying a profile optimizer does not establish these bounds.

\paragraph{Full-ratio inference and partial misspecification.}
 For a fully cross-fitted SIME-F estimator satisfying the corresponding mediation/AIPW nuisance bounds, the full-ratio population index gradient is zero and the usual full-model effect influence functions apply. Their efficiency claim concerns the unrestricted mediator model. No efficiency claim is made for the augmented SIME-I estimator in the smaller additive/Gaussian structural model.

 The unaugmented cross-world estimator retains union-model consistency under the limiting-nuisance and laws-of-large-numbers conditions in Section~\ref{sec:semiparam}. The augmented estimator inherits this consistency only if its validation-fold augmentation averages are $o_p(1)$ under the specified partially misspecified law. Both contrasts also require the corresponding AIPW consistency condition. Union-model consistency alone does not establish root-$n$ inference or the efficient variance under partial misspecification.

\begin{remark}[Legacy fixed-index intervals and stationarity]
 \label{rem:index_variance}
 If the unaugmented estimator uses a full-sample asymptotically linear profile index and satisfies the appropriate local score expansion, its first-order influence functions are also $\varphi_{\NDE,\mathrm I}+\bG_T^\top\bpsi_T$ and $\varphi_{\NIE,\mathrm I}-\bG_T^\top\bpsi_T$. The variance includes the covariance of the effect and index scores. Adding only a separate index variance is insufficient. The fixed-index variance can instead be used only when the first-order index contribution vanishes, for example with a known index or a zero tangent gradient. Cross-fitting alone does not make that gradient zero. The simulation and A4 unaugmented benchmarks use training-fold profile indices and analytic integration. At the same fitted links, their comparison with corrected scores isolates the effect of augmentation. Their fixed-index variances are benchmarks rather than a guarantee of valid learned-index inference.
\end{remark}

\subsection{Full and Reduced Influence Functions}
\label{sub:app_eif}

For either $W=\bM$ or fixed $W=V_{\btheta_0}$, the observed law is factored as $p(\bz)p(x\mid\bz)p(w\mid x,\bz)p(y\mid x,w,\bz)$. The target is $\E_{\bZ}\int\mu(1,w,\bZ)p(w\mid0,\bZ)\,dw$. Along a regular submodel, the derivative contains variation in the treated outcome mean, the control mediator law, and the marginal covariate law. The corresponding representers are
\[
 \frac{X}{\pi}r\{Y-\mu(1,W,\bZ)\},\quad
 \frac{1-X}{1-\pi}\{\mu(1,W,\bZ)-\phi(0,\bZ)\},\quad
 \phi(0,\bZ)-\zeta_0.
\]
Each term has mean zero, and their sum represents the target derivative against every component of the observed-law score. The propensity component has derivative zero. Under the unrestricted law, the tangent space is the mean-zero $L^2$ space, so the square-integrable representer is the canonical gradient. Theorems~\ref{thm:eif_full} and~\ref{thm:eif} follow, with the causal interpretation given by the identification conditions, as in \citet{tchetgen2012semiparametric}. The AIPW representer is $\ind(X=x)\{Y-\xi_x(\bZ)\}/p_x(\bZ)+\xi_x(\bZ)-\delta_x$. The NDE and NIE gradients follow by linearity. For the unaugmented fixed-index score, Proposition~\ref{prop:R1} and the analogous AIPW expansion give the asymptotically linear expansion at a fully correct nuisance law, establishing Theorem~\ref{thm:localeff}. Theorem~\ref{thm:causal_an} gives the augmented learned-index expansion separately. The argument above concerns the unrestricted law rather than a projection onto the smaller additive/Gaussian structural-model tangent space.

\subsection{Local Coordinates and the Constrained Score}
\label{sub:app_lemmas}

At an exact constrained optimum $\tilde{\btheta}$ on the active unit sphere, the first-order condition is
\[
 (\matI_{s_0}-\tilde{\btheta}\tilde{\btheta}^{\top})
       \nabla Q_n(\tilde{\btheta})=\matzero.
\]
The projection is evaluated at the optimizer rather than the unknown true direction. Under the chart $\bchi$ in Appendix~\ref{sub:app_scad_asym}, the exact score is
\[
 \nabla q_n(\bt)=D\bchi(\bt)^\top
                       \nabla Q_n\{\bchi(\bt)\}.
\]
Only at $\bt=\matzero$ does $D\bchi(\bt)=\bU$. The chart Hessian also includes constraint curvature and adjustments for profiling out $(\gamma_0,\gamma,f,\bdelta)$. Residualizing the mediators against $V$ alone generally does not provide the full profile adjustment.

\subsection{Proof of Theorem~\ref{thm:selection}}
\label{sub:app_proof_selection}

\begin{proof}
 On the active chart ball, the nonzero coordinates retain their true signs because $r_n=o(c_{\min,n})$. The restricted penalty is therefore linear. The gradient of the smooth anchor term is bounded by a constant, as its denominator is bounded away from zero. Thus, the gradient of the restricted objective at $\bu_0$ is $O_p(\sqrt{s_0}\lambda_n)$. Its Hessian has minimum eigenvalue at least $m$ throughout the active ball. By Taylor's inequality, its boundary value exceeds its value at $\bu_0$ with probability tending to one, since $\sqrt{s_0}\lambda_n=o(r_n)$. The restricted minimizer $\bar\bu$ is consequently interior and unique on the active ball. Strong convexity gives
 \[
  \|\bar\bu-\bu_0\|_2
       \leq m^{-1}\|\nabla_A L_n(\bu_0)
                         +\lambda_n\sign(\bu_{0A})\|_2
       =O_p(\sqrt{s_0}\lambda_n).
 \]
 The minimizer has support $A$. At $\bar\bu$, the active penalty subgradient cancels the active gradient. Under Assumption~\ref{ass:irrep}, each inactive subgradient has a margin of at least $\tau\lambda_n$. For a sufficiently small perturbation $\bdeta$, Taylor expansion gives a lower bound on the objective increase consisting of a positive active quadratic term and $\tau\lambda_n\|\bdeta_{A^c}\|_1$, up to bounded cross and inactive quadratic terms. The cross term is absorbed into half the active quadratic term and a constant multiple of $\|\bdeta_{A^c}\|_2^2$. In a sufficiently small neighborhood, the linear inactive margin dominates the remaining quadratic term. Thus, $\bar\bu$ is a strict local minimum of the full chart criterion. The chart is locally Lipschitz and preserves inactive zeros, which establishes the rate and support of its image. When $A$ is empty, the inactive-margin argument suffices.
\end{proof}

\subsection{Proof of Theorem~\ref{thm:oracle}}
\label{sub:app_proof_oracle}

\begin{proof}
 On the event of correct screening, whose probability tends to one, consider the two local solutions. Under fixed dimension, nonzero true active coefficients and consistency of both solutions, their coordinates exceed $a\mu_n$ in absolute value with probability tending to one. The SCAD penalty is then locally constant, so the SCAD and oracle solutions satisfy the same chart score equation. Positive chart curvature gives uniqueness in the stated neighborhood. Taylor expansion gives
 \[
  \sqrt n\,\hat{\bt}
   =-\bH_T^{-1}\frac1{\sqrt n}\sum_{i=1}^n
                              \bs_T(\mathcal O_i)+o_p(1).
 \]
 The fixed-law multivariate central limit theorem and
 $\bchi(\hat{\bt})-\btheta_{0,\mathcal S_0}
   =\bU\hat{\bt}+O_p(\|\hat{\bt}\|_2^2)$ yield
 \eqref{eq:clt_theta}--\eqref{eq:avar_theta}. Incorrect screening has vanishing probability and leaves this distributional limit unchanged. Under the additional numerical-equivalence condition, the final statement follows from Slutsky's theorem. Verification of that condition for normalized working regressions is beyond this argument.
\end{proof}

\subsection{Proof of Theorem~\ref{thm:causal_an}}
\label{sub:app_proof_causal}

\begin{proposition}[Cross-fitting remainder]
  \label{prop:R1}
  Under Assumption~\ref{ass:cov_rates}, at the true index $\btheta_0$ the fold-assembled difference $R_1=\Pn\{\hat h_{\btheta_0}-h_{\btheta_0}\}$ is $\op(n^{-1/2})$.
\end{proposition}

\begin{proof}
  Conditional on the training data for fold $k$, suppress the fold superscript and let $\E_x$ denote expectation over $W\mid X=x,\bZ$. Under the error notation in Appendix~\ref{sub:app_cov_asym}, the conditional mean drift of the uncentered mediation score is
  \begin{equation}
    \begin{split}
    D_k&=P_0(\hat h-h)\\
       &=\E_{\bZ}\!\left[
          \frac{\hat\pi-\pi}{1-\hat\pi}
          \left\{\frac{\E_0\Delta\mu}{\hat\pi}-\Delta\phi\right\}
          -\frac{\pi}{\hat\pi}\E_1\{\Delta r\,\Delta\mu\}
        \right].
    \end{split}
    \label{eq:cov_drift}
  \end{equation}
  The identity follows by conditioning the three score terms on $(X,\bZ)$ and using $\E_1(r\Delta\mu)=\E_0\Delta\mu$ and $\phi=\E_0\mu$. Under the propensity bounds, conditional Jensen's inequality and Cauchy--Schwarz give
  \[
     |D_k|\leq C\{e_\pi(e_\mu+e_\phi)+e_r e_\mu\}=\op(n^{-1/2}).
  \]
  The drift terms need not vanish because cross-fitting controls centered fluctuations rather than conditional means. The centered term $(P_{n,k}-P_0)(\hat h-h)$ has conditional variance at most $\|\hat h-h\|_{L^2(P_0)}^2/|I_k|=\op(n^{-1})$ and is therefore $\op(n^{-1/2})$. Summing over the fixed number of folds establishes the result. Covariance estimation enters through $\Delta r$ and $\Delta\phi$, including residual-mean and shrinkage errors. Dependence between these errors within a training fold is allowed.
\end{proof}

\paragraph{Profile derivatives.}
At the truth, perturbing the index along a tangent vector changes the unprofiled mean by $f_0'(V_0)\bU^\top\bM_{\mathcal S_0}$. Differentiation of the population nuisance normal equations removes its projection onto $\mathcal H_0$, so the derivative of the profiled mean is $\bD$. The envelope derivative of the squared-error loss is $-2\eta\bD$. The Hessian is
\[
 2\E\{\bD\bD^\top\}-2\E\{\eta\nabla^2_{\bt}m_{\matzero}\}
       =2\E\{\bD\bD^\top\},
\]
because the second mean derivative is a function of $(X,\bM,\bZ)$ and $\E(\eta\mid X,\bM,\bZ)=0$. 

At the true index, if $\hat m-m_0\in\mathcal H_0$ and the fitted projection score is $\hat\bs=-2(Y-\hat m)\hat\bD$, then conditioning on the training data gives
\[
 P_0(\hat\bs-\bs_T)
       =2P_0\{(\hat m-m_0)(\hat\bD-\bD)\}.
\]
The terms $P_0(\eta\hat\bD)$ and $P_0\{(\hat m-m_0)\bD\}$ vanish by the conditional mean and projection properties. Thus a sufficient true-index drift bound is $\|\hat m-m_0\|_2\|\hat\bD-\bD\|_2=o_p(n^{-1/2})$. Approximation errors in the nuisance span or its projection, and local perturbations from a learned index, require the additional uniform bounds in \eqref{eq:joint_score_remainders}.

\paragraph{The mediation gradient.}
Consider a smooth path $\btheta(t)$ through $\btheta_0$, with derivative $\bu$ supported on $\mathcal S_0$ and tangent to the unit sphere. Let $a_x(\bZ)=\btheta_0^\top\bmu_x(\bZ)$, $b_{x,t}(\bZ)=\btheta(t)^\top\bmu_x(\bZ)$, and $\tau_0=\btheta_0^\top\bSigma_\epsilon\btheta_0$. Gaussian conditioning in the treated arm gives
\begin{align*}
 V_0\mid V_{\btheta(t)}=v,X=1,\bZ
   \sim N\{a_1+\kappa_t(v-b_{1,t}),\,\omega_t^2\},\\
 \kappa_t=\frac{\btheta_0^\top\bSigma_\epsilon\btheta(t)}
                   {\btheta(t)^\top\bSigma_\epsilon\btheta(t)}, \qquad
 \omega_t^2=\tau_0-
  \frac{\{\btheta_0^\top\bSigma_\epsilon\btheta(t)\}^2}
                  {\btheta(t)^\top\bSigma_\epsilon\btheta(t)}.
\end{align*}
Consequently, the true reduced conditional outcome mean is $\gamma_0+\gamma+g_2(\bZ)$ plus the expectation of $f_0$ under this conditional Gaussian law. At zero, $\kappa_0=1$, $\omega_0^2=0$, $\dot\kappa_0=-\bu^\top\bSigma_\epsilon\btheta_0/\tau_0$, and $\left.\partial_t\omega_t^2\right|_0=0$. Under the stated smoothness and envelope conditions,
\begin{equation}
 \left.\partial_t\mu_{\btheta(t)}
       \{1,V_{\btheta(t)},\bZ\}\right|_0
    = f_0'(V_0)\bu^\top
         [\bM-\E\{\bM\mid V_0,X=1,\bZ\}].
 \label{eq:reduced_regression_derivative}
\end{equation}
The ratio derivative in the mediation score has zero expectation because the outcome residual has mean zero given $(X,\bM,\bZ)$ at the truth. The derivative of the conditional cross-world mean cancels between the control and final score terms. The conditional-mean component subtracted in \eqref{eq:reduced_regression_derivative} depends only on $(V_0,\bZ)$ and thus also cancels under index-ratio reweighting. Hence
\[
 \bu^\top\bG=\bu^\top\E_{\bZ}\left[
   \E_0\{f_0'(V_0)\bM\}
    -\E_1\{r_{\btheta_0}(V_0,\bZ)f_0'(V_0)\bM\}\right].
\]
For the same Gaussian model,
\[
 \E(\bM\mid V_0=v,X=x,\bZ)
   =\bmu_x(\bZ)+\frac{\bSigma_\epsilon\btheta_0}{\tau_0}
          \{v-a_x(\bZ)\}.
\]
The control-minus-treated difference is $-\balpha+\bSigma_\epsilon\btheta_0(\btheta_0^\top\balpha)/\tau_0$. Reweighting the treated index distribution to the control distribution gives \eqref{eq:index_gradient_gaussian} on the active tangent space, as required by the inference theorem. For the full ratio, the reweighting identity applies to every function of the entire mediator vector. The population index derivative is therefore zero.

\begin{proof}[Proof of Theorem~\ref{thm:causal_an}]
Let $\hat a_k=\hat h_k-(\hat\bA_T^{(k)})^\top\hat\bs_k$ and $a_0=h_{\matzero}-\bA_T^\top\bs_T$. For the argument, use oracle-support versions of each training fit that agree with the actual fit whenever that complement selects correctly. Their simultaneous agreement has probability tending to one. Condition only on the training data for fold $k$, without conditioning on the simultaneous selection event. Taylor expansion of the population reference scores gives
\begin{equation*}
 P_0h_{\hat\bt_k}-\zeta_0
    =\bG_T^\top\hat\bt_k+O_p(\|\hat\bt_k\|^2), \qquad
 P_0\bs_{\hat\bt_k}
    =\bH_T\hat\bt_k+O_p(\|\hat\bt_k\|^2).
\end{equation*}
Combining these expressions with \eqref{eq:joint_score_remainders} yields
\[
 P_0\hat a_k-\zeta_0
  =\{\bG_T^\top-(\hat\bA_T^{(k)})^\top\bH_T\}\hat\bt_k
     +O_p(\|\hat\bt_k\|^2)+o_p(n^{-1/2})
  =o_p(n^{-1/2}).
\]
The first-order training-index contribution therefore cancels within each fold. Independence of the training samples is not required.

By the $L^2$ conditions and coefficient consistency, $\|\hat a_k-a_0\|_{L^2(P_0)}=o_p(1)$. Conditional on the training data, the centered evaluation term satisfies
\[
 (P_{n,k}-P_0)(\hat a_k-a_0)=o_p(n^{-1/2}),
\]
by its conditional variance bound $\|\hat a_k-a_0\|_{L^2(P_0)}^2/|I_k|$. A finite sum of these foldwise bounds is still $o_p(n^{-1/2})$, although the training complements overlap. Therefore
\begin{align}
 \hat\zeta_{0,\mathrm I}^{\mathrm{aug}}-\zeta_0
 &=\Pn(a_0-\zeta_0)
    +\sum_k\frac{|I_k|}{n}
       \{(P_{n,k}-P_0)(\hat a_k-a_0)
                              +P_0\hat a_k-\zeta_0\}\nonumber\\
 &=\Pn\{\varphi_{\zeta_0,\mathrm I}
                  -\bG_T^\top\bH_T^{-1}\bs_T\}
       +o_p(n^{-1/2}).
 \label{eq:decomposition}
\end{align}
The evaluation-fold profile score has already canceled the first-order training-index drift. No additional training-index influence function is required. The index-estimation contribution appears once, in $-\bG_T^\top\bH_T^{-1}\bs_T$.

The analogous cross-fitted AIPW expansions give
\[
 \hat\delta_x^{\mathrm{dr}}-\delta_x
       =\Pn\varphi_{\delta_x}+o_p(n^{-1/2}),\qquad x=0,1.
\]
Taking the NDE and NIE differences gives \eqref{eq:joint_effect_if}. Their sum has the TE influence function $\varphi_{\delta_1}-\varphi_{\delta_0}$. The assumed joint central limit theorem gives the distributional result. Consistent screening makes the exceptional event negligible.

For SIME-F, the true full density ratio reweights every function of $(\bM,\bZ)$, giving zero population index gradient. The fixed-index nuisance drift, local remainder bounds and cross-fitting give the usual full-model influence-function expansion. Consistency under partially misspecified nuisance limits follows separately from the exact drift identity and score laws of large numbers. An augmented estimate additionally requires a correction with vanishing average. 
\end{proof}

\subsection{Variance Estimation and Confidence Intervals}
\label{sub:app_variance_consistency}

\begin{lemma}[Consistency of the estimated score covariance]
 \label{lem:score_covariance}
 Let $\hat{\mathbf q}_i\in\R^d$ be fitted complete effect score contributions and let $\mathbf q_i^0=\boldsymbol{\tau}_n+\boldsymbol{\Phi}_i^0$ be their true counterparts, where $d$ is fixed, $\boldsymbol{\tau}_n$ is the true effect vector and $\E\boldsymbol{\Phi}_i^0=\matzero$. Suppose
 \begin{equation}
  \frac1n\sum_{i=1}^n\|\hat{\mathbf q}_i-\mathbf q_i^0\|_2^2=o_p(1),\quad
  \frac1n\sum_{i=1}^n\boldsymbol{\Phi}_i^0=o_p(1),\quad 
  \frac1n\sum_{i=1}^n\boldsymbol{\Phi}_i^0\boldsymbol{\Phi}_i^{0\top}
       \xrightarrow{p}\bV,
 \label{eq:score_covariance_conditions}
 \end{equation}
 for a finite matrix $\bV$. Then the centered sample covariance of the fitted vectors, defined as in~\eqref{eq:effect_covariance}, satisfies $\hat{\bV}\xrightarrow{p}\bV$.
\end{lemma}

\begin{proof}
Write $\be_i=\hat{\mathbf q}_i-\mathbf q_i^0$ and let $\Pn$ denote averaging over observations. The conditions give $\Pn\|\boldsymbol{\Phi}^0\|_2^2=O_p(1)$ and $\Pn\|\be\|_2^2=o_p(1)$. By Cauchy--Schwarz,
\[
 \|\Pn(\boldsymbol{\Phi}^0\be^\top)\|_{\mathrm{op}}
 \leq\{\Pn\|\boldsymbol{\Phi}^0\|_2^2\,\Pn\|\be\|_2^2\}^{1/2}
 =o_p(1).
\]
The transposed cross-product has the same bound, and $\|\Pn(\be\be^\top)\|_{\mathrm{op}}\leq\Pn\|\be\|_2^2=o_p(1)$. Thus the empirical second moment of $\hat{\mathbf q}_i-\boldsymbol{\tau}_n$ converges to $\bV$. Also $\|\Pn\be\|_2\leq(\Pn\|\be\|_2^2)^{1/2}=o_p(1)$, so $\Pn(\hat{\mathbf q}-\boldsymbol{\tau}_n)=\Pn\boldsymbol{\Phi}^0+\Pn\be=o_p(1)$. Subtracting the outer product of that mean and multiplying by $n/(n-1)$ proves the result.
\end{proof}

For the complete effect vector in~\eqref{eq:complete_effect_scores}, $d=3$. The centered first-moment condition in~\eqref{eq:score_covariance_conditions} follows from the joint true-score central limit theorem assumed for Theorem~\ref{thm:causal_an}.

Under the conditions of Corollary~\ref{cor:ci}, the nominal 95\% Wald intervals are
\[
 \widehat{\NDE}_{\mathrm I}^{\mathrm{aug}}\pm1.96\hat\sigma_{\NDE,\mathrm I}/\sqrt n,\qquad\text{and}\qquad
 \widehat{\NIE}_{\mathrm I}^{\mathrm{aug}}\pm1.96\hat\sigma_{\NIE,\mathrm I}/\sqrt n.
\]

\begin{proof}[Proof of Corollary~\ref{cor:ci}]
Lemma~\ref{lem:score_covariance} gives consistency of the covariance estimate in~\eqref{eq:effect_covariance}. Theorem~\ref{thm:causal_an} and the positive NDE and NIE limiting variances then yield the stated Wald intervals by Slutsky's theorem. The complete effect scores incorporate the effect--index covariance.
\end{proof}

\subsection{Sensitivity to Mediator--Outcome Error Dependence}
\label{sub:app_sensitivity_model}

Identification of natural effects requires untestable mediator--outcome and cross-world restrictions \citep{andrews2021insights}. The sensitivity analysis retains the invariant potential-outcome equations \eqref{eq:potential_structural}, joint exposure--error independence \eqref{eq:joint_error_exposure}, and centered error means. The baseline error-mean restriction is replaced by the aligned dependence model
\begin{align}
 \rho&=\operatorname{Corr}(\eta,\btheta_0^\top\bepsilon\mid\bZ),
                       \qquad |\rho|<1,\label{eq:rho}\\
 \eta&=c_\rho\btheta_0^\top\bepsilon+u,\qquad
 c_\rho=\rho\frac{\sigma_\eta}{\sigma_\epsilon},\qquad
 u\indep\bepsilon\mid\bZ,\nonumber\\
 u\mid\bZ&\sim\mathcal N\{0,\sigma_\eta^2(1-\rho^2)\},\qquad
 \sigma_\epsilon^2=\btheta_0^\top\bSigma_\epsilon\btheta_0.
 \label{eq:sensitivity_alignment}
\end{align}
The mediator errors are assumed to be conditionally Gaussian with a common covariance. The variances and $\rho$ are constant across $\bZ$. Under this model,
\begin{equation}
 \E(\eta\mid\bepsilon,X,\bZ)
 =\E(\eta\mid\btheta_0^\top\bepsilon,X,\bZ)
 =c_\rho\btheta_0^\top\bepsilon.
 \label{eq:eta_cond}
\end{equation}
The first equality requires an additional alignment assumption, which does not follow from bivariate normality of $(\eta,\btheta_0^\top\bepsilon)$ alone. Under joint Gaussianity, this assumption corresponds to $\operatorname{Cov}(\bepsilon,\eta\mid\bZ)=c_\rho\bSigma_\epsilon\btheta_0$. A nonzero $\rho$ generally violates mediator--outcome exchangeability (Assumption~\ref{ass:conf2}) as well as cross-world independence (Assumption~\ref{ass:cross}). Thus, this model relaxes both assumptions.

Let $\bmu_x(\bz)=\balpha_0+\balpha x+\bg_1(\bz)$. The observed outcome regression is then
\begin{equation}
 \mu^{\mathrm{obs}}(x,\bfm,\bz)
 =\gamma_0+\gamma x+f(\btheta_0^\top\bfm)+g_2(\bz)+c_\rho\{\btheta_0^\top\bfm-\btheta_0^\top\bmu_x(\bz)\}.
 \label{eq:biased_regression}
\end{equation}
Under this model, the observed link is $f(v)+c_\rho v$, and the observed exposure coefficient is $\gamma-c_\rho\btheta_0^\top\balpha$. The sensitivity calculation requires the observed link to identify the same index direction. This condition may fail if the transformed link is constant.

\begin{samepage}
\begin{proposition}[Population sensitivity discrepancy]
 \label{prop:sensitivity}
 Under \eqref{eq:potential_structural}, \eqref{eq:joint_error_exposure}, and~\eqref{eq:sensitivity_alignment}, with consistency and positivity, let $\zeta_0^{\mathrm{gf}}$ denote the observed-data mediation g-formula, and let $\NDE^{\mathrm{gf}}$ and $\NIE^{\mathrm{gf}}$ denote its contrasts with the identified single-treatment means $\delta_x$. Then
 \begin{align}
  \zeta_0^{\mathrm{gf}}-\zeta_0
      &=-c_\rho\btheta_0^\top\balpha,\nonumber\\
  \NDE^{\mathrm{gf}}-\NDE
      &=-\rho\frac{\sigma_\eta}{\sigma_\epsilon}
                                \btheta_0^\top\balpha,\label{eq:bias_NDE}\\
  \NIE^{\mathrm{gf}}-\NIE
      &=\rho\frac{\sigma_\eta}{\sigma_\epsilon}
                                \btheta_0^\top\balpha.\label{eq:bias_NIE}
 \end{align}
 The single-treatment g-formulas equal $\delta_x$, and the total-effect contrast remains identified. 
\end{proposition}
\end{samepage}

The proof is presented in Appendix~\ref{sub:app_proof_sensitivity}. The direct and indirect discrepancies have opposite signs and cancel in the total effect. When $\btheta_0^\top\balpha=0$, there is no discrepancy under this aligned sensitivity model, although other forms of unmeasured confounding may still affect identification.

\paragraph{Calibration from observed residuals.}
The fitted outcome residual estimates $u$, with standard deviation $\sigma_u=\sigma_\eta\sqrt{1-\rho^2}$, rather than the structural error $\eta$. At a fixed sensitivity value $\rho$, residual-based calibration therefore requires $\sigma_\eta(\rho)=\sigma_u/\sqrt{1-\rho^2}$. Let $\hat c_u=(\hat\sigma_u/\hat\sigma_\epsilon)\hat{\btheta}^\top\hat{\balpha}$, where $\hat\sigma_\epsilon^2=\hat{\btheta}^\top\hat{\bSigma}_\epsilon\hat{\btheta}$. The residual-calibrated point-estimate adjustments are
\begin{align}
 \widehat{\NDE}(\rho)&=\widehat{\NDE}
           +\frac{\rho}{\sqrt{1-\rho^2}}\hat c_u,
           \label{eq:NDE_corrected}\\
 \widehat{\NIE}(\rho)&=\widehat{\NIE}
           -\frac{\rho}{\sqrt{1-\rho^2}}\hat c_u.
           \label{eq:NIE_corrected}
\end{align}
If the structural $\sigma_\eta$ is independently known, the adjustment is instead linear in $\rho$, as given in Proposition~\ref{prop:sensitivity}. The two adjustments correspond to different calibrations. Calculation of the outcome residuals and mediator coefficients requires the corresponding fitted models. These quantities are not all included in the standard SIME effect summary.

For $\hat c_u\ne0$, the absolute sensitivity value at which the adjusted NIE point estimate equals zero is
\begin{equation}
 |\rho^*|=\frac{|\widehat{\NIE}|}
      {\sqrt{\widehat{\NIE}^{\,2}+\hat c_u^{\,2}}},\qquad
 \operatorname{sign}(\rho^*)=\operatorname{sign}(\widehat{\NIE}/\hat c_u).
 \label{eq:breakdown}
\end{equation}
This value is a point-estimate threshold rather than the sensitivity value at which a confidence interval first contains zero. Shifting an existing interval treats the estimated adjustment as known. Valid interval inference must incorporate adjustment uncertainty, its covariance with the original effect estimate, and index-estimation uncertainty. 

\paragraph{Comparison with the legacy adjustment.}
The sensitivity experiment compares \eqref{eq:NIE_corrected} with the legacy residual-SD adjustment
\begin{equation}
 \widehat{\NIE}_{\mathrm{legacy}}(\rho)
          =\widehat{\NIE}-\rho\hat c_u,
 \label{eq:legacy_sensitivity}
\end{equation}
which omits $1/\sqrt{1-\rho^2}$. This legacy adjustment uses the same fitted ingredients as residual calibration. Even with perfectly estimated nuisance functions and a known index direction, its limiting residual discrepancy is
\begin{equation}
 \rho\frac{\sigma_\eta}{\sigma_\epsilon}
       \btheta_0^\top\balpha\{1-\sqrt{1-\rho^2}\}.
 \label{eq:legacy_bias}
\end{equation}
For $\sigma_\eta=\sigma_\epsilon=1$ and $\btheta_0^\top\balpha=0.5\sqrt5$, the discrepancy is approximately $0.075$ at $\rho=0.5$ and $0.568$ at $\rho=0.9$. These values agree with the persistent large-sample discrepancies in Appendix~\ref{sub:app_sensitivity}. Thus, the remaining bias cannot be explained solely by finite-sample nuisance estimation. In the simulation study, residual-calibrated adjustment keeps absolute NIE bias below $0.005$ at $n=5{,}000$ across the tested sensitivity values for combined SIME-I and the active-space and full-ratio benchmarks. At $n=500$, the combined strategy reduces the negative-correlation bias at $\rho=-0.1$ but increases it at $\rho=-0.5$. Additional small-sample bias remains at some negative correlations, and adjustment uncertainty still requires separate inference.

\subsection{Proof of Proposition~\ref{prop:sensitivity}}
\label{sub:app_proof_sensitivity}

\begin{proof}
 Under \eqref{eq:potential_structural} and the centered structural error model, $\E(\eta\mid\bZ)=0$. Joint exposure--error independence \eqref{eq:joint_error_exposure} ensures exposure exchangeability for the complete potential-outcome collection induced by these equations. For fixed $x$, integration of \eqref{eq:biased_regression} under the mediator law in the same exposure arm removes the additional term, since $\E\{\btheta_0^\top\bM-\btheta_0^\top\bmu_x(\bZ)\mid X=x,\bZ\}=0$. Therefore consistency and exposure exchangeability identify $\delta_x$ despite mediator--outcome dependence.

 For the cross-world mean, integrating the treated observed regression against the control mediator law gives
 \[
  \zeta_0^{\mathrm{gf}}-\zeta_0
   =c_\rho\E\{\btheta_0^\top\bmu_0(\bZ)
                     -\btheta_0^\top\bmu_1(\bZ)\}
   =-c_\rho\btheta_0^\top\balpha.
 \]
 The structural potential-outcome mean has no additional error term because $\E\eta=0$. Under confounding, the observed conditional regression differs from the structural response. Subtracting $\delta_0$ from the cross-world mean, or subtracting the cross-world mean from $\delta_1$, gives \eqref{eq:bias_NDE}--\eqref{eq:bias_NIE}. The discrepancies cancel in TE. Finally, $\operatorname{sd}(u)=\sigma_\eta\sqrt{1-\rho^2}$ gives the residual calibration and \eqref{eq:legacy_bias}.
\end{proof}

\section{Additional Simulation Results}
\label{app:simulation}

\subsection{Simulation Settings and Data Generation}
\label{sub:app_simulation_settings}

The simulations compare the combined SIME-I estimator in Appendix~\ref{sub:app_combined_strategy} with its active-space benchmark, full-ratio estimators and linear mediation methods. The comparisons examine the effects of mediator correlation, the outcome link and the choice of covariance estimator.
For each observation, data are generated as
\begin{align*}
  \bZ &\sim \mathcal{N}(\matzero,\matI_q), \\
  X\mid\bZ &\sim \mathrm{Bernoulli}\{\mathrm{expit}(0.5Z_1+0.5Z_2)\}, \\
  \bM &= \balpha X+0.3Z_1\matone_p+\bepsilon,
  \qquad \bepsilon\sim\mathcal{N}(\matzero,\bSigma_{\varrho}), \\
  Y &= 0.5X+f_0(\btheta_0^\top\bM)+0.5Z_1+\eta,
  \qquad \eta\sim\mathcal{N}(0,1),
\end{align*}
where $(\bSigma_{\varrho})_{jk}=\varrho^{|j-k|}$, $\btheta_0=(1,1,1,1,1,0,\ldots,0)^\top/\sqrt{5}$, $\alpha_j=0.5\ind(j\leq5)$, and $q=5$. The mediator errors are independent of $(X,\bZ)$, and the outcome error is independent of $(X,\bZ,\bepsilon)$. In this design, $\varrho$ determines the conditional correlation among mediators, while the marginal error variances remain one and cross-world independence holds. The outcome--mediator error correlation $\rho$ considered in Appendix~\ref{sub:app_sensitivity_model} is distinct from the mediator correlation.

Two outcome links are considered: \textbf{Setting~1 (linear)}, $f_0(t)=t/\sqrt{5}$, and \textbf{Setting~2 (nonlinear)}, $f_0(t)=\sin(t)$. The combinations of $p\in\{50,500\}$, $n\in\{500,5{,}000\}$, and $\varrho\in\{0,0.4,0.8\}$ yield 24 settings, each with 500 replications. Covariates and exposures are generated in each replication. The evaluation concerns the population natural effects. 
At a given $(p,n)$, base random draws are shared across the outcome links and correlation levels.

The true NDE is $0.5$ in all settings. In Setting~1, the true NIE is $0.5$. For Setting~2, let $a=\btheta_0^\top\balpha=0.5\sqrt{5}$, $b=0.3\sqrt{5}$, and
\[
  v_{\varrho}=\btheta_0^\top\bSigma_{\varrho}\btheta_0
  =1+\frac{2}{5}\sum_{h=1}^{4}(5-h)\varrho^h.
\]
The population NIE is then
\[
  \NIE=\sin(a)\exp\{-(v_{\varrho}+b^2)/2\},
\]
which takes values $0.435525$, $0.278615$, and $0.117429$ at $\varrho=0$, $0.4$, and $0.8$, respectively. The true effects are the same across $p$. Increasing mediator correlation therefore changes both the covariance structure and the magnitude of the nonlinear indirect effect. 

\subsection{Implementation Details}
\label{sub:app_simulation_implementation}

The principal SIME-I uses projected residual variance, analytic Gaussian integration and the combined strategy in Appendix~\ref{sub:app_combined_strategy}. Index fitting, preprocessing and tuning use the training complement of each of five evaluation folds. Corrected standard errors use the complete effect scores. The unaugmented benchmarks use fixed-index standard errors, and SIME-F uses the full-ratio scores.

The link is modeled by a natural cubic spline in the training-standardized index. It uses equally spaced knots on $[-2.5,2.5]$ with linear extrapolation outside this range. The number of knots is $\lceil3N^{1/5}\rceil$, bounded below by six and above by 30, where $N$ is the training-sample size. Generalized cross-validation selects the roughness penalty at the initial direction. The selected value is held fixed during directional optimization and reconsidered at the selected direction. A second optimization is performed when the absolute log-ratio of the new and previous penalties is at least $0.05$. Otherwise, the previous penalty and solution are retained. Mediators are standardized using training means and standard deviations. The Stage~1 outcome is standardized and rescaled so that the profile criterion has unit average curvature at the starting direction. Working-LASSO screening uses five-fold inner cross-validation with minimum-error tuning, at most 20 updates, and at most $\lfloor N/\log N\rfloor$ retained mediators. If the initial linear LASSO selects nothing, the $\lceil\sqrt N\rceil$ largest marginal partial correlations initialize screening. These implementation settings do not guarantee support recovery.

The SCAD shape parameter is $a=3.7$~\cite[recommended in][]{fan2001variable}. On the screened support of size $s$, tuning is $\mu=\tau\sqrt{\log(s+1)/N}$, where $\tau$ is the root-mean-square standard deviation of the profile-score coordinates at the starting direction. Optimization uses chart proximal-gradient updates followed by quasi-Newton refinement, with initializations at the working-LASSO and least-squares directions. The solution with the smaller penalized criterion is retained. The profile score uses index-residualized mediators before the link and covariate nuisance projection.

Prediction uses $\max\{1001,\lceil100N^{1/3}\rceil\}$ equally spaced grid points spanning the training indices and fitted arm means plus or minus six projected standard deviations, with constant tails. The same interpolant is integrated analytically. Linear mediator-mean regressions and logistic propensity regressions are fitted on the training data, and held-out propensities are clipped to $[0.01,0.99]$. Density ratios are not truncated. 

Three covariance specifications are compared for each ratio route: fixed identity, unadjusted residual sample covariance, and residual-contrast Ledoit--Wolf (LW) shrinkage. With $d=7$ mean-model coefficients and residual covariance $\bS$ with divisor $N$, the sample option is $\bS$, and LW uses the degrees-of-freedom-adjusted covariance $N\bS/(N-d)$, a scaled-identity target matching its average variance, and a data-adaptive shrinkage weight. The residual degrees of freedom are 393 and 3,993 at $n=500$ and $5{,}000$, respectively. The principal projected estimator fits the index mean directly and divides its residual sum of squares by $(N-d)$. It does not estimate the full covariance. Active residual cross-products supply the covariance action used by the augmentation. Each covariance specification is used consistently for the density ratio and outcome integration, and the index-ratio comparisons use augmentation.

At $p=n=500$, the full/sample covariance is singular because each training fold has 400 observations and at most 393 residual degrees of freedom. These fits are recorded as unavailable. No ridge or generalized inverse is substituted. The index/sample route uses only a positive projected variance and remains available. The support-Oracle counterparts restrict the candidate index support to the first five mediators and run the same constrained SCAD profile refit. The true index coefficients, link, and covariance are not supplied. SCAD may still omit active mediators. Within each SIME or support-Oracle family, all eight variants reuse one set of Stage~1 training-fold index fits and evaluation folds. The combined strategy refits the links while retaining those directions and supports. Its correction may use additional mediator coordinates without changing the fitted index support.

The eight SIME, eight support-Oracle and three competitor configurations give 19 fits per dataset. For the combined fits, the Stage~1 spline basis is retained and the one-standard-error rule uses five inner folds and twelve roughness values extending from the GCV choice toward an almost linear fit. Five-fold cross-validation selects the broader correction penalty from 15 logarithmically spaced values between $\lambda_{\max}$ and $0.5\lambda_{\max}$. The curvature and mediator-gradient moments are recomputed within the inner training and validation subsets, while the link, nuisance projection and mean integrated link slope are fitted on the outer training complement. The final correction refit uses a $10^{-8}$ ridge and at most $2\lfloor\sqrt N\rfloor$ added coordinates.

\paragraph{Linear mediation comparisons.}
\textbf{LinMed}, \textbf{simplified HIMA} and \textbf{JonesPRD} are fitted to each full simulated dataset. For these methods, let $\hat\alpha_j$ denote the fitted exposure coefficient in the mediator regression and $\hat\beta_j$ the corresponding coefficient in the linear outcome regression. With $\hat\gamma$ denoting the outcome exposure coefficient, the effect estimates are
\[
 \widehat{\NDE}=\hat\gamma,\qquad
 \widehat{\NIE}=\sum_{j=1}^p\hat\alpha_j\hat\beta_j,\qquad
 \widehat{\TE}=\widehat{\NDE}+\widehat{\NIE}.
\]

\paragraph{LinMed.}
The outcome is regressed on exposure, mediators and covariates using LASSO, with only the mediator coefficients penalized. Five-fold cross-validation selects the penalty minimizing prediction error. Separate ordinary least-squares regressions of each mediator on exposure and covariates estimate the exposure--mediator coefficients. Point estimates retain the penalized outcome coefficients. An ordinary least-squares refit on exposure, covariates and the selected mediators provides the NDE standard error and the outcome-coefficient variances used in a component-wise NIE delta approximation. The mediator-coefficient variance approximation uses its regression residual variance divided by $\sum_i(X_i-\bar X)^2$. Normal 95\% intervals are centered on the penalized effect estimates. These intervals are working approximations because coefficient cross-covariances and selection uncertainty are omitted.

\paragraph{Simplified HIMA.}
Following the screening and MCP components of \citet{zhang2016estimating}, mediators are ranked by absolute outcome-coefficient $t$-statistics from separate regressions adjusting for exposure and covariates. The top $\min\{p,\lceil n/\log n\rceil\}$ mediators enter a joint MCP outcome regression, with exposure and covariates unpenalized. A BIC-type criterion selects the penalty along the MCP path. Exposure--mediator coefficients are estimated by ordinary least squares for the selected mediators. The simplified implementation reports point estimates and the selected support, without standard errors or confidence intervals.

\paragraph{JonesPRD.}
The product-weight adaptive LASSO comparison follows \citet{jones2025causal}. Robinson residualization uses five-fold cross-fitting: linear working regressions of the outcome and each mediator on $\bZ$, and logistic regression of exposure on $\bZ$, provide held-out residuals. Fitted propensities are clipped to $[0.01,0.99]$. Each exposure--mediator coefficient is estimated by regressing the residualized mediator on residualized exposure. A ridge regression of residualized outcome on residualized exposure and mediators provides pilot outcome coefficients. The subsequent adaptive LASSO uses mediator weights
\[
 w_j=\min\!\left\{
       \max\!\left(|\hat\alpha_j\hat\beta_{j,\mathrm{pilot}}|,n^{-1/2}\right)^{-\kappa},
       10^6\right\},\qquad \kappa\in\{1,2,3,4\},
\]
with exposure unpenalized. Five-fold minimum-error cross-validation selects the ridge and LASSO penalties and the exponent $\kappa$, using the same tuning folds. The final point estimates retain the penalized coefficients after setting mediator coefficients below $10^{-8}$ in magnitude to zero.

Inference is based on 500 perturbation draws using independent exponential observation weights with mean one. Each draw recomputes the mediator coefficients, ridge pilot and adaptive weights, and refits the adaptive LASSO. The residualized data and selected tuning parameters are held fixed. Perturbation standard deviations estimate standard errors, and nominal 95\% basic intervals are formed from quantiles of the centered perturbation errors.


\subsection{Evaluation Criteria}
\label{sub:app_simulation_evaluation}

Empirical bias, root mean squared error (RMSE), and coverage probability (CP) of nominal 95\% intervals are reported for NDE and NIE. For a selected support $\hat{\mathcal S}$, $\mathrm{TPR}=|\hat{\mathcal S}\cap\mathcal S_0|/5$ and $\mathrm{FPR}=|\hat{\mathcal S}\setminus\mathcal S_0|/(p-5)$. 
Appendix~\ref{sub:app_covariance} separately reports selection averaged over the five training-fold indices used for effect estimation. 
CP is unavailable for HIMA and the singular full/sample configurations. 

\subsection{Simulation Results}
\label{sub:app_simulation_results}

Tables~\ref{tab:sim_linear_p50} and~\ref{tab:sim_nonlinear_p500_n5000} compare effect estimation, interval coverage and mediator selection for the two outcome links. Bold SIME-I and Oracle-I entries use the combined strategy. Full-ratio entries use LW covariance. A dash indicates an unavailable fit or interval, and $a\mathrm{e}{+b}$ denotes $a\times10^b$.

\begin{figure}
\centering
\includegraphics[width=0.95\textwidth]{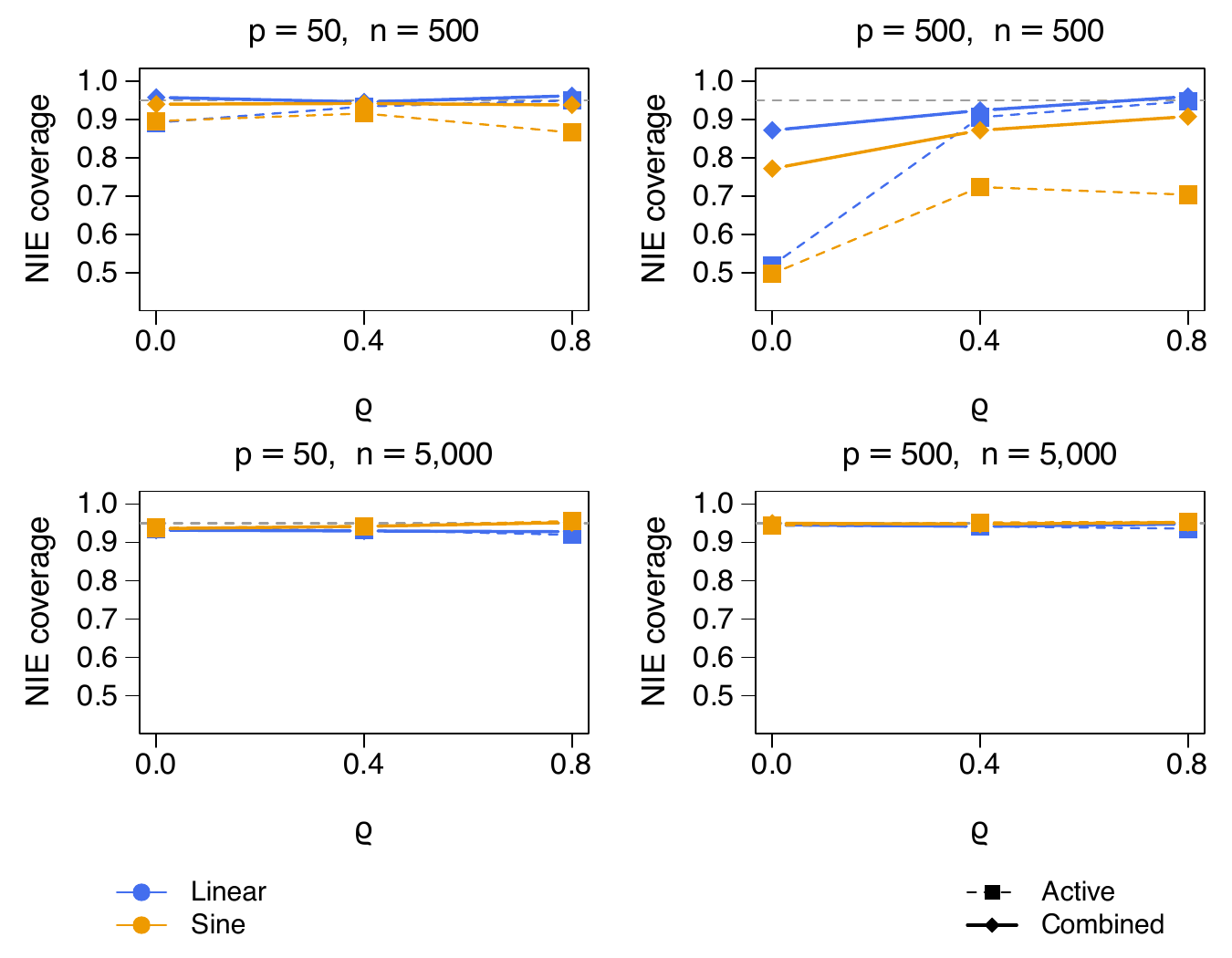}
\caption{NIE interval coverage for active-space and combined SIME-I over 500 replications per setting. Blue: linear link; orange: nonlinear link. Dashed squares: active-space correction; solid diamonds: combined strategy. Gray dashed lines mark nominal 95\% coverage.}
\label{fig:inference_comparison}
\end{figure}

Combined SIME-I gives small bias and coverage close to nominal in larger samples. In smaller samples, Figure~\ref{fig:inference_comparison} shows improved NIE coverage relative to the active-space benchmark, although some undercoverage remains when the mediator dimension equals the sample size. The smaller bias for support-Oracle-I in the more difficult settings suggests that support and index estimation contribute to the remaining error.

HIMA and JonesPRD are competitive in RMSE under the linear link. Under the nonlinear link, the linear methods show larger NIE bias, and the intervals reported by LinMed and JonesPRD often have lower coverage than SIME-I. Mediator selection improves with sample size and becomes more difficult with stronger correlation in the smaller samples.

\begingroup
\setlength{\tabcolsep}{3pt}
\renewcommand{\arraystretch}{0.95}
\newcommand{\simresultheaders}{%
  \hline
  & & & & \multicolumn{3}{c}{NDE} & \multicolumn{3}{c}{NIE} & \multicolumn{2}{c}{Selection} \\
  \cmidrule(lr){5-7} \cmidrule(lr){8-10} \cmidrule(lr){11-12}
  \multirow{-2}{*}{\makebox[1.6em]{$p$}} &
  \multirow{-2}{*}{$n$} &
  \multirow{-2}{*}{$\varrho$} &
  \multicolumn{1}{c}{\multirow{-2}{*}{Method}} &
  \multicolumn{1}{c}{\makebox[3.6em]{Bias}} &
  \multicolumn{1}{c}{\makebox[3.6em]{RMSE}} &
  \multicolumn{1}{c}{\makebox[2.1em]{CP}} &
  \multicolumn{1}{c}{\makebox[3.6em]{Bias}} &
  \multicolumn{1}{c}{\makebox[3.6em]{RMSE}} &
  \multicolumn{1}{c}{\makebox[2.1em]{CP}} &
  \multicolumn{1}{c}{\makebox[2.5em]{TPR}} &
  \multicolumn{1}{c}{\makebox[2.5em]{FPR}} \\
  \hline
}

\begin{table}
\caption{Simulation Setting~1 (linear). Effect estimation and selection in $\btheta$ over 500 replications per $(p,n,\varrho)$ combination. RMSE: root mean squared error. CP: coverage probability of nominal 95\% intervals. CP is given in \%. TPR/FPR: true/false positive rates.}
\label{tab:sim_linear_p50}
\centering
{\renewcommand{\baselinestretch}{1}\normalsize
\simtablefit{%
\begin{tabular}{@{}c c c l r r r r r r r r@{}}
\simresultheaders
&  &  & \textbf{SIME-I} & 0.002 & 0.142 & 95.4 & -0.003 & 0.117 & 95.8 & 0.982 & 0.066 \\ 
&  &  & \textbf{Oracle-I} & -0.002 & 0.143 & 96.2 & 0.001 & 0.118 & 97.0 & 0.999 & 0.000 \\ 
&  &  & SIME-F/LW & -0.026 & 0.285 & 98.0 & 0.026 & 0.267 & 97.4 & 0.982 & 0.066 \\ 
&  &  & Oracle-F/LW & -0.016 & 0.277 & 98.2 & 0.015 & 0.260 & 97.6 & 0.999 & 0.000 \\ 
&  &  & LinMed & 0.150 & 0.184 & 73.0 & -0.150 & 0.165 & 33.6 & 0.998 & 0.216 \\ 
&  &  & HIMA & 0.080 & 0.143 & -- & -0.080 & 0.122 & -- & 0.975 & 0.023 \\ 
&  & \multirow{-7}{*}{0.0} & JonesPRD & 0.087 & 0.148 & 86.8 & -0.089 & 0.132 & 70.4 & 0.989 & 0.110 \\ 
\cline{3-12}
&  &  & \textbf{SIME-I} & 0.005 & 0.123 & 96.0 & -0.005 & 0.100 & 94.6 & 0.965 & 0.085 \\ 
&  &  & \textbf{Oracle-I} & 0.000 & 0.119 & 95.4 & 0.000 & 0.098 & 94.8 & 0.994 & 0.000 \\ 
&  &  & SIME-F/LW & -0.028 & 0.341 & 97.2 & 0.028 & 0.329 & 97.6 & 0.965 & 0.085 \\ 
&  &  & Oracle-F/LW & -0.018 & 0.341 & 97.6 & 0.018 & 0.329 & 97.2 & 0.994 & 0.000 \\ 
&  &  & LinMed & 0.095 & 0.136 & 86.2 & -0.094 & 0.115 & 69.2 & 0.998 & 0.144 \\ 
&  &  & HIMA & 0.048 & 0.116 & -- & -0.048 & 0.094 & -- & 0.860 & 0.016 \\ 
&  & \multirow{-7}{*}{0.4} & JonesPRD & 0.029 & 0.105 & 93.8 & -0.030 & 0.089 & 88.8 & 0.972 & 0.054 \\ 
\cline{3-12}
&  &  & \textbf{SIME-I} & 0.009 & 0.113 & 96.0 & -0.007 & 0.104 & 96.2 & 0.706 & 0.074 \\ 
&  &  & \textbf{Oracle-I} & 0.005 & 0.112 & 95.6 & -0.002 & 0.104 & 94.8 & 0.856 & 0.000 \\ 
&  &  & SIME-F/LW & -0.012 & 0.329 & 97.2 & 0.015 & 0.319 & 97.0 & 0.706 & 0.074 \\ 
&  &  & Oracle-F/LW & -0.005 & 0.319 & 97.4 & 0.008 & 0.310 & 96.4 & 0.856 & 0.000 \\ 
&  &  & LinMed & 0.058 & 0.112 & 93.4 & -0.055 & 0.096 & 94.0 & 0.975 & 0.101 \\ 
&  &  & HIMA & 0.023 & 0.104 & -- & -0.020 & 0.089 & -- & 0.543 & 0.018 \\ 
& \multirow{-21}{*}{500} & \multirow{-7}{*}{0.8} & JonesPRD & 0.014 & 0.096 & 94.4 & -0.013 & 0.091 & 95.0 & 0.886 & 0.049 \\ 
\cline{2-12}
&  &  & \textbf{SIME-I} & 0.002 & 0.048 & 96.2 & -0.001 & 0.039 & 93.2 & 1.000 & 0.085 \\ 
&  &  & \textbf{Oracle-I} & 0.002 & 0.047 & 96.0 & -0.001 & 0.039 & 93.2 & 1.000 & 0.000 \\ 
&  &  & SIME-F/LW & 0.002 & 0.048 & 97.0 & -0.001 & 0.040 & 94.0 & 1.000 & 0.085 \\ 
&  &  & Oracle-F/LW & 0.002 & 0.048 & 97.0 & -0.001 & 0.040 & 93.8 & 1.000 & 0.000 \\ 
&  &  & LinMed & 0.047 & 0.058 & 72.0 & -0.046 & 0.051 & 36.6 & 1.000 & 0.226 \\ 
&  &  & HIMA & 0.001 & 0.034 & -- & -0.001 & 0.023 & -- & 1.000 & 0.001 \\ 
&  & \multirow{-7}{*}{0.0} & JonesPRD & 0.007 & 0.034 & 94.0 & -0.006 & 0.024 & 91.0 & 1.000 & 0.036 \\ 
\cline{3-12}
&  &  & \textbf{SIME-I} & 0.001 & 0.038 & 95.4 & 0.000 & 0.030 & 93.0 & 1.000 & 0.085 \\ 
&  &  & \textbf{Oracle-I} & 0.001 & 0.037 & 95.2 & 0.000 & 0.030 & 93.2 & 1.000 & 0.000 \\ 
&  &  & SIME-F/LW & 0.002 & 0.040 & 96.2 & -0.001 & 0.033 & 93.4 & 1.000 & 0.085 \\ 
&  &  & Oracle-F/LW & 0.002 & 0.040 & 96.4 & -0.001 & 0.033 & 93.4 & 1.000 & 0.000 \\ 
&  &  & LinMed & 0.030 & 0.044 & 83.4 & -0.030 & 0.037 & 70.4 & 1.000 & 0.150 \\ 
&  &  & HIMA & 0.001 & 0.032 & -- & -0.001 & 0.023 & -- & 1.000 & 0.001 \\ 
&  & \multirow{-7}{*}{0.4} & JonesPRD & 0.005 & 0.032 & 94.8 & -0.005 & 0.024 & 92.8 & 1.000 & 0.036 \\ 
\cline{3-12}
&  &  & \textbf{SIME-I} & 0.002 & 0.035 & 95.6 & 0.000 & 0.033 & 92.8 & 1.000 & 0.062 \\ 
&  &  & \textbf{Oracle-I} & 0.001 & 0.035 & 95.6 & 0.000 & 0.033 & 93.0 & 1.000 & 0.000 \\ 
&  &  & SIME-F/LW & 0.002 & 0.040 & 95.0 & -0.001 & 0.039 & 92.6 & 1.000 & 0.062 \\ 
&  &  & Oracle-F/LW & 0.002 & 0.040 & 95.0 & -0.001 & 0.039 & 92.8 & 1.000 & 0.000 \\ 
&  &  & LinMed & 0.019 & 0.037 & 91.0 & -0.018 & 0.033 & 93.8 & 1.000 & 0.106 \\ 
&  &  & HIMA & 0.001 & 0.031 & -- & 0.000 & 0.028 & -- & 0.996 & 0.007 \\ 
\multirow{-42}{*}{50} & \multirow{-21}{*}{$5{,}000$} & \multirow{-7}{*}{0.8} & JonesPRD & 0.004 & 0.031 & 94.8 & -0.003 & 0.028 & 93.0 & 1.000 & 0.026 \\ 
\hline
\end{tabular}%
}%
}
\end{table}
\clearpage
\begin{table}
\centering
\vspace{\abovecaptionskip}
\tablename~\ref*{tab:sim_linear_p50}: Simulation Setting~1 (linear), continued.\par
\smallskip
{\renewcommand{\baselinestretch}{1}\normalsize
\simtablefit{%
\begin{tabular}{@{}c c c l r r r r r r r r@{}}
\simresultheaders
&  &  & \textbf{SIME-I} & 0.058 & 0.165 & 91.4 & -0.056 & 0.140 & 87.2 & 0.844 & 0.005 \\ 
&  &  & \textbf{Oracle-I} & 0.001 & 0.151 & 95.6 & 0.001 & 0.127 & 95.0 & 0.997 & 0.000 \\ 
&  &  & SIME-F/LW & -1.033 & 44.493 & 98.4 & 1.035 & 44.489 & 98.0 & 0.844 & 0.005 \\ 
&  &  & Oracle-F/LW & 0.597 & 44.256 & 98.8 & -0.595 & 44.252 & 98.6 & 0.997 & 0.000 \\ 
&  &  & LinMed & 0.263 & 0.287 & 27.0 & -0.260 & 0.268 & 2.0 & 0.967 & 0.039 \\ 
&  &  & HIMA & 0.213 & 0.256 & -- & -0.210 & 0.235 & -- & 0.869 & 0.005 \\ 
&  & \multirow{-7}{*}{0.0} & JonesPRD & 0.254 & 0.279 & 37.2 & -0.252 & 0.260 & 9.0 & 0.972 & 0.043 \\ 
\cline{3-12}
&  &  & \textbf{SIME-I} & 0.042 & 0.142 & 95.0 & -0.039 & 0.116 & 92.4 & 0.897 & 0.018 \\ 
&  &  & \textbf{Oracle-I} & 0.003 & 0.125 & 95.4 & 0.000 & 0.103 & 94.4 & 0.992 & 0.000 \\ 
&  &  & SIME-F/LW & -5.368 & 167.011 & 98.0 & 5.370 & 167.013 & 97.8 & 0.897 & 0.018 \\ 
&  &  & Oracle-F/LW & -0.093 & 159.281 & 98.6 & 0.095 & 159.284 & 98.6 & 0.992 & 0.000 \\ 
&  &  & LinMed & 0.157 & 0.190 & 64.0 & -0.153 & 0.163 & 33.4 & 0.993 & 0.024 \\ 
&  &  & HIMA & 0.113 & 0.159 & -- & -0.109 & 0.130 & -- & 0.764 & 0.002 \\ 
&  & \multirow{-7}{*}{0.4} & JonesPRD & 0.152 & 0.185 & 70.0 & -0.149 & 0.160 & 38.6 & 0.992 & 0.028 \\ 
\cline{3-12}
&  &  & \textbf{SIME-I} & 0.026 & 0.120 & 95.4 & -0.021 & 0.106 & 96.0 & 0.633 & 0.015 \\ 
&  &  & \textbf{Oracle-I} & 0.009 & 0.114 & 95.6 & -0.004 & 0.104 & 94.4 & 0.858 & 0.000 \\ 
&  &  & SIME-F/LW & 50.719 & 967.279 & 98.8 & -50.714 & 967.284 & 98.8 & 0.633 & 0.015 \\ 
&  &  & Oracle-F/LW & 55.013 & 974.692 & 99.2 & -55.008 & 974.696 & 99.2 & 0.858 & 0.000 \\ 
&  &  & LinMed & 0.092 & 0.137 & 84.0 & -0.087 & 0.113 & 88.2 & 0.962 & 0.015 \\ 
&  &  & HIMA & 0.040 & 0.111 & -- & -0.034 & 0.089 & -- & 0.488 & 0.004 \\ 
& \multirow{-21}{*}{500} & \multirow{-7}{*}{0.8} & JonesPRD & 0.089 & 0.134 & 84.2 & -0.084 & 0.111 & 80.6 & 0.964 & 0.017 \\ 
\cline{2-12}
&  &  & \textbf{SIME-I} & 0.002 & 0.045 & 96.0 & -0.002 & 0.037 & 94.6 & 1.000 & 0.023 \\ 
&  &  & \textbf{Oracle-I} & 0.002 & 0.045 & 96.0 & -0.002 & 0.037 & 94.2 & 1.000 & 0.000 \\ 
&  &  & SIME-F/LW & 0.004 & 0.072 & 96.8 & -0.004 & 0.068 & 94.8 & 1.000 & 0.023 \\ 
&  &  & Oracle-F/LW & 0.002 & 0.072 & 96.8 & -0.002 & 0.067 & 95.0 & 1.000 & 0.000 \\ 
&  &  & LinMed & 0.078 & 0.085 & 35.8 & -0.078 & 0.081 & 2.6 & 1.000 & 0.037 \\ 
&  &  & HIMA & 0.000 & 0.033 & -- & 0.000 & 0.020 & -- & 1.000 & 0.000 \\ 
&  & \multirow{-7}{*}{0.0} & JonesPRD & 0.005 & 0.034 & 94.8 & -0.005 & 0.021 & 94.0 & 1.000 & 0.004 \\ 
\cline{3-12}
&  &  & \textbf{SIME-I} & 0.001 & 0.037 & 95.8 & -0.001 & 0.029 & 94.2 & 1.000 & 0.022 \\ 
&  &  & \textbf{Oracle-I} & 0.001 & 0.037 & 96.0 & -0.001 & 0.028 & 94.2 & 1.000 & 0.000 \\ 
&  &  & SIME-F/LW & 0.005 & 0.077 & 97.2 & -0.006 & 0.075 & 95.4 & 1.000 & 0.022 \\ 
&  &  & Oracle-F/LW & 0.002 & 0.077 & 97.2 & -0.002 & 0.075 & 95.4 & 1.000 & 0.000 \\ 
&  &  & LinMed & 0.047 & 0.057 & 68.0 & -0.047 & 0.051 & 39.2 & 1.000 & 0.023 \\ 
&  &  & HIMA & 0.000 & 0.031 & -- & 0.000 & 0.022 & -- & 1.000 & 0.000 \\ 
&  & \multirow{-7}{*}{0.4} & JonesPRD & 0.005 & 0.032 & 94.4 & -0.005 & 0.023 & 92.6 & 1.000 & 0.004 \\ 
\cline{3-12}
&  &  & \textbf{SIME-I} & 0.001 & 0.035 & 94.8 & -0.001 & 0.032 & 94.8 & 1.000 & 0.013 \\ 
&  &  & \textbf{Oracle-I} & 0.001 & 0.035 & 94.6 & -0.001 & 0.032 & 95.0 & 1.000 & 0.000 \\ 
&  &  & SIME-F/LW & 0.000 & 0.070 & 96.6 & 0.000 & 0.072 & 95.6 & 1.000 & 0.013 \\ 
&  &  & Oracle-F/LW & 0.000 & 0.070 & 96.2 & 0.000 & 0.072 & 95.4 & 1.000 & 0.000 \\ 
&  &  & LinMed & 0.028 & 0.041 & 86.2 & -0.028 & 0.038 & 89.2 & 1.000 & 0.015 \\ 
&  &  & HIMA & 0.001 & 0.031 & -- & 0.000 & 0.028 & -- & 0.946 & 0.004 \\ 
\multirow{-42}{*}{500} & \multirow{-21}{*}{$5{,}000$} & \multirow{-7}{*}{0.8} & JonesPRD & 0.003 & 0.031 & 94.6 & -0.003 & 0.028 & 93.8 & 1.000 & 0.004 \\ 
\hline
\end{tabular}%
}%
}
\end{table}

\begin{table}
\caption{Simulation Setting~2 (nonlinear). Effect estimation and selection in $\btheta$ over 500 replications per $(p,n,\varrho)$ combination. RMSE: root mean squared error. CP: coverage probability of nominal 95\% intervals. CP is given in \%. TPR/FPR: true/false positive rates.}
\label{tab:sim_nonlinear_p500_n5000}
\centering
{\renewcommand{\baselinestretch}{1}\normalsize
\simtablefit{%
\begin{tabular}{@{}c c c l r r r r r r r r@{}}
\simresultheaders
&  &  & \textbf{SIME-I} & -0.004 & 0.165 & 94.4 & 0.000 & 0.145 & 94.0 & 0.991 & 0.070 \\ 
&  &  & \textbf{Oracle-I} & -0.010 & 0.165 & 95.4 & 0.006 & 0.140 & 94.6 & 1.000 & 0.000 \\ 
&  &  & SIME-F/LW & -0.021 & 0.299 & 98.2 & 0.018 & 0.284 & 97.4 & 0.991 & 0.070 \\ 
&  &  & Oracle-F/LW & -0.014 & 0.292 & 98.2 & 0.011 & 0.276 & 97.2 & 1.000 & 0.000 \\ 
&  &  & LinMed & 0.273 & 0.300 & 38.2 & -0.264 & 0.276 & 4.8 & 0.842 & 0.168 \\ 
&  &  & HIMA & 0.304 & 0.342 & -- & -0.296 & 0.320 & -- & 0.473 & 0.015 \\ 
&  & \multirow{-7}{*}{0.0} & JonesPRD & 0.258 & 0.290 & 37.2 & -0.253 & 0.269 & 12.0 & 0.836 & 0.162 \\ 
\cline{3-12}
&  &  & \textbf{SIME-I} & -0.003 & 0.133 & 95.4 & -0.001 & 0.099 & 94.2 & 0.994 & 0.078 \\ 
&  &  & \textbf{Oracle-I} & -0.007 & 0.131 & 95.0 & 0.003 & 0.096 & 95.0 & 1.000 & 0.000 \\ 
&  &  & SIME-F/LW & -0.013 & 0.367 & 97.6 & 0.008 & 0.352 & 97.4 & 0.994 & 0.078 \\ 
&  &  & Oracle-F/LW & -0.005 & 0.355 & 97.6 & 0.000 & 0.340 & 97.0 & 1.000 & 0.000 \\ 
&  &  & LinMed & 0.169 & 0.208 & 67.2 & -0.165 & 0.174 & 21.6 & 0.692 & 0.109 \\ 
&  &  & HIMA & 0.178 & 0.220 & -- & -0.175 & 0.187 & -- & 0.338 & 0.010 \\ 
&  & \multirow{-7}{*}{0.4} & JonesPRD & 0.162 & 0.202 & 66.0 & -0.161 & 0.171 & 14.2 & 0.687 & 0.110 \\ 
\cline{3-12}
&  &  & \textbf{SIME-I} & 0.002 & 0.123 & 94.4 & -0.006 & 0.081 & 93.8 & 0.723 & 0.075 \\ 
&  &  & \textbf{Oracle-I} & -0.005 & 0.117 & 95.2 & 0.001 & 0.083 & 94.8 & 0.869 & 0.000 \\ 
&  &  & SIME-F/LW & -0.008 & 0.366 & 96.0 & 0.004 & 0.350 & 96.0 & 0.723 & 0.075 \\ 
&  &  & Oracle-F/LW & -0.001 & 0.320 & 97.8 & -0.003 & 0.307 & 96.8 & 0.869 & 0.000 \\ 
&  &  & LinMed & 0.079 & 0.140 & 90.4 & -0.079 & 0.086 & 51.6 & 0.287 & 0.059 \\ 
&  &  & HIMA & 0.069 & 0.140 & -- & -0.070 & 0.084 & -- & 0.125 & 0.012 \\ 
& \multirow{-21}{*}{500} & \multirow{-7}{*}{0.8} & JonesPRD & 0.074 & 0.138 & 88.0 & -0.077 & 0.084 & 30.0 & 0.292 & 0.064 \\ 
\cline{2-12}
&  &  & \textbf{SIME-I} & 0.002 & 0.050 & 96.4 & -0.001 & 0.043 & 93.6 & 1.000 & 0.095 \\ 
&  &  & \textbf{Oracle-I} & 0.002 & 0.050 & 96.4 & -0.001 & 0.043 & 93.2 & 1.000 & 0.000 \\ 
&  &  & SIME-F/LW & 0.002 & 0.048 & 96.6 & -0.002 & 0.042 & 93.4 & 1.000 & 0.095 \\ 
&  &  & Oracle-F/LW & 0.002 & 0.048 & 96.4 & -0.002 & 0.042 & 93.2 & 1.000 & 0.000 \\ 
&  &  & LinMed & 0.145 & 0.150 & 4.2 & -0.134 & 0.136 & 0.0 & 1.000 & 0.224 \\ 
&  &  & HIMA & 0.095 & 0.102 & -- & -0.084 & 0.087 & -- & 1.000 & 0.001 \\ 
&  & \multirow{-7}{*}{0.0} & JonesPRD & 0.096 & 0.103 & 25.4 & -0.086 & 0.089 & 1.6 & 1.000 & 0.033 \\ 
\cline{3-12}
&  &  & \textbf{SIME-I} & 0.001 & 0.038 & 95.8 & 0.000 & 0.030 & 94.2 & 1.000 & 0.092 \\ 
&  &  & \textbf{Oracle-I} & 0.001 & 0.038 & 95.8 & 0.000 & 0.030 & 93.6 & 1.000 & 0.000 \\ 
&  &  & SIME-F/LW & 0.002 & 0.040 & 96.0 & -0.001 & 0.034 & 94.0 & 1.000 & 0.092 \\ 
&  &  & Oracle-F/LW & 0.002 & 0.040 & 95.8 & -0.001 & 0.034 & 94.4 & 1.000 & 0.000 \\ 
&  &  & LinMed & 0.094 & 0.101 & 25.4 & -0.087 & 0.089 & 0.0 & 1.000 & 0.145 \\ 
&  &  & HIMA & 0.073 & 0.082 & -- & -0.065 & 0.069 & -- & 0.943 & 0.008 \\ 
&  & \multirow{-7}{*}{0.4} & JonesPRD & 0.064 & 0.074 & 58.8 & -0.058 & 0.061 & 10.4 & 0.999 & 0.035 \\ 
\cline{3-12}
&  &  & \textbf{SIME-I} & 0.001 & 0.034 & 96.4 & 0.000 & 0.024 & 95.2 & 1.000 & 0.064 \\ 
&  &  & \textbf{Oracle-I} & 0.001 & 0.034 & 96.4 & 0.000 & 0.024 & 95.4 & 1.000 & 0.000 \\ 
&  &  & SIME-F/LW & 0.002 & 0.041 & 94.8 & 0.000 & 0.034 & 94.6 & 1.000 & 0.064 \\ 
&  &  & Oracle-F/LW & 0.002 & 0.041 & 94.8 & -0.001 & 0.034 & 94.6 & 1.000 & 0.000 \\ 
&  &  & LinMed & 0.047 & 0.060 & 77.2 & -0.043 & 0.045 & 81.8 & 0.790 & 0.087 \\ 
&  &  & HIMA & 0.041 & 0.055 & -- & -0.037 & 0.039 & -- & 0.272 & 0.004 \\ 
\multirow{-42}{*}{50} & \multirow{-21}{*}{$5{,}000$} & \multirow{-7}{*}{0.8} & JonesPRD & 0.043 & 0.057 & 76.6 & -0.040 & 0.043 & 21.0 & 0.652 & 0.063 \\ 
\hline
\end{tabular}%
}%
}
\end{table}
\clearpage
\begin{table}
\centering
\vspace{\abovecaptionskip}
\tablename~\ref*{tab:sim_nonlinear_p500_n5000}: Simulation Setting~2 (nonlinear), continued.\par
\smallskip
{\renewcommand{\baselinestretch}{1}\normalsize
\simtablefit{%
\begin{tabular}{@{}c c c l r r r r r r r r@{}}
\simresultheaders
&  &  & \textbf{SIME-I} & 0.069 & 0.179 & 90.6 & -0.071 & 0.163 & 77.2 & 0.837 & 0.013 \\ 
&  &  & \textbf{Oracle-I} & -0.007 & 0.158 & 96.6 & 0.005 & 0.139 & 94.6 & 1.000 & 0.000 \\ 
&  &  & SIME-F/LW & -2.024 & 48.678 & 98.2 & 2.023 & 48.676 & 98.0 & 0.837 & 0.013 \\ 
&  &  & Oracle-F/LW & 1.291 & 50.531 & 99.2 & -1.291 & 50.526 & 98.8 & 1.000 & 0.000 \\ 
&  &  & LinMed & 0.383 & 0.404 & 8.2 & -0.371 & 0.376 & 0.2 & 0.488 & 0.019 \\ 
&  &  & HIMA & 0.404 & 0.424 & -- & -0.392 & 0.397 & -- & 0.193 & 0.001 \\ 
&  & \multirow{-7}{*}{0.0} & JonesPRD & 0.373 & 0.394 & 10.2 & -0.364 & 0.369 & 0.6 & 0.518 & 0.023 \\ 
\cline{3-12}
&  &  & \textbf{SIME-I} & 0.033 & 0.139 & 94.0 & -0.035 & 0.110 & 87.2 & 0.869 & 0.016 \\ 
&  &  & \textbf{Oracle-I} & -0.003 & 0.126 & 96.8 & 0.001 & 0.099 & 92.8 & 1.000 & 0.000 \\ 
&  &  & SIME-F/LW & -8.110 & 233.572 & 97.8 & 8.107 & 233.569 & 97.8 & 0.869 & 0.016 \\ 
&  &  & Oracle-F/LW & 2.078 & 184.486 & 98.8 & -2.081 & 184.485 & 98.6 & 1.000 & 0.000 \\ 
&  &  & LinMed & 0.227 & 0.259 & 47.6 & -0.220 & 0.225 & 2.6 & 0.473 & 0.014 \\ 
&  &  & HIMA & 0.234 & 0.267 & -- & -0.227 & 0.232 & -- & 0.210 & 0.001 \\ 
&  & \multirow{-7}{*}{0.4} & JonesPRD & 0.222 & 0.254 & 45.0 & -0.217 & 0.222 & 2.6 & 0.477 & 0.017 \\ 
\cline{3-12}
&  &  & \textbf{SIME-I} & 0.024 & 0.133 & 94.8 & -0.026 & 0.082 & 90.8 & 0.430 & 0.015 \\ 
&  &  & \textbf{Oracle-I} & 0.001 & 0.118 & 95.8 & -0.002 & 0.083 & 94.8 & 0.878 & 0.000 \\ 
&  &  & SIME-F/LW & 20.095 & $1.34\mathrm{e}{+3}$ & 99.4 & -20.097 & $1.34\mathrm{e}{+3}$ & 99.4 & 0.430 & 0.015 \\ 
&  &  & Oracle-F/LW & 58.290 & $1.10\mathrm{e}{+3}$ & 99.0 & -58.291 & $1.10\mathrm{e}{+3}$ & 98.8 & 0.878 & 0.000 \\ 
&  &  & LinMed & 0.104 & 0.157 & 85.6 & -0.101 & 0.103 & 22.8 & 0.160 & 0.008 \\ 
&  &  & HIMA & 0.102 & 0.157 & -- & -0.099 & 0.104 & -- & 0.068 & 0.001 \\ 
& \multirow{-21}{*}{500} & \multirow{-7}{*}{0.8} & JonesPRD & 0.101 & 0.155 & 83.6 & -0.100 & 0.103 & 13.8 & 0.162 & 0.009 \\ 
\cline{2-12}
&  &  & \textbf{SIME-I} & 0.002 & 0.046 & 95.6 & -0.002 & 0.040 & 95.0 & 1.000 & 0.030 \\ 
&  &  & \textbf{Oracle-I} & 0.002 & 0.047 & 96.2 & -0.002 & 0.040 & 94.8 & 1.000 & 0.000 \\ 
&  &  & SIME-F/LW & 0.002 & 0.072 & 96.8 & -0.002 & 0.069 & 96.0 & 1.000 & 0.030 \\ 
&  &  & Oracle-F/LW & 0.001 & 0.072 & 96.6 & -0.001 & 0.069 & 96.2 & 1.000 & 0.000 \\ 
&  &  & LinMed & 0.180 & 0.184 & 0.0 & -0.170 & 0.171 & 0.0 & 1.000 & 0.036 \\ 
&  &  & HIMA & 0.100 & 0.107 & -- & -0.089 & 0.092 & -- & 1.000 & 0.000 \\ 
&  & \multirow{-7}{*}{0.0} & JonesPRD & 0.114 & 0.121 & 17.2 & -0.105 & 0.108 & 1.4 & 1.000 & 0.006 \\ 
\cline{3-12}
&  &  & \textbf{SIME-I} & 0.000 & 0.036 & 95.0 & 0.000 & 0.028 & 94.8 & 1.000 & 0.023 \\ 
&  &  & \textbf{Oracle-I} & 0.000 & 0.037 & 95.0 & -0.001 & 0.028 & 95.0 & 1.000 & 0.000 \\ 
&  &  & SIME-F/LW & 0.002 & 0.079 & 96.8 & -0.002 & 0.078 & 96.8 & 1.000 & 0.023 \\ 
&  &  & Oracle-F/LW & 0.001 & 0.079 & 96.6 & -0.001 & 0.077 & 96.4 & 1.000 & 0.000 \\ 
&  &  & LinMed & 0.113 & 0.119 & 11.6 & -0.107 & 0.108 & 0.0 & 1.000 & 0.023 \\ 
&  &  & HIMA & 0.092 & 0.100 & -- & -0.086 & 0.088 & -- & 0.862 & 0.001 \\ 
&  & \multirow{-7}{*}{0.4} & JonesPRD & 0.086 & 0.094 & 40.6 & -0.080 & 0.083 & 5.0 & 0.994 & 0.011 \\ 
\cline{3-12}
&  &  & \textbf{SIME-I} & 0.001 & 0.034 & 95.2 & 0.000 & 0.024 & 95.2 & 1.000 & 0.014 \\ 
&  &  & \textbf{Oracle-I} & 0.001 & 0.034 & 94.4 & 0.000 & 0.024 & 95.0 & 1.000 & 0.000 \\ 
&  &  & SIME-F/LW & 0.000 & 0.070 & 96.8 & 0.000 & 0.068 & 96.0 & 1.000 & 0.014 \\ 
&  &  & Oracle-F/LW & 0.000 & 0.070 & 96.8 & 0.000 & 0.068 & 96.4 & 1.000 & 0.000 \\ 
&  &  & LinMed & 0.057 & 0.067 & 66.4 & -0.054 & 0.055 & 61.6 & 0.744 & 0.013 \\ 
&  &  & HIMA & 0.045 & 0.058 & -- & -0.042 & 0.043 & -- & 0.221 & 0.000 \\ 
\multirow{-42}{*}{500} & \multirow{-21}{*}{$5{,}000$} & \multirow{-7}{*}{0.8} & JonesPRD & 0.056 & 0.067 & 65.2 & -0.053 & 0.054 & 0.8 & 0.750 & 0.014 \\ 
\hline
\end{tabular}%
}%
}
\end{table}
\endgroup

\clearpage

\subsection{Sensitivity Analysis}
\label{sub:app_sensitivity}

The sensitivity study evaluates the adjustments in Appendix~\ref{sub:app_sensitivity_model} under the nonlinear outcome link and independent mediator errors. The outcome error model varies the correlation $\rho$ with the mediator index error.

\[
\eta = \rho\frac{\sigma_\eta}{\sigma_\epsilon}\btheta_0^\top\bepsilon + \sqrt{1-\rho^2}\,\sigma_\eta\nu, \qquad \nu\sim N(0,1),
\]

\begin{figure}
\centering
\subfloat[]{%
  \includegraphics[width=0.3\textwidth]{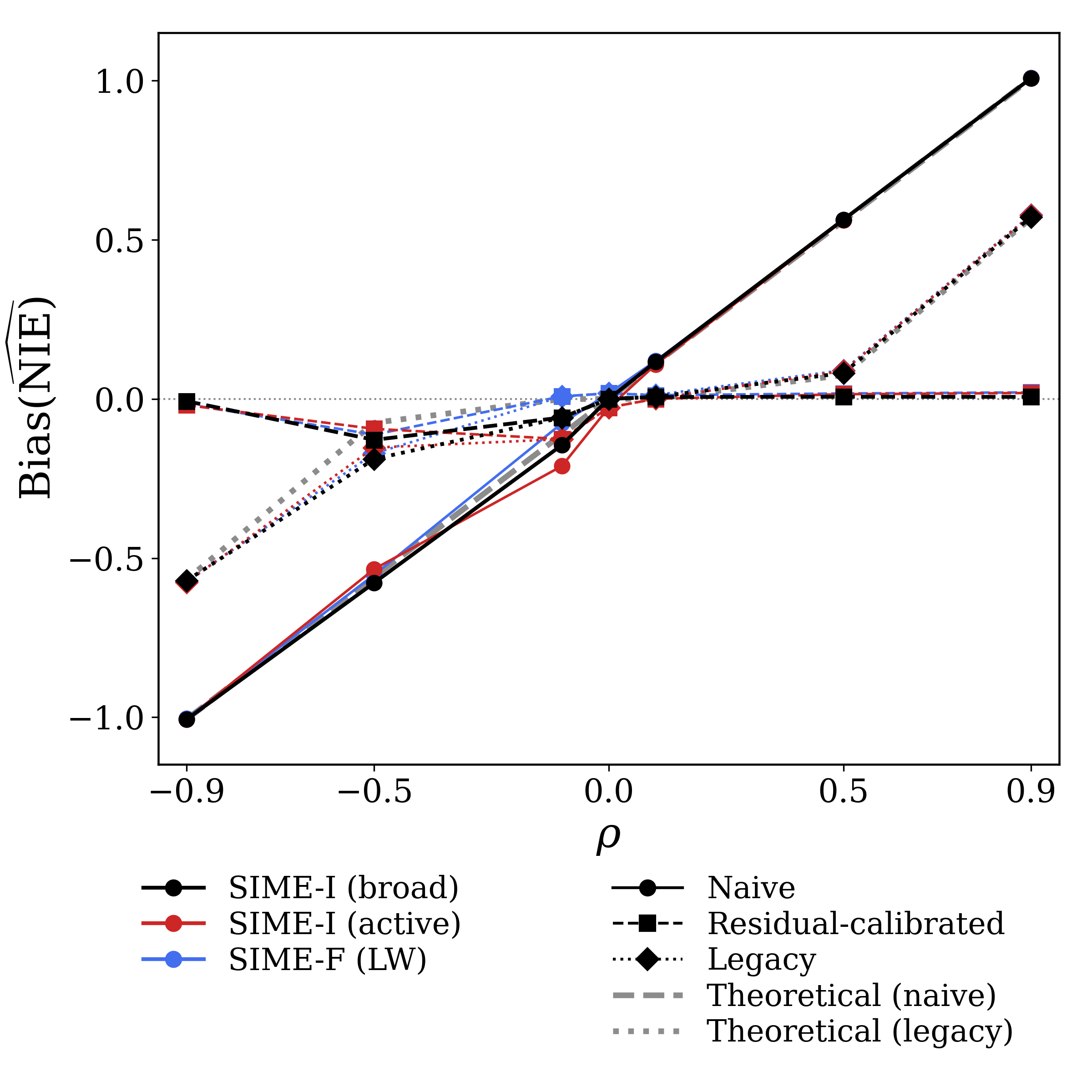}%
  \label{fig:sensitivity_p50}%
}
\enskip{}
\subfloat[]{%
  \includegraphics[width=0.3\textwidth]{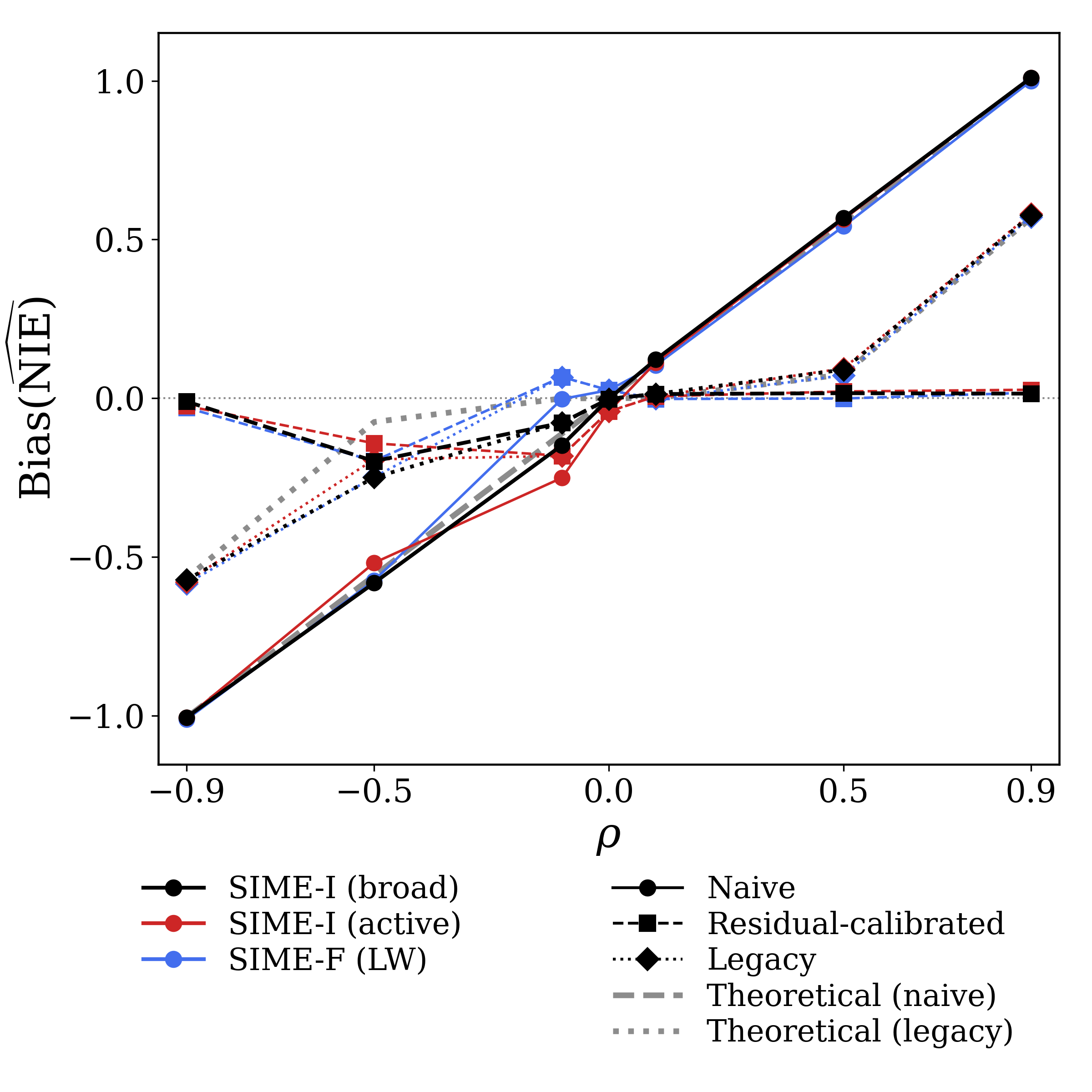}%
  \label{fig:sensitivity_p100}%
}
\enskip{}
\subfloat[]{%
  \includegraphics[width=0.3\textwidth]{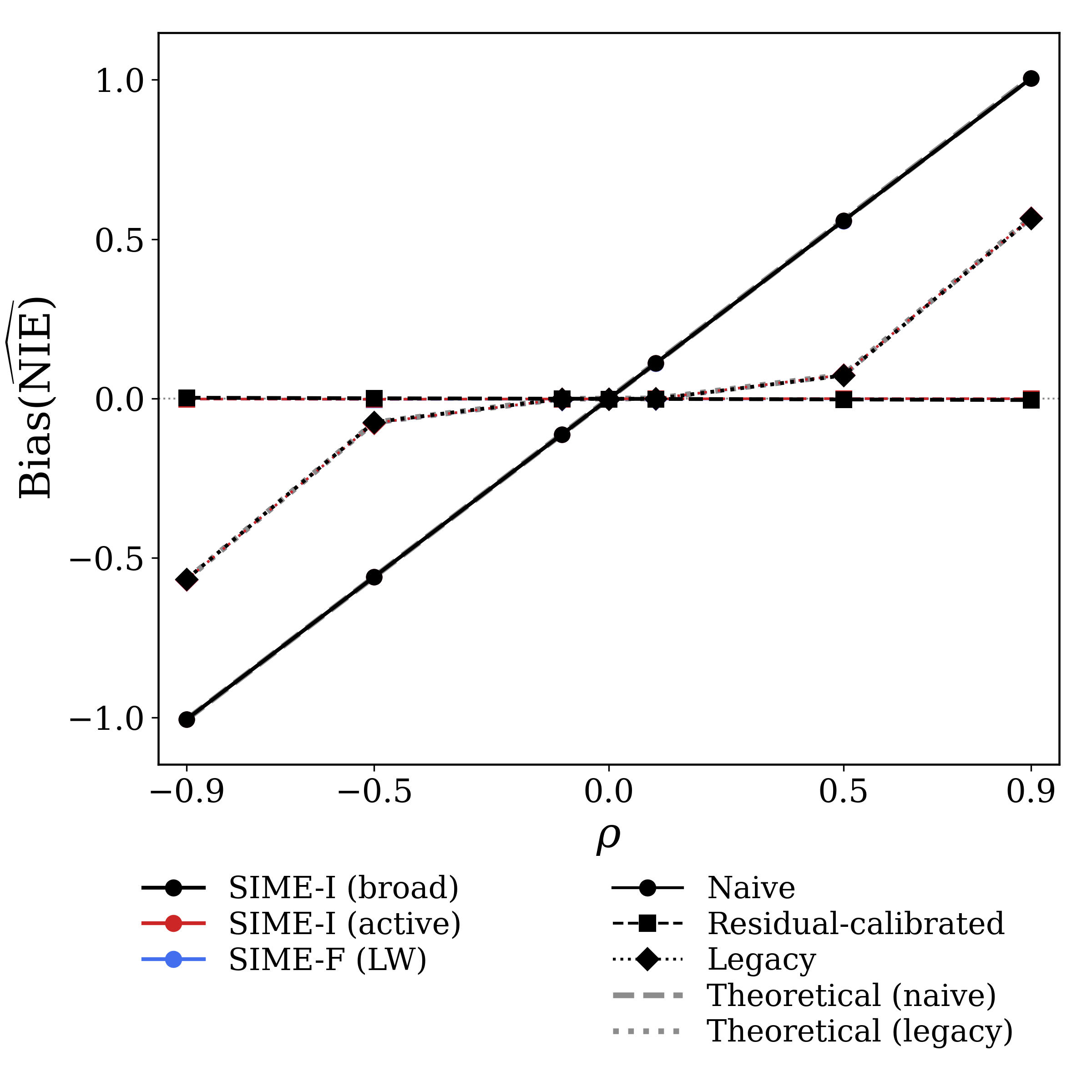}%
  \label{fig:sensitivity_p50_n5000}%
}
\caption{NIE bias under mediator--outcome error dependence over 500 replications per setting. (a) $p=50$, $n=500$, (b) $p=100$, $n=500$, and (c) $p=50$, $n=5000$. Black, red and blue denote combined SIME-I, active-space SIME-I and SIME-F/LW, respectively. Circles, squares and diamonds denote naive, residual-calibrated and legacy-adjusted estimates, respectively. Thick gray curves give the population discrepancies, and the thin gray line marks zero.}
\label{fig:sensitivity}
\end{figure}

Here $\nu$ is independent of the mediator errors, and $\sigma_\eta=\sigma_\epsilon=1$. The settings use $\rho\in\{-0.9,-0.5,-0.1,0,0.1,0.5,0.9\}$. The $(p,n)$ combinations in Figure~\ref{fig:sensitivity} each use 500 replications per value of $\rho$, with the structural effects held fixed.

Combined SIME-I, its active-space benchmark and SIME-F/LW are compared using residual calibration in~\eqref{eq:NIE_corrected}, the legacy adjustment and the adjustment with known structural error standard deviation. Each fitted index is oriented toward $\btheta_0$. The adjustment uses the true $\rho$ and averages the training-fold quantities with evaluation-fold size weights.

\begin{table}
\caption{Residual-calibrated NIE bias and RMSE over 500 replications per setting. Column blocks identify $(p,n)$. RMSE: root mean squared error.}
\label{tab:sensitivity}
\begin{center}
{\renewcommand{\baselinestretch}{1}\normalsize
\renewcommand{\arraystretch}{1}
\begin{tabular}{l r r r c r r c r r}
\hline
 &  & \multicolumn{2}{c}{$(50,500)$} &  & \multicolumn{2}{c}{$(100,500)$} &  & \multicolumn{2}{c}{$(50,5{,}000)$} \\
\cline{3-4} \cline{6-7} \cline{9-10}
\multicolumn{1}{c}{\multirow{-2}{*}{Method}} & \multicolumn{1}{c}{\multirow{-2}{*}{$\rho$}} & \multicolumn{1}{c}{Bias} & \multicolumn{1}{c}{RMSE} &  & \multicolumn{1}{c}{Bias} & \multicolumn{1}{c}{RMSE} &  & \multicolumn{1}{c}{Bias} & \multicolumn{1}{c}{RMSE} \\
\hline
 & $-$0.9 & $-$0.007 & 0.097 &  & $-$0.010 & 0.099 &  & 0.003 & 0.031 \\ 
 & $-$0.5 & $-$0.129 & 0.208 &  & $-$0.198 & 0.269 &  & 0.001 & 0.039 \\ 
 & $-$0.1 & $-$0.058 & 0.298 &  & $-$0.077 & 0.171 &  & 0.000 & 0.043 \\ 
 & 0.0 & 0.000 & 0.145 &  & $-$0.002 & 0.156 &  & $-$0.001 & 0.043 \\ 
 & 0.1 & 0.006 & 0.145 &  & 0.013 & 0.142 &  & $-$0.001 & 0.042 \\ 
 & 0.5 & 0.007 & 0.131 &  & 0.016 & 0.136 &  & $-$0.003 & 0.038 \\ 
\multirow{-7}{*}{SIME-I (broad)} & 0.9 & 0.006 & 0.096 &  & 0.014 & 0.097 &  & $-$0.005 & 0.028 \\ 
\hline
 & $-$0.9 & $-$0.019 & 0.097 &  & $-$0.024 & 0.100 &  & $-$0.001 & 0.031 \\ 
 & $-$0.5 & $-$0.094 & 0.171 &  & $-$0.142 & 0.203 &  & $-$0.002 & 0.038 \\ 
 & $-$0.1 & $-$0.126 & 0.284 &  & $-$0.181 & 0.262 &  & $-$0.001 & 0.041 \\ 
 & 0.0 & $-$0.027 & 0.154 &  & $-$0.042 & 0.174 &  & $-$0.001 & 0.041 \\ 
 & 0.1 & 0.000 & 0.139 &  & 0.006 & 0.154 &  & $-$0.001 & 0.041 \\ 
 & 0.5 & 0.015 & 0.126 &  & 0.021 & 0.134 &  & 0.000 & 0.037 \\ 
\multirow{-7}{*}{SIME-I (active)} & 0.9 & 0.019 & 0.094 &  & 0.027 & 0.099 &  & 0.000 & 0.028 \\ 
\hline
 & $-$0.9 & $-$0.016 & 0.141 &  & $-$0.030 & 0.246 &  & $-$0.002 & 0.031 \\ 
 & $-$0.5 & $-$0.112 & 0.284 &  & $-$0.198 & 0.543 &  & $-$0.003 & 0.039 \\ 
 & $-$0.1 & 0.008 & 0.282 &  & 0.067 & 0.626 &  & $-$0.002 & 0.042 \\ 
 & 0.0 & 0.018 & 0.284 &  & 0.025 & 0.565 &  & $-$0.002 & 0.042 \\ 
 & 0.1 & 0.012 & 0.278 &  & $-$0.002 & 0.532 &  & $-$0.001 & 0.042 \\ 
 & 0.5 & 0.017 & 0.238 &  & $-$0.001 & 0.471 &  & $-$0.001 & 0.038 \\ 
\multirow{-7}{*}{SIME-F (LW)} & 0.9 & 0.021 & 0.138 &  & 0.017 & 0.249 &  & $-$0.001 & 0.028 \\ 
\hline
\end{tabular}
}
\end{center}
\end{table}

Table~\ref{tab:sensitivity} and Figure~\ref{fig:sensitivity} summarize the point-estimate comparisons. In larger samples, the naive NIE bias follows the population discrepancy. Residual calibration and adjustment with the known structural error standard deviation reduce this bias, while the legacy adjustment retains systematic bias. For negative error correlations in smaller samples, the relative performance of combined and active-space SIME-I varies with $\rho$.

\subsection{Covariance Comparisons}
\label{sub:app_covariance}

Tables~\ref{tab:cov_linear_p50} and~\ref{tab:cov_nonlinear_p500} compare covariance choices for NIE estimation. Proj denotes projected residual variance. Id, S and LW denote identity, sample and Ledoit--Wolf covariance, respectively. Broad uses the combined strategy, while Active uses the original link with active-space correction.

When the mediator dimension equals the sample size, the singular sample covariance makes the corresponding full-ratio fits unavailable. The index-ratio fits remain available through the projected variance.

\begin{figure}
\centering
\includegraphics[width=0.95\textwidth]{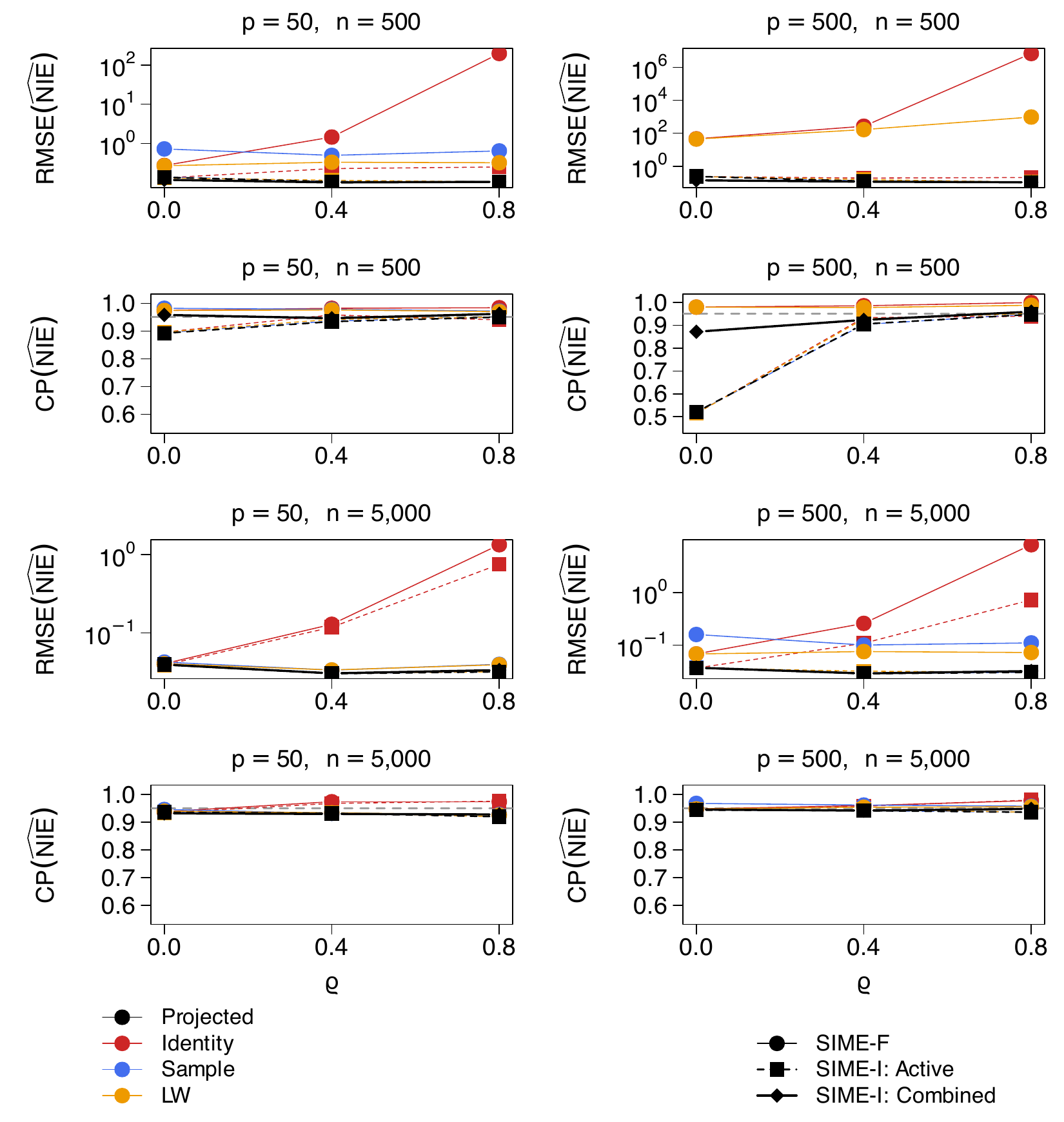}
\caption{Simulation Setting~1 (linear): NIE RMSE and nominal 95\% interval coverage over 500 replications per setting. Black: projected variance; red: identity; blue: sample; orange: LW. Circles: SIME-F; squares: active-space SIME-I; diamonds: combined SIME-I. RMSE axes are logarithmic. Gray dashed lines mark 95\% coverage. Singular full/sample fits are omitted. RMSE: root mean squared error. CP: coverage probability of nominal 95\% intervals.}
\label{fig:sim_linear}
\end{figure}

\begin{figure}
\centering
\includegraphics[width=0.95\textwidth]{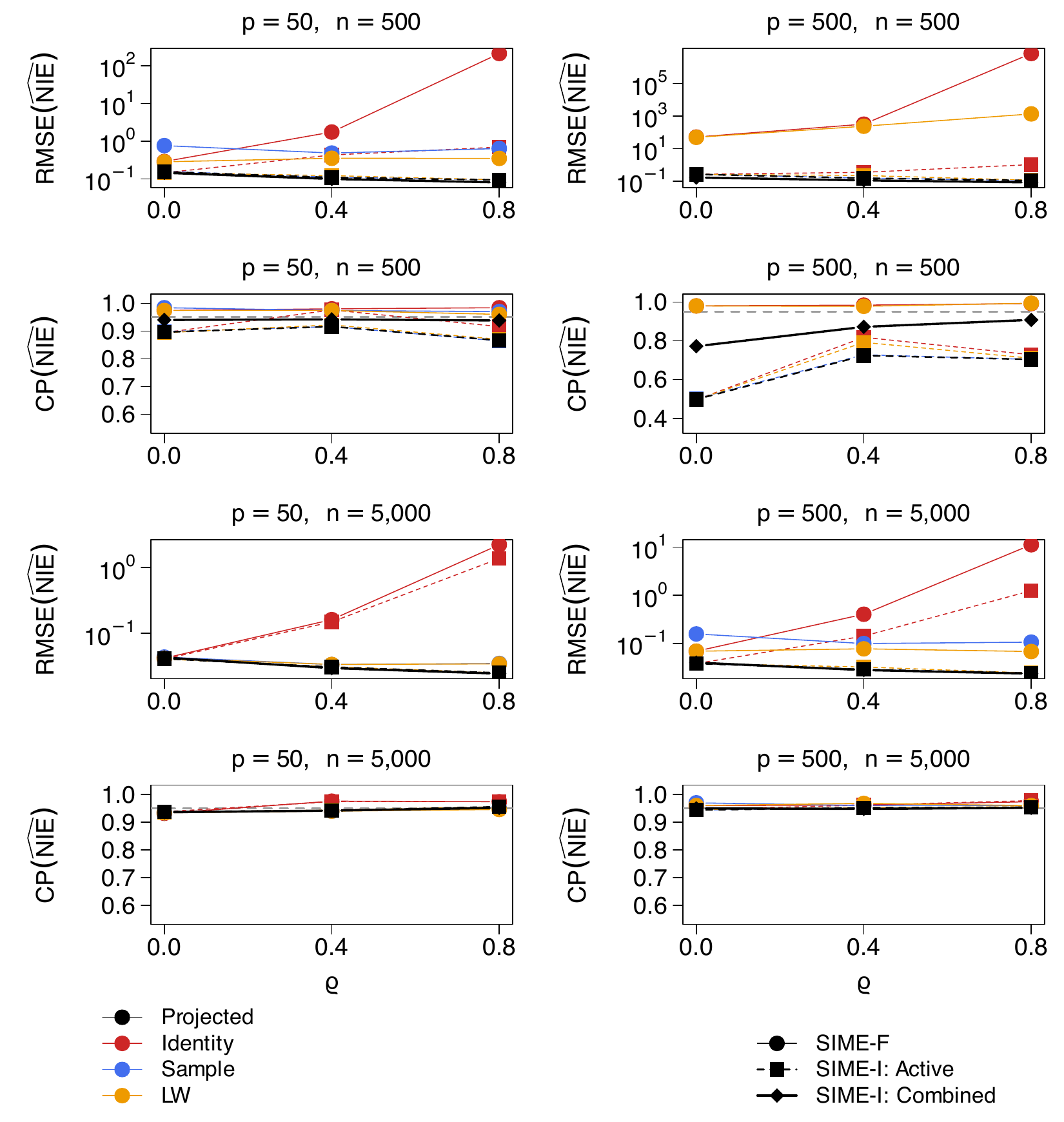}
\caption{Simulation Setting~2 (nonlinear): NIE RMSE and nominal 95\% interval coverage over 500 replications per setting. Display conventions follow Figure~\ref{fig:sim_linear}. RMSE: root mean squared error. CP: coverage probability of nominal 95\% intervals.}
\label{fig:sim_nonlinear}
\end{figure}

For correlated mediators, LW shrinkage improves on identity covariance, particularly for full-ratio estimation. The projected and sample-covariance index estimators give similar results. Figures~\ref{fig:sim_linear} and~\ref{fig:sim_nonlinear} show that the full ratio can remain unstable with shrinkage when the mediator dimension is large relative to the sample size. Its high coverage in these settings comes from wide intervals. Restricting the candidate support yields a similar pattern, and the differences between ratio approaches diminish with larger samples.

\begin{table}
\caption{Simulation Setting~1 (linear). NIE estimation under various covariance estimation strategies over 500 replications. RMSE: root mean squared error. CP: coverage probability of nominal 95\% intervals. CP is given in \%.}
\label{tab:cov_linear_p50}
\centering
{\renewcommand{\baselinestretch}{1}\normalsize
\renewcommand{\arraystretch}{0.86}
\simtablefit{%
\begin{tabular}{c c l c c r r r c r r r}
\hline
 &  &  &  &  & \multicolumn{3}{c}{$n=500$} &  & \multicolumn{3}{c}{$n=5{,}000$} \\
\cline{6-8} \cline{10-12}
\multicolumn{1}{c}{\multirow{-2}{*}{$p$}} & \multicolumn{1}{c}{\multirow{-2}{*}{$\varrho$}} & \multicolumn{1}{c}{\multirow{-2}{*}{Method}} & \multicolumn{1}{c}{\multirow{-2}{*}{Cov.}} & \multicolumn{1}{c}{\multirow{-2}{*}{Inference}} & \multicolumn{1}{c}{Bias} & \multicolumn{1}{c}{RMSE} & \multicolumn{1}{c}{CP} &  & \multicolumn{1}{c}{Bias} & \multicolumn{1}{c}{RMSE} & \multicolumn{1}{c}{CP} \\
\hline
 &  &  &  & Broad & -0.003 & 0.117 & 95.8 &  & -0.001 & 0.039 & 93.2 \\ 
\cline{5-12}
 &  &  & \multirow{-2}{*}{Proj} &  & -0.040 & 0.135 & 89.2 &  & -0.001 & 0.039 & 93.6 \\ 
 &  &  & LW &  & -0.034 & 0.130 & 89.6 &  & -0.001 & 0.039 & 93.4 \\ 
 &  &  & S &  & -0.042 & 0.139 & 89.4 &  & -0.001 & 0.039 & 93.6 \\ 
 &  & \multirow{-5}{*}{SIME-I} & Id & \multirow{-4}{*}{Active} & -0.034 & 0.131 & 89.6 &  & -0.001 & 0.039 & 93.4 \\ 
\cline{3-12}
 &  &  & LW &  & 0.026 & 0.267 & 97.4 &  & -0.001 & 0.040 & 94.0 \\ 
 &  &  & S &  & 0.079 & 0.720 & 98.2 &  & -0.001 & 0.042 & 94.6 \\ 
 & \multirow{-8}{*}{0.0} & \multirow{-3}{*}{SIME-F} & Id & \multirow{-3}{*}{--} & 0.026 & 0.271 & 97.4 &  & -0.001 & 0.040 & 93.8 \\ 
\cline{2-12}
 &  &  &  & Broad & -0.005 & 0.100 & 94.6 &  & 0.000 & 0.030 & 93.0 \\ 
\cline{5-12}
 &  &  & \multirow{-2}{*}{Proj} &  & -0.014 & 0.105 & 93.4 &  & 0.000 & 0.030 & 93.2 \\ 
 &  &  & LW &  & -0.005 & 0.114 & 94.2 &  & 0.000 & 0.030 & 93.4 \\ 
 &  &  & S &  & -0.015 & 0.106 & 93.6 &  & 0.000 & 0.030 & 93.4 \\ 
 &  & \multirow{-5}{*}{SIME-I} & Id & \multirow{-4}{*}{Active} & 0.028 & 0.226 & 95.8 &  & -0.001 & 0.118 & 96.8 \\ 
\cline{3-12}
 &  &  & LW &  & 0.028 & 0.329 & 97.6 &  & -0.001 & 0.033 & 93.4 \\ 
 &  &  & S &  & 0.047 & 0.493 & 97.8 &  & -0.002 & 0.033 & 93.2 \\ 
 & \multirow{-8}{*}{0.4} & \multirow{-3}{*}{SIME-F} & Id & \multirow{-3}{*}{--} & 0.143 & 1.448 & 98.2 &  & 0.001 & 0.129 & 97.4 \\ 
\cline{2-12}
 &  &  &  & Broad & -0.007 & 0.104 & 96.2 &  & 0.000 & 0.033 & 92.8 \\ 
\cline{5-12}
 &  &  & \multirow{-2}{*}{Proj} &  & -0.015 & 0.102 & 95.0 &  & 0.000 & 0.032 & 92.0 \\ 
 &  &  & LW &  & -0.013 & 0.103 & 95.0 &  & 0.000 & 0.032 & 92.0 \\ 
 &  &  & S &  & -0.016 & 0.103 & 95.0 &  & 0.000 & 0.032 & 92.0 \\ 
 &  & \multirow{-5}{*}{SIME-I} & Id & \multirow{-4}{*}{Active} & 0.063 & 0.249 & 94.0 &  & 0.009 & 0.754 & 97.6 \\ 
\cline{3-12}
 &  &  & LW &  & 0.015 & 0.319 & 97.0 &  & -0.001 & 0.039 & 92.6 \\ 
 &  &  & S &  & 0.047 & 0.640 & 97.2 &  & -0.001 & 0.039 & 92.8 \\ 
\multirow{-24}{*}{50} & \multirow{-8}{*}{0.8} & \multirow{-3}{*}{SIME-F} & Id & \multirow{-3}{*}{--} & 8.859 & 197.736 & 98.4 &  & 0.024 & 1.341 & 97.4 \\ 
\hline
 &  &  &  & Broad & -0.056 & 0.140 & 87.2 &  & -0.002 & 0.037 & 94.6 \\ 
\cline{5-12}
 &  &  & \multirow{-2}{*}{Proj} &  & -0.185 & 0.241 & 52.0 &  & -0.002 & 0.037 & 94.4 \\ 
 &  &  & LW &  & -0.183 & 0.236 & 51.4 &  & -0.002 & 0.036 & 94.6 \\ 
 &  &  & S &  & -0.188 & 0.244 & 51.8 &  & -0.002 & 0.037 & 94.4 \\ 
 &  & \multirow{-5}{*}{SIME-I} & Id & \multirow{-4}{*}{Active} & -0.183 & 0.236 & 51.4 &  & -0.002 & 0.036 & 94.6 \\ 
\cline{3-12}
 &  &  & LW &  & 1.035 & 44.489 & 98.0 &  & -0.004 & 0.068 & 94.8 \\ 
 &  &  & S &  & -- & -- & -- &  & -0.016 & 0.159 & 96.8 \\ 
 & \multirow{-8}{*}{0.0} & \multirow{-3}{*}{SIME-F} & Id & \multirow{-3}{*}{--} & 0.956 & 46.114 & 98.0 &  & -0.004 & 0.068 & 94.8 \\ 
\cline{2-12}
 &  &  &  & Broad & -0.039 & 0.116 & 92.4 &  & -0.001 & 0.029 & 94.2 \\ 
\cline{5-12}
 &  &  & \multirow{-2}{*}{Proj} &  & -0.055 & 0.123 & 90.6 &  & -0.001 & 0.028 & 94.2 \\ 
 &  &  & LW &  & -0.042 & 0.156 & 92.8 &  & -0.002 & 0.032 & 94.6 \\ 
 &  &  & S &  & -0.057 & 0.125 & 90.4 &  & -0.001 & 0.029 & 94.2 \\ 
 &  & \multirow{-5}{*}{SIME-I} & Id & \multirow{-4}{*}{Active} & -0.042 & 0.191 & 93.4 &  & -0.019 & 0.108 & 95.8 \\ 
\cline{3-12}
 &  &  & LW &  & 5.370 & 167.013 & 97.8 &  & -0.006 & 0.075 & 95.4 \\ 
 &  &  & S &  & -- & -- & -- &  & -0.008 & 0.100 & 96.2 \\ 
 & \multirow{-8}{*}{0.4} & \multirow{-3}{*}{SIME-F} & Id & \multirow{-3}{*}{--} & 17.284 & 266.704 & 98.6 &  & -0.018 & 0.258 & 96.0 \\ 
\cline{2-12}
 &  &  &  & Broad & -0.021 & 0.106 & 96.0 &  & -0.001 & 0.032 & 94.8 \\ 
\cline{5-12}
 &  &  & \multirow{-2}{*}{Proj} &  & -0.022 & 0.104 & 94.8 &  & -0.001 & 0.030 & 93.6 \\ 
 &  &  & LW &  & -0.011 & 0.108 & 95.2 &  & -0.001 & 0.030 & 93.6 \\ 
 &  &  & S &  & -0.023 & 0.105 & 95.0 &  & -0.001 & 0.030 & 93.6 \\ 
 &  & \multirow{-5}{*}{SIME-I} & Id & \multirow{-4}{*}{Active} & 0.030 & 0.206 & 94.0 &  & -0.046 & 0.725 & 98.0 \\ 
\cline{3-12}
 &  &  & LW &  & -50.714 & 967.284 & 98.8 &  & 0.000 & 0.072 & 95.6 \\ 
 &  &  & S &  & -- & -- & -- &  & -0.001 & 0.110 & 95.8 \\ 
\multirow{-24}{*}{500} & \multirow{-8}{*}{0.8} & \multirow{-3}{*}{SIME-F} & Id & \multirow{-3}{*}{--} & $-2.47\mathrm{e}{+5}$ & $6.92\mathrm{e}{+6}$ & 100.0 &  & 0.344 & 8.140 & 97.8 \\ 
\hline
\end{tabular}%
}%
}
\end{table}

\begin{table}
\caption{Simulation Setting~2 (nonlinear). NIE estimation under various covariance estimation strategies over 500 replications. RMSE: root mean squared error. CP: coverage probability of nominal 95\% intervals. CP is given in \%.}
\label{tab:cov_nonlinear_p500}
\centering
{\renewcommand{\baselinestretch}{1}\normalsize
\renewcommand{\arraystretch}{0.86}
\simtablefit{%
\begin{tabular}{c c l c c r r r c r r r}
\hline
 &  &  &  &  & \multicolumn{3}{c}{$n=500$} &  & \multicolumn{3}{c}{$n=5{,}000$} \\
\cline{6-8} \cline{10-12}
\multicolumn{1}{c}{\multirow{-2}{*}{$p$}} & \multicolumn{1}{c}{\multirow{-2}{*}{$\varrho$}} & \multicolumn{1}{c}{\multirow{-2}{*}{Method}} & \multicolumn{1}{c}{\multirow{-2}{*}{Cov.}} & \multicolumn{1}{c}{\multirow{-2}{*}{Inference}} & \multicolumn{1}{c}{Bias} & \multicolumn{1}{c}{RMSE} & \multicolumn{1}{c}{CP} &  & \multicolumn{1}{c}{Bias} & \multicolumn{1}{c}{RMSE} & \multicolumn{1}{c}{CP} \\
\hline
 &  &  &  & Broad & 0.000 & 0.145 & 94.0 &  & -0.001 & 0.043 & 93.6 \\ 
\cline{5-12}
 &  &  & \multirow{-2}{*}{Proj} &  & -0.027 & 0.154 & 89.6 &  & -0.001 & 0.041 & 93.8 \\ 
 &  &  & LW &  & -0.024 & 0.145 & 89.4 &  & -0.001 & 0.041 & 93.6 \\ 
 &  &  & S &  & -0.028 & 0.158 & 89.8 &  & -0.001 & 0.041 & 93.8 \\ 
 &  & \multirow{-5}{*}{SIME-I} & Id & \multirow{-4}{*}{Active} & -0.024 & 0.145 & 89.4 &  & -0.001 & 0.041 & 93.8 \\ 
\cline{3-12}
 &  &  & LW &  & 0.018 & 0.284 & 97.4 &  & -0.002 & 0.042 & 93.4 \\ 
 &  &  & S &  & 0.044 & 0.762 & 98.4 &  & -0.002 & 0.044 & 93.4 \\ 
 & \multirow{-8}{*}{0.0} & \multirow{-3}{*}{SIME-F} & Id & \multirow{-3}{*}{--} & 0.018 & 0.289 & 97.4 &  & -0.002 & 0.042 & 93.2 \\ 
\cline{2-12}
 &  &  &  & Broad & -0.001 & 0.099 & 94.2 &  & 0.000 & 0.030 & 94.2 \\ 
\cline{5-12}
 &  &  & \multirow{-2}{*}{Proj} &  & -0.012 & 0.109 & 91.6 &  & 0.000 & 0.031 & 94.2 \\ 
 &  &  & LW &  & -0.008 & 0.122 & 92.2 &  & 0.000 & 0.031 & 94.4 \\ 
 &  &  & S &  & -0.012 & 0.110 & 91.6 &  & 0.000 & 0.031 & 94.2 \\ 
 &  & \multirow{-5}{*}{SIME-I} & Id & \multirow{-4}{*}{Active} & 0.031 & 0.430 & 97.6 &  & 0.009 & 0.149 & 97.4 \\ 
\cline{3-12}
 &  &  & LW &  & 0.008 & 0.352 & 97.4 &  & -0.001 & 0.034 & 94.0 \\ 
 &  &  & S &  & 0.014 & 0.484 & 97.4 &  & -0.001 & 0.034 & 94.0 \\ 
 & \multirow{-8}{*}{0.4} & \multirow{-3}{*}{SIME-F} & Id & \multirow{-3}{*}{--} & 0.092 & 1.746 & 98.0 &  & 0.008 & 0.162 & 97.6 \\ 
\cline{2-12}
 &  &  &  & Broad & -0.006 & 0.081 & 93.8 &  & 0.000 & 0.024 & 95.2 \\ 
\cline{5-12}
 &  &  & \multirow{-2}{*}{Proj} &  & -0.020 & 0.094 & 86.6 &  & 0.000 & 0.025 & 95.6 \\ 
 &  &  & LW &  & -0.019 & 0.095 & 87.0 &  & 0.000 & 0.025 & 95.6 \\ 
 &  &  & S &  & -0.020 & 0.095 & 86.4 &  & 0.000 & 0.025 & 95.6 \\ 
 &  & \multirow{-5}{*}{SIME-I} & Id & \multirow{-4}{*}{Active} & 0.082 & 0.696 & 91.6 &  & 0.220 & 1.375 & 97.4 \\ 
\cline{3-12}
 &  &  & LW &  & 0.004 & 0.350 & 96.0 &  & 0.000 & 0.034 & 94.6 \\ 
 &  &  & S &  & 0.016 & 0.638 & 97.0 &  & -0.001 & 0.035 & 95.0 \\ 
\multirow{-24}{*}{50} & \multirow{-8}{*}{0.8} & \multirow{-3}{*}{SIME-F} & Id & \multirow{-3}{*}{--} & 9.866 & 210.869 & 98.4 &  & 0.257 & 2.227 & 97.4 \\ 
\hline
 &  &  &  & Broad & -0.071 & 0.163 & 77.2 &  & -0.002 & 0.040 & 95.0 \\ 
\cline{5-12}
 &  &  & \multirow{-2}{*}{Proj} &  & -0.182 & 0.259 & 49.8 &  & -0.001 & 0.039 & 94.4 \\ 
 &  &  & LW &  & -0.181 & 0.256 & 49.6 &  & -0.001 & 0.038 & 94.6 \\ 
 &  &  & S &  & -0.184 & 0.262 & 50.2 &  & -0.001 & 0.039 & 94.4 \\ 
 &  & \multirow{-5}{*}{SIME-I} & Id & \multirow{-4}{*}{Active} & -0.181 & 0.256 & 49.6 &  & -0.001 & 0.038 & 94.6 \\ 
\cline{3-12}
 &  &  & LW &  & 2.023 & 48.676 & 98.0 &  & -0.002 & 0.069 & 96.0 \\ 
 &  &  & S &  & -- & -- & -- &  & -0.010 & 0.158 & 97.0 \\ 
 & \multirow{-8}{*}{0.0} & \multirow{-3}{*}{SIME-F} & Id & \multirow{-3}{*}{--} & 1.933 & 50.576 & 98.0 &  & -0.002 & 0.069 & 96.0 \\ 
\cline{2-12}
 &  &  &  & Broad & -0.035 & 0.110 & 87.2 &  & 0.000 & 0.028 & 94.8 \\ 
\cline{5-12}
 &  &  & \multirow{-2}{*}{Proj} &  & -0.079 & 0.150 & 72.4 &  & -0.001 & 0.029 & 95.2 \\ 
 &  &  & LW &  & -0.077 & 0.218 & 79.2 &  & -0.001 & 0.033 & 95.2 \\ 
 &  &  & S &  & -0.079 & 0.152 & 72.8 &  & -0.001 & 0.029 & 95.2 \\ 
 &  & \multirow{-5}{*}{SIME-I} & Id & \multirow{-4}{*}{Active} & -0.072 & 0.344 & 81.8 &  & -0.001 & 0.142 & 96.2 \\ 
\cline{3-12}
 &  &  & LW &  & 8.107 & 233.569 & 97.8 &  & -0.002 & 0.078 & 96.8 \\ 
 &  &  & S &  & -- & -- & -- &  & -0.004 & 0.100 & 96.0 \\ 
 & \multirow{-8}{*}{0.4} & \multirow{-3}{*}{SIME-F} & Id & \multirow{-3}{*}{--} & 22.328 & 313.802 & 98.4 &  & 0.005 & 0.405 & 96.2 \\ 
\cline{2-12}
 &  &  &  & Broad & -0.026 & 0.082 & 90.8 &  & 0.000 & 0.024 & 95.2 \\ 
\cline{5-12}
 &  &  & \multirow{-2}{*}{Proj} &  & -0.062 & 0.107 & 70.4 &  & -0.001 & 0.024 & 95.4 \\ 
 &  &  & LW &  & -0.062 & 0.113 & 71.0 &  & -0.001 & 0.024 & 95.6 \\ 
 &  &  & S &  & -0.062 & 0.107 & 70.4 &  & -0.001 & 0.024 & 95.4 \\ 
 &  & \multirow{-5}{*}{SIME-I} & Id & \multirow{-4}{*}{Active} & -0.040 & 1.022 & 72.8 &  & 0.132 & 1.249 & 97.8 \\ 
\cline{3-12}
 &  &  & LW &  & -20.097 & $1.34\mathrm{e}{+3}$ & 99.4 &  & 0.000 & 0.068 & 96.0 \\ 
 &  &  & S &  & -- & -- & -- &  & -0.001 & 0.107 & 96.0 \\ 
\multirow{-24}{*}{500} & \multirow{-8}{*}{0.8} & \multirow{-3}{*}{SIME-F} & Id & \multirow{-3}{*}{--} & $1.41\mathrm{e}{+5}$ & $6.92\mathrm{e}{+6}$ & 99.2 &  & 0.917 & 11.198 & 97.4 \\ 
\hline
\end{tabular}%
}%
}
\end{table}

\clearpage

\paragraph{Augmentation comparison.}
Tables~\ref{tab:aug_linear} and~\ref{tab:aug_nonlinear} compare four projected SIME-I scores in the smaller samples. With the refitted links held fixed, broader correction improves coverage and generally reduces RMSE relative to the unaugmented score. The comparison with the original active-space score additionally reflects the effect of link refitting.

\begin{table}
\caption{Simulation Setting~1 (linear): SIME-I augmentation comparisons at $n=500$, with 500 paired replications per setting. RMSE: root mean squared error. CP: coverage probability of nominal 95\% intervals. CP is given in \%.}
\label{tab:aug_linear}
\begin{center}
{\renewcommand{\baselinestretch}{1}\normalsize
\renewcommand{\arraystretch}{1}
\begin{tabular}{r c l r c r r r}
\hline
 &  &  & \multicolumn{1}{c}{NDE} &  & \multicolumn{3}{c}{NIE} \\
\cline{4-4} \cline{6-8}
\multicolumn{1}{c}{\multirow{-2}{*}{$p$}} & \multicolumn{1}{c}{\multirow{-2}{*}{$\varrho$}} & \multicolumn{1}{c}{\multirow{-2}{*}{Procedure}} & \multicolumn{1}{c}{CP} &  & \multicolumn{1}{c}{Bias} & \multicolumn{1}{c}{RMSE} & \multicolumn{1}{c}{CP} \\
\hline
 &  & Original active & 92.6 &  & -0.040 & 0.135 & 89.2 \\ 
 &  & Refitted active & 93.4 &  & -0.031 & 0.131 & 90.2 \\ 
 &  & Broad & 95.4 &  & -0.003 & 0.117 & 95.8 \\ 
 & \multirow{-4}{*}{0.0} & Refitted unaug. & 84.0 &  & -0.122 & 0.176 & 72.2 \\ 
\cline{2-8}
 &  & Original active & 95.2 &  & -0.014 & 0.105 & 93.4 \\ 
 &  & Refitted active & 95.6 &  & -0.013 & 0.103 & 94.2 \\ 
 &  & Broad & 96.0 &  & -0.005 & 0.100 & 94.6 \\ 
 & \multirow{-4}{*}{0.4} & Refitted unaug. & 92.4 &  & -0.071 & 0.119 & 86.2 \\ 
\cline{2-8}
 &  & Original active & 95.4 &  & -0.015 & 0.102 & 95.0 \\ 
 &  & Refitted active & 95.2 &  & -0.015 & 0.101 & 95.8 \\ 
 &  & Broad & 96.0 &  & -0.007 & 0.104 & 96.2 \\ 
\multirow{-12}{*}{50} & \multirow{-4}{*}{0.8} & Refitted unaug. & 94.2 &  & -0.040 & 0.104 & 93.6 \\ 
\hline
 &  & Original active & 67.4 &  & -0.185 & 0.241 & 52.0 \\ 
 &  & Refitted active & 71.6 &  & -0.166 & 0.222 & 55.2 \\ 
 &  & Broad & 91.4 &  & -0.056 & 0.140 & 87.2 \\ 
 & \multirow{-4}{*}{0.0} & Refitted unaug. & 62.2 &  & -0.214 & 0.254 & 43.8 \\ 
\cline{2-8}
 &  & Original active & 92.8 &  & -0.055 & 0.123 & 90.6 \\ 
 &  & Refitted active & 93.4 &  & -0.046 & 0.119 & 91.0 \\ 
 &  & Broad & 95.0 &  & -0.039 & 0.116 & 92.4 \\ 
 & \multirow{-4}{*}{0.4} & Refitted unaug. & 84.0 &  & -0.116 & 0.154 & 75.2 \\ 
\cline{2-8}
 &  & Original active & 94.4 &  & -0.022 & 0.104 & 94.8 \\ 
 &  & Refitted active & 94.6 &  & -0.021 & 0.104 & 94.6 \\ 
 &  & Broad & 95.4 &  & -0.021 & 0.106 & 96.0 \\ 
\multirow{-12}{*}{500} & \multirow{-4}{*}{0.8} & Refitted unaug. & 90.4 &  & -0.061 & 0.112 & 91.4 \\ 
\hline
\end{tabular}
}
\end{center}
\end{table}

\begin{table}
\caption{Simulation Setting~2 (nonlinear): SIME-I augmentation comparisons at $n=500$, with 500 paired replications per setting. RMSE: root mean squared error. CP: coverage probability of nominal 95\% intervals. CP is given in \%.}
\label{tab:aug_nonlinear}
\begin{center}
{\renewcommand{\baselinestretch}{1}\normalsize
\renewcommand{\arraystretch}{1}
\begin{tabular}{r c l r c r r r}
\hline
 &  &  & \multicolumn{1}{c}{NDE} &  & \multicolumn{3}{c}{NIE} \\
\cline{4-4} \cline{6-8}
\multicolumn{1}{c}{\multirow{-2}{*}{$p$}} & \multicolumn{1}{c}{\multirow{-2}{*}{$\varrho$}} & \multicolumn{1}{c}{\multirow{-2}{*}{Procedure}} & \multicolumn{1}{c}{CP} &  & \multicolumn{1}{c}{Bias} & \multicolumn{1}{c}{RMSE} & \multicolumn{1}{c}{CP} \\
\hline
 &  & Original active & 92.2 &  & -0.027 & 0.154 & 89.6 \\ 
 &  & Refitted active & 91.6 &  & -0.016 & 0.157 & 89.2 \\ 
 &  & Broad & 94.4 &  & 0.000 & 0.145 & 94.0 \\ 
 & \multirow{-4}{*}{0.0} & Refitted unaug. & 92.6 &  & -0.045 & 0.162 & 87.4 \\ 
\cline{2-8}
 &  & Original active & 95.2 &  & -0.012 & 0.109 & 91.6 \\ 
 &  & Refitted active & 94.2 &  & -0.008 & 0.105 & 91.2 \\ 
 &  & Broad & 95.4 &  & -0.001 & 0.099 & 94.2 \\ 
 & \multirow{-4}{*}{0.4} & Refitted unaug. & 93.8 &  & -0.023 & 0.105 & 89.4 \\ 
\cline{2-8}
 &  & Original active & 95.0 &  & -0.020 & 0.094 & 86.6 \\ 
 &  & Refitted active & 94.6 &  & -0.019 & 0.086 & 87.6 \\ 
 &  & Broad & 94.4 &  & -0.006 & 0.081 & 93.8 \\ 
\multirow{-12}{*}{50} & \multirow{-4}{*}{0.8} & Refitted unaug. & 94.0 &  & -0.029 & 0.090 & 81.8 \\ 
\hline
 &  & Original active & 65.6 &  & -0.182 & 0.259 & 49.8 \\ 
 &  & Refitted active & 68.0 &  & -0.163 & 0.246 & 50.8 \\ 
 &  & Broad & 90.6 &  & -0.071 & 0.163 & 77.2 \\ 
 & \multirow{-4}{*}{0.0} & Refitted unaug. & 62.0 &  & -0.205 & 0.268 & 44.4 \\ 
\cline{2-8}
 &  & Original active & 86.4 &  & -0.079 & 0.150 & 72.4 \\ 
 &  & Refitted active & 87.0 &  & -0.068 & 0.140 & 72.2 \\ 
 &  & Broad & 94.0 &  & -0.035 & 0.110 & 87.2 \\ 
 & \multirow{-4}{*}{0.4} & Refitted unaug. & 84.2 &  & -0.094 & 0.152 & 67.8 \\ 
\cline{2-8}
 &  & Original active & 91.0 &  & -0.062 & 0.107 & 70.4 \\ 
 &  & Refitted active & 90.6 &  & -0.059 & 0.101 & 69.6 \\ 
 &  & Broad & 94.8 &  & -0.026 & 0.082 & 90.8 \\ 
\multirow{-12}{*}{500} & \multirow{-4}{*}{0.8} & Refitted unaug. & 88.8 &  & -0.071 & 0.106 & 59.4 \\ 
\hline
\end{tabular}
}
\end{center}
\end{table}

\paragraph{Training-fold and descriptive selection.}
Tables~\ref{tab:selection_linear} and~\ref{tab:selection_nonlinear} show that selection improves with sample size, although false positives remain. In smaller samples, training-fold fits can recover fewer true mediators than the descriptive full-sample fit. Restricting the candidate support still allows the SCAD refit to omit true mediators. The index error is the mean training-fold squared distance to $\btheta_0$, minimized over sign.

\begin{table}
\caption{Simulation Setting~1 (linear): descriptive full-sample and training-fold index selection over 500 replications. TPR/FPR: true/false positive rates.}
\label{tab:selection_linear}
\begin{center}
{\renewcommand{\baselinestretch}{1}\normalsize
\renewcommand{\arraystretch}{1}
\begin{tabular}{r r c l r r c r r c r}
\hline
 &  &  &  & \multicolumn{2}{c}{Full sample} &  & \multicolumn{2}{c}{Training folds} &  &  \\
\cline{5-6} \cline{8-9}
\multicolumn{1}{c}{\multirow{-2}{*}{$p$}} & \multicolumn{1}{c}{\multirow{-2}{*}{$n$}} & \multicolumn{1}{c}{\multirow{-2}{*}{$\varrho$}} & \multicolumn{1}{c}{\multirow{-2}{*}{Method}} & \multicolumn{1}{c}{TPR} & \multicolumn{1}{c}{FPR} &  & \multicolumn{1}{c}{TPR} & \multicolumn{1}{c}{FPR} &  & \multicolumn{1}{c}{\multirow{-2}{*}{Index error}} \\
\hline
 &  &  & SIME & 0.982 & 0.066 &  & 0.946 & 0.061 &  & 0.253 \\ 
 &  & \multirow{-2}{*}{0.0} & Oracle & 0.999 & 0.000 &  & 0.995 & 0.000 &  & 0.127 \\ 
\cline{3-11}
 &  &  & SIME & 0.965 & 0.085 &  & 0.935 & 0.088 &  & 0.353 \\ 
 &  & \multirow{-2}{*}{0.4} & Oracle & 0.994 & 0.000 &  & 0.983 & 0.000 &  & 0.214 \\ 
\cline{3-11}
 &  &  & SIME & 0.706 & 0.074 &  & 0.668 & 0.073 &  & 0.655 \\ 
 & \multirow{-6}{*}{500} & \multirow{-2}{*}{0.8} & Oracle & 0.856 & 0.000 &  & 0.823 & 0.000 &  & 0.497 \\ 
\cline{2-11}
 &  &  & SIME & 1.000 & 0.085 &  & 1.000 & 0.083 &  & 0.007 \\ 
 &  & \multirow{-2}{*}{0.0} & Oracle & 1.000 & 0.000 &  & 1.000 & 0.000 &  & 0.005 \\ 
\cline{3-11}
 &  &  & SIME & 1.000 & 0.085 &  & 1.000 & 0.086 &  & 0.010 \\ 
 &  & \multirow{-2}{*}{0.4} & Oracle & 1.000 & 0.000 &  & 1.000 & 0.000 &  & 0.007 \\ 
\cline{3-11}
 &  &  & SIME & 1.000 & 0.062 &  & 0.999 & 0.061 &  & 0.029 \\ 
\multirow{-12}{*}{50} & \multirow{-6}{*}{5,000} & \multirow{-2}{*}{0.8} & Oracle & 1.000 & 0.000 &  & 1.000 & 0.000 &  & 0.027 \\ 
\hline
 &  &  & SIME & 0.844 & 0.005 &  & 0.725 & 0.006 &  & 0.474 \\ 
 &  & \multirow{-2}{*}{0.0} & Oracle & 0.997 & 0.000 &  & 0.994 & 0.000 &  & 0.133 \\ 
\cline{3-11}
 &  &  & SIME & 0.897 & 0.018 &  & 0.828 & 0.018 &  & 0.491 \\ 
 &  & \multirow{-2}{*}{0.4} & Oracle & 0.992 & 0.000 &  & 0.985 & 0.000 &  & 0.215 \\ 
\cline{3-11}
 &  &  & SIME & 0.633 & 0.015 &  & 0.599 & 0.014 &  & 0.736 \\ 
 & \multirow{-6}{*}{500} & \multirow{-2}{*}{0.8} & Oracle & 0.858 & 0.000 &  & 0.826 & 0.000 &  & 0.499 \\ 
\cline{2-11}
 &  &  & SIME & 1.000 & 0.023 &  & 1.000 & 0.022 &  & 0.009 \\ 
 &  & \multirow{-2}{*}{0.0} & Oracle & 1.000 & 0.000 &  & 1.000 & 0.000 &  & 0.005 \\ 
\cline{3-11}
 &  &  & SIME & 1.000 & 0.022 &  & 1.000 & 0.023 &  & 0.017 \\ 
 &  & \multirow{-2}{*}{0.4} & Oracle & 1.000 & 0.000 &  & 1.000 & 0.000 &  & 0.007 \\ 
\cline{3-11}
 &  &  & SIME & 1.000 & 0.013 &  & 0.998 & 0.013 &  & 0.037 \\ 
\multirow{-12}{*}{500} & \multirow{-6}{*}{5,000} & \multirow{-2}{*}{0.8} & Oracle & 1.000 & 0.000 &  & 1.000 & 0.000 &  & 0.025 \\ 
\hline
\end{tabular}
}
\end{center}
\end{table}

\begin{table}
\caption{Simulation Setting~2 (nonlinear): descriptive full-sample and training-fold index selection over 500 replications. TPR/FPR: true/false positive rates.}
\label{tab:selection_nonlinear}
\begin{center}
{\renewcommand{\baselinestretch}{1}\normalsize
\renewcommand{\arraystretch}{1}
\begin{tabular}{r r c l r r c r r c r}
\hline
 &  &  &  & \multicolumn{2}{c}{Full sample} &  & \multicolumn{2}{c}{Training folds} &  &  \\
\cline{5-6} \cline{8-9}
\multicolumn{1}{c}{\multirow{-2}{*}{$p$}} & \multicolumn{1}{c}{\multirow{-2}{*}{$n$}} & \multicolumn{1}{c}{\multirow{-2}{*}{$\varrho$}} & \multicolumn{1}{c}{\multirow{-2}{*}{Method}} & \multicolumn{1}{c}{TPR} & \multicolumn{1}{c}{FPR} &  & \multicolumn{1}{c}{TPR} & \multicolumn{1}{c}{FPR} &  & \multicolumn{1}{c}{\multirow{-2}{*}{Index error}} \\
\hline
 &  &  & SIME & 0.991 & 0.070 &  & 0.960 & 0.066 &  & 0.112 \\ 
 &  & \multirow{-2}{*}{0.0} & Oracle & 1.000 & 0.000 &  & 1.000 & 0.000 &  & 0.034 \\ 
\cline{3-11}
 &  &  & SIME & 0.994 & 0.078 &  & 0.962 & 0.078 &  & 0.155 \\ 
 &  & \multirow{-2}{*}{0.4} & Oracle & 1.000 & 0.000 &  & 0.999 & 0.000 &  & 0.075 \\ 
\cline{3-11}
 &  &  & SIME & 0.723 & 0.075 &  & 0.649 & 0.074 &  & 0.755 \\ 
 & \multirow{-6}{*}{500} & \multirow{-2}{*}{0.8} & Oracle & 0.869 & 0.000 &  & 0.842 & 0.000 &  & 0.449 \\ 
\cline{2-11}
 &  &  & SIME & 1.000 & 0.095 &  & 1.000 & 0.094 &  & 0.003 \\ 
 &  & \multirow{-2}{*}{0.0} & Oracle & 1.000 & 0.000 &  & 1.000 & 0.000 &  & 0.002 \\ 
\cline{3-11}
 &  &  & SIME & 1.000 & 0.092 &  & 1.000 & 0.091 &  & 0.004 \\ 
 &  & \multirow{-2}{*}{0.4} & Oracle & 1.000 & 0.000 &  & 1.000 & 0.000 &  & 0.003 \\ 
\cline{3-11}
 &  &  & SIME & 1.000 & 0.064 &  & 1.000 & 0.063 &  & 0.011 \\ 
\multirow{-12}{*}{50} & \multirow{-6}{*}{5,000} & \multirow{-2}{*}{0.8} & Oracle & 1.000 & 0.000 &  & 1.000 & 0.000 &  & 0.010 \\ 
\hline
 &  &  & SIME & 0.837 & 0.013 &  & 0.684 & 0.012 &  & 0.666 \\ 
 &  & \multirow{-2}{*}{0.0} & Oracle & 1.000 & 0.000 &  & 1.000 & 0.000 &  & 0.035 \\ 
\cline{3-11}
 &  &  & SIME & 0.869 & 0.016 &  & 0.737 & 0.016 &  & 0.600 \\ 
 &  & \multirow{-2}{*}{0.4} & Oracle & 1.000 & 0.000 &  & 0.999 & 0.000 &  & 0.081 \\ 
\cline{3-11}
 &  &  & SIME & 0.430 & 0.015 &  & 0.343 & 0.015 &  & 1.309 \\ 
 & \multirow{-6}{*}{500} & \multirow{-2}{*}{0.8} & Oracle & 0.878 & 0.000 &  & 0.846 & 0.000 &  & 0.441 \\ 
\cline{2-11}
 &  &  & SIME & 1.000 & 0.030 &  & 1.000 & 0.030 &  & 0.006 \\ 
 &  & \multirow{-2}{*}{0.0} & Oracle & 1.000 & 0.000 &  & 1.000 & 0.000 &  & 0.002 \\ 
\cline{3-11}
 &  &  & SIME & 1.000 & 0.023 &  & 1.000 & 0.022 &  & 0.007 \\ 
 &  & \multirow{-2}{*}{0.4} & Oracle & 1.000 & 0.000 &  & 1.000 & 0.000 &  & 0.003 \\ 
\cline{3-11}
 &  &  & SIME & 1.000 & 0.014 &  & 1.000 & 0.014 &  & 0.014 \\ 
\multirow{-12}{*}{500} & \multirow{-6}{*}{5,000} & \multirow{-2}{*}{0.8} & Oracle & 1.000 & 0.000 &  & 1.000 & 0.000 &  & 0.010 \\ 
\hline
\end{tabular}
}
\end{center}
\end{table}


\subsection{Combined Link Refitting and Broader Correction}
\label{sub:app_combined_inference}

Relative to the original active-space estimator, the combined strategy reduces NIE bias and improves coverage in smaller samples. RMSE generally decreases, and differences between the two procedures diminish with larger samples.

Table~\ref{tab:combined_inference_ablation} compares active-space correction, link refitting alone, broader correction alone and both changes together at fixed Stage~1 directions and supports. Combining the two changes gives the largest improvement in coverage with both outcome links in the settings shown. The NIE standard deviations are empirical, and the standard errors are averaged across the replications.
\begin{table}[H]
\caption{Paired SIME-I component diagnostics at $p=n=500$ and $\varrho=0$, with 500 replications per link. SD: standard deviation. SE: standard error. CP: coverage probability of nominal 95\% intervals. CP is given in \%.}
\label{tab:combined_inference_ablation}
\begin{center}
{\renewcommand{\baselinestretch}{1}\normalsize
\renewcommand{\arraystretch}{1}
\begin{tabular}{l l r r r r r}
\hline
\multicolumn{1}{c}{Link} & \multicolumn{1}{c}{Procedure} & \multicolumn{1}{c}{NIE bias} & \multicolumn{1}{c}{SD} & \multicolumn{1}{c}{SE} & \multicolumn{1}{c}{NDE CP} & \multicolumn{1}{c}{NIE CP} \\
\hline
 & Active & -0.185 & 0.153 & 0.105 & 67.4 & 52.0 \\ 
 & Refit & -0.166 & 0.148 & 0.103 & 71.6 & 55.2 \\ 
 & Broad & -0.099 & 0.141 & 0.110 & 87.0 & 75.8 \\ 
\multirow{-4}{*}{Linear} & Both & -0.056 & 0.129 & 0.108 & 91.6 & 86.6 \\ 
\hline
 & Active & -0.182 & 0.185 & 0.106 & 65.6 & 49.8 \\ 
 & Refit & -0.163 & 0.185 & 0.101 & 68.0 & 50.8 \\ 
 & Broad & -0.118 & 0.162 & 0.112 & 82.4 & 64.8 \\ 
\multirow{-4}{*}{Sine} & Both & -0.069 & 0.147 & 0.110 & 90.8 & 77.6 \\ 
\hline
\end{tabular}
}
\end{center}
\end{table}


\section{Additional A4 Study Results}
\label{app:a4}

This section provides the data preparation, method comparisons and diagnostic results for the A4 analyses in Section~\ref{sec:a4}.


\subsection{Data Preparation and Analysis}
\label{sub:app_a4_analysis}

The exposure ($X$) is randomized assignment to solanezumab or placebo, and the outcome ($Y$) is PACC change from treatment baseline to the 240-week visit. The 43 candidate mediators ($\bM$) are MRI volume changes from screening to endpoint, normalized by intracranial volume and divided by elapsed days. The covariates ($\bZ$) are age, sex, years of education and body mass index.

The complete-case analysis includes 487 carriers (237 treated and 250 controls) and 304 non-carriers (150 treated and 154 controls). One treated non-carrier with extreme MRI changes was excluded before standardization based on the influence diagnostics in Appendix~\ref{sub:app_a4_stability}. Mediators, outcome and covariates are standardized within subgroups before creating the folds. Effect units are subgroup-specific standard deviations of PACC change, corresponding to 4.618 points for carriers and 3.474 for non-carriers.

Index, link and nuisance models are fitted on the training complements of five evaluation folds. The principal SIME-I uses projected residual variance, Gaussian integration and the combined strategy in Appendix~\ref{sub:app_combined_strategy}. Its Wald intervals use the complete held-out effect scores. 

LinMed, HIMA and JonesPRD use the same subgroup samples. Inference for JonesPRD uses 500 perturbation draws. In Table~\ref{tab:a4_effects}, Selected gives the number of mediators in the descriptive full-sample SIME index or in the competitor's full-sample selected set. The SIME effect estimates use the training-fold indices. LinMed intervals are based on exploratory post-selection ordinary least-squares refits and delta-method calculations. HIMA intervals and competitor TE intervals are unavailable.

\subsection{Density-Ratio Stability}
\label{sub:app_a4_ratios}

Table~\ref{tab:a4_ratios} reports the maximum, 99th percentile and maximum natural logarithm of the held-out density ratios. The fitted ratios are finite and used as estimated.

\begin{table}
\centering
\setlength{\abovecaptionskip}{0pt}
\setlength{\belowcaptionskip}{3pt}
\caption{Held-out density-ratio diagnostics in the A4 study. Proj: projected residual variance. LW: Ledoit--Wolf shrinkage. D0.1: fixed diagonal shrinkage.}
\label{tab:a4_ratios}
{\renewcommand{\baselinestretch}{1}\normalsize
\renewcommand{\arraystretch}{1}
\resizebox{\textwidth}{!}{%
\begin{tabular}{l c c r r r c r r r}
\hline
 & & & \multicolumn{3}{c}{Carrier ($n=487$)} & & \multicolumn{3}{c}{Non-carrier ($n=304$)} \\
\cline{4-6}\cline{8-10}
\multicolumn{1}{c}{\multirow{-2}{*}{Method}} & \multicolumn{1}{c}{\multirow{-2}{*}{Cov.}} & \multicolumn{1}{c}{\multirow{-2}{*}{Inference}} & \multicolumn{1}{c}{Max $\hat r$} & \multicolumn{1}{c}{99th pct.} & \multicolumn{1}{c}{Max $\log\hat r$} &   & \multicolumn{1}{c}{Max $\hat r$} & \multicolumn{1}{c}{99th pct.} & \multicolumn{1}{c}{Max $\log\hat r$} \\
\hline
 &  & Combined & 1.395 & 1.285 & 0.333 &  & 1.163 & 1.118 & 0.151 \\ 
\cline{3-10}
 & \multirow{-2}{*}{Proj} &  & 1.395 & 1.285 & 0.333 &  & 1.163 & 1.118 & 0.151 \\ 
\cline{2-2}\cline{4-10}
 & LW &  & 1.410 & 1.296 & 0.344 &  & 1.175 & 1.127 & 0.161 \\ 
\cline{2-2}\cline{4-10}
 & D0.1 &  & 1.427 & 1.307 & 0.356 &  & 1.178 & 1.129 & 0.164 \\ 
\cline{2-2}\cline{4-10}
\multirow{-5}{*}{SIME-I} & Sample & \multirow{-4}{*}{Active} & 1.402 & 1.290 & 0.338 &  & 1.167 & 1.121 & 0.154 \\ 
\hline
 & LW &  & 28.049 & 5.917 & 3.334 &  & 20.625 & 13.048 & 3.026 \\ 
\cline{2-2}\cline{4-10}
 & D0.1 &  & 27.734 & 5.914 & 3.323 &  & 20.966 & 15.374 & 3.043 \\ 
\cline{2-2}\cline{4-10}
\multirow{-3}{*}{SIME-F} & Sample & \multirow{-3}{*}{--} & 36.687 & 7.450 & 3.602 &  & 41.785 & 34.699 & 3.733 \\ 
\hline
\end{tabular}
}
}
\end{table}

\begin{samepage}
The index ratios give less extreme reweighting than the full-mediator ratios in both subgroups. The projected ratios are identical for the combined and original active-space estimates because both use the same index directions and mediator models.\par
\end{samepage}

\subsection{Covariance and Ratio Comparisons}
\label{sub:app_a4_covariance}

Tables~\ref{tab:a4_covariance} and~\ref{tab:a4_covariance_noncarrier} compare the covariance choices at the same Stage~1 directions. LW shrinkage uses residual contrasts, a scaled-identity target and $N-6$ residual degrees of freedom, as in Appendix~\ref{sub:app_covariance_estimation}. Fixed diagonal shrinkage uses weight 0.1, and sample covariance uses divisor $N$, where $N$ is the training-sample size.

\begin{table}
\centering
\setlength{\abovecaptionskip}{0pt}
\setlength{\belowcaptionskip}{3pt}
\caption{Estimated effects and 95\% CIs under the covariance comparisons for APOE-$\varepsilon4$ carriers ($n=487$). CI: confidence interval. Proj: projected residual variance. LW: Ledoit--Wolf shrinkage. D0.1: fixed diagonal shrinkage.}
\label{tab:a4_covariance}
{\renewcommand{\baselinestretch}{1}\normalsize
\renewcommand{\arraystretch}{1}
\begin{tabular}{l c c l r r}
\hline
\multicolumn{1}{c}{Method} & \multicolumn{1}{c}{Cov.} & \multicolumn{1}{c}{Inference} & \multicolumn{1}{c}{Quantity} & \multicolumn{1}{c}{Estimate} & \multicolumn{1}{c}{95\% CI} \\
\hline
 &  &  & NDE & $-0.0329$ & $[-0.1729,\,0.1071]$ \\ 
 &  &  & NIE & $-0.0525$ & $[-0.1877,\,0.0826]$ \\ 
 &  & \multirow{-3}{*}{Combined} & TE & $-0.0854$ & $[-0.2632,\,0.0924]$ \\ 
\cline{3-6}
 &  &  & NDE & $-0.0447$ & $[-0.1820,\,0.0925]$ \\ 
 &  &  & NIE & $-0.0418$ & $[-0.1794,\,0.0958]$ \\ 
 & \multirow{-6}{*}{Proj} &  & TE & $-0.0865$ & $[-0.2644,\,0.0913]$ \\ 
\cline{2-2}\cline{4-6}
 &  &  & NDE & $-0.0444$ & $[-0.1816,\,0.0929]$ \\ 
 &  &  & NIE & $-0.0422$ & $[-0.1798,\,0.0955]$ \\ 
 & \multirow{-3}{*}{LW} &  & TE & $-0.0865$ & $[-0.2644,\,0.0913]$ \\ 
\cline{2-2}\cline{4-6}
 &  &  & NDE & $-0.0441$ & $[-0.1813,\,0.0930]$ \\ 
 &  &  & NIE & $-0.0424$ & $[-0.1800,\,0.0953]$ \\ 
 & \multirow{-3}{*}{D0.1} &  & TE & $-0.0865$ & $[-0.2644,\,0.0914]$ \\ 
\cline{2-2}\cline{4-6}
 &  &  & NDE & $-0.0447$ & $[-0.1819,\,0.0925]$ \\ 
 &  &  & NIE & $-0.0419$ & $[-0.1795,\,0.0958]$ \\ 
\multirow{-15}{*}{SIME-I} & \multirow{-3}{*}{Sample} & \multirow{-12}{*}{Active} & TE & $-0.0866$ & $[-0.2644,\,0.0913]$ \\ 
\hline
 &  &  & NDE & $-0.2265$ & $[-0.6959,\,0.2429]$ \\ 
 &  &  & NIE & $0.1400$ & $[-0.3229,\,0.6029]$ \\ 
 & \multirow{-3}{*}{LW} &  & TE & $-0.0865$ & $[-0.2644,\,0.0913]$ \\ 
\cline{2-2}\cline{4-6}
 &  &  & NDE & $-0.2288$ & $[-0.6929,\,0.2352]$ \\ 
 &  &  & NIE & $0.1423$ & $[-0.3154,\,0.6000]$ \\ 
 & \multirow{-3}{*}{D0.1} &  & TE & $-0.0865$ & $[-0.2644,\,0.0914]$ \\ 
\cline{2-2}\cline{4-6}
 &  &  & NDE & $-0.2729$ & $[-0.8747,\,0.3289]$ \\ 
 &  &  & NIE & $0.1864$ & $[-0.4082,\,0.7809]$ \\ 
\multirow{-9}{*}{SIME-F} & \multirow{-3}{*}{Sample} & \multirow{-9}{*}{--} & TE & $-0.0866$ & $[-0.2644,\,0.0913]$ \\ 
\hline
\end{tabular}
}
\end{table}

\begin{table}
\centering
\setlength{\abovecaptionskip}{0pt}
\setlength{\belowcaptionskip}{3pt}
\caption{Estimated effects and 95\% CIs under the covariance comparisons for APOE-$\varepsilon4$ non-carriers ($n=304$). CI: confidence interval. Proj: projected residual variance. LW: Ledoit--Wolf shrinkage. D0.1: fixed diagonal shrinkage.}
\label{tab:a4_covariance_noncarrier}
{\renewcommand{\baselinestretch}{1}\normalsize
\renewcommand{\arraystretch}{1}
\begin{tabular}{l c c l r r}
\hline
\multicolumn{1}{c}{Method} & \multicolumn{1}{c}{Cov.} & \multicolumn{1}{c}{Inference} & \multicolumn{1}{c}{Quantity} & \multicolumn{1}{c}{Estimate} & \multicolumn{1}{c}{95\% CI} \\
\hline
 &  &  & NDE & $0.0676$ & $[-0.1560,\,0.2912]$ \\ 
 &  &  & NIE & $0.0448$ & $[-0.1464,\,0.2360]$ \\ 
 &  & \multirow{-3}{*}{Combined} & TE & $0.1124$ & $[-0.1156,\,0.3403]$ \\ 
\cline{3-6}
 &  &  & NDE & $0.0458$ & $[-0.2163,\,0.3079]$ \\ 
 &  &  & NIE & $0.0768$ & $[-0.1648,\,0.3185]$ \\ 
 & \multirow{-6}{*}{Proj} &  & TE & $0.1226$ & $[-0.1053,\,0.3506]$ \\ 
\cline{2-2}\cline{4-6}
 &  &  & NDE & $0.0469$ & $[-0.2148,\,0.3087]$ \\ 
 &  &  & NIE & $0.0749$ & $[-0.1665,\,0.3163]$ \\ 
 & \multirow{-3}{*}{LW} &  & TE & $0.1218$ & $[-0.1061,\,0.3497]$ \\ 
\cline{2-2}\cline{4-6}
 &  &  & NDE & $0.0472$ & $[-0.2144,\,0.3088]$ \\ 
 &  &  & NIE & $0.0742$ & $[-0.1670,\,0.3155]$ \\ 
 & \multirow{-3}{*}{D0.1} &  & TE & $0.1214$ & $[-0.1065,\,0.3494]$ \\ 
\cline{2-2}\cline{4-6}
 &  &  & NDE & $0.0462$ & $[-0.2157,\,0.3081]$ \\ 
 &  &  & NIE & $0.0762$ & $[-0.1653,\,0.3176]$ \\ 
\multirow{-15}{*}{SIME-I} & \multirow{-3}{*}{Sample} & \multirow{-12}{*}{Active} & TE & $0.1224$ & $[-0.1056,\,0.3503]$ \\ 
\hline
 &  &  & NDE & $0.3057$ & $[-0.2740,\,0.8854]$ \\ 
 &  &  & NIE & $-0.1838$ & $[-0.7192,\,0.3516]$ \\ 
 & \multirow{-3}{*}{LW} &  & TE & $0.1218$ & $[-0.1061,\,0.3497]$ \\ 
\cline{2-2}\cline{4-6}
 &  &  & NDE & $0.3120$ & $[-0.2810,\,0.9051]$ \\ 
 &  &  & NIE & $-0.1906$ & $[-0.7392,\,0.3580]$ \\ 
 & \multirow{-3}{*}{D0.1} &  & TE & $0.1214$ & $[-0.1065,\,0.3494]$ \\ 
\cline{2-2}\cline{4-6}
 &  &  & NDE & $0.7255$ & $[-0.3082,\,1.7593]$ \\ 
 &  &  & NIE & $-0.6032$ & $[-1.6021,\,0.3957]$ \\ 
\multirow{-9}{*}{SIME-F} & \multirow{-3}{*}{Sample} & \multirow{-9}{*}{--} & TE & $0.1224$ & $[-0.1056,\,0.3503]$ \\ 
\hline
\end{tabular}
}
\end{table}

With the original links, the index-ratio estimates vary little across covariance choices. The full-ratio estimates are more sensitive, especially among non-carriers, and their intervals are wider. All direct- and indirect-effect intervals contain zero. Within each covariance choice, the original-link index- and full-ratio fits give the same TE because the conditional cross-world mean cancels. Comparisons with the combined SIME-I also reflect link refitting and broader correction.

\subsection{Link Refitting and Score Correction}
\label{sub:app_a4_components}

Table~\ref{tab:a4_components} compares active-space correction before and after refitting, broader correction at the refitted links and the unaugmented score at the same links. The three refitted procedures share their outcome predictions. Corrected intervals use complete effect scores, and the unaugmented fixed-index intervals serve as descriptive benchmarks.

\begin{table}
\centering
\setlength{\abovecaptionskip}{0pt}
\setlength{\belowcaptionskip}{3pt}
\caption{Estimated SIME-I effects and 95\% CIs for the link-refitting and score comparisons in the A4 study. CI: confidence interval.}
\label{tab:a4_components}
{\renewcommand{\baselinestretch}{1}\normalsize
\renewcommand{\arraystretch}{1}
\begin{tabular}{l l r r c r r}
\hline
 & & \multicolumn{2}{c}{Carrier ($n=487$)} & & \multicolumn{2}{c}{Non-carrier ($n=304$)} \\
\cline{3-4}\cline{6-7}
\multicolumn{1}{c}{\multirow{-2}{*}{Procedure}} & \multicolumn{1}{c}{\multirow{-2}{*}{Quantity}} & \multicolumn{1}{c}{Estimate} & \multicolumn{1}{c}{95\% CI} &   & \multicolumn{1}{c}{Estimate} & \multicolumn{1}{c}{95\% CI} \\
\hline
 & NDE & $-0.045$ & $[-0.182,\,0.093]$ &  & $0.046$ & $[-0.216,\,0.308]$ \\ 
\multirow{-2}{*}{Original active} & NIE & $-0.042$ & $[-0.179,\,0.096]$ &  & $0.077$ & $[-0.165,\,0.318]$ \\ 
\hline
 & NDE & $-0.034$ & $[-0.173,\,0.104]$ &  & $0.068$ & $[-0.156,\,0.291]$ \\ 
\multirow{-2}{*}{Refitted active} & NIE & $-0.051$ & $[-0.186,\,0.084]$ &  & $0.045$ & $[-0.146,\,0.236]$ \\ 
\hline
 & NDE & $-0.033$ & $[-0.173,\,0.107]$ &  & $0.068$ & $[-0.156,\,0.291]$ \\ 
\multirow{-2}{*}{Combined} & NIE & $-0.053$ & $[-0.188,\,0.083]$ &  & $0.045$ & $[-0.146,\,0.236]$ \\ 
\hline
 & NDE & $-0.039$ & $[-0.173,\,0.095]$ &  & $0.065$ & $[-0.150,\,0.280]$ \\ 
\multirow{-2}{*}{Refitted unaug.} & NIE & $-0.047$ & $[-0.180,\,0.086]$ &  & $0.047$ & $[-0.142,\,0.236]$ \\ 
\hline
\end{tabular}
}
\end{table}

Link refitting changes the estimates more than adding the broader correction. Among non-carriers, the two refitted corrections give identical estimates. At fixed links, the correction terms for the direct and indirect effects cancel in the TE. Link refitting changes the arm-specific outcome predictions and can change the TE.

\subsection{Model Diagnostics}
\label{sub:app_a4_stability}

Tables~\ref{tab:a4_folds} and~\ref{tab:a4_folds_noncarrier} report the training-fold diagnostics. Screened and Active count Stage~1 mediators, and Added counts correction coordinates outside the active set. A certificate ratio above one indicates a screened-out direction whose gradient exceeds the SCAD level.

Selected mediators vary across training folds in both subgroups, with a small subset selected in every fold. The broader correction adds a coordinate in one carrier fold. The weights in Table~\ref{tab:a4_weights} and maps in Figure~\ref{fig:a4_glass} describe the separate full-sample fits.

\begin{figure}
\centering
\subfloat[APOE-$\varepsilon4$ carriers ($n=487$)]{%
  \includegraphics[width=\textwidth]{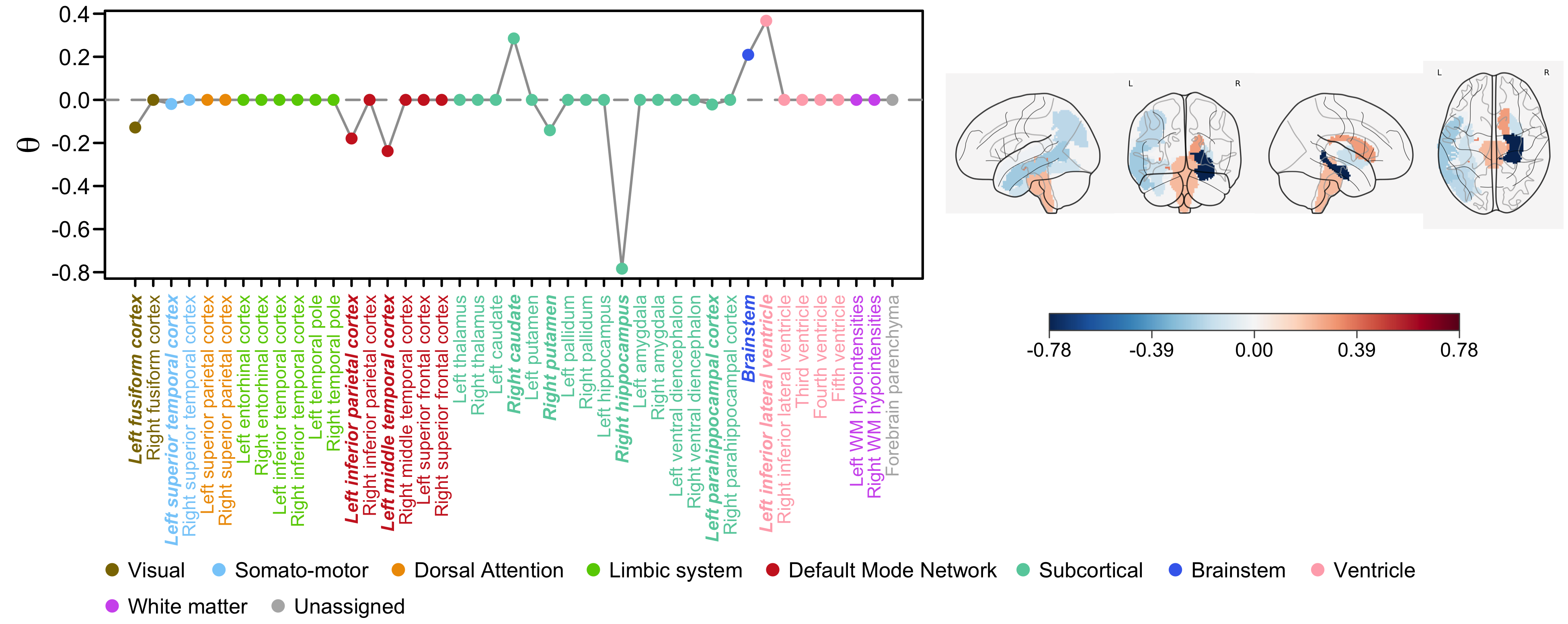}%
}\par\smallskip
\subfloat[APOE-$\varepsilon4$ non-carriers ($n=304$)]{%
  \includegraphics[width=\textwidth]{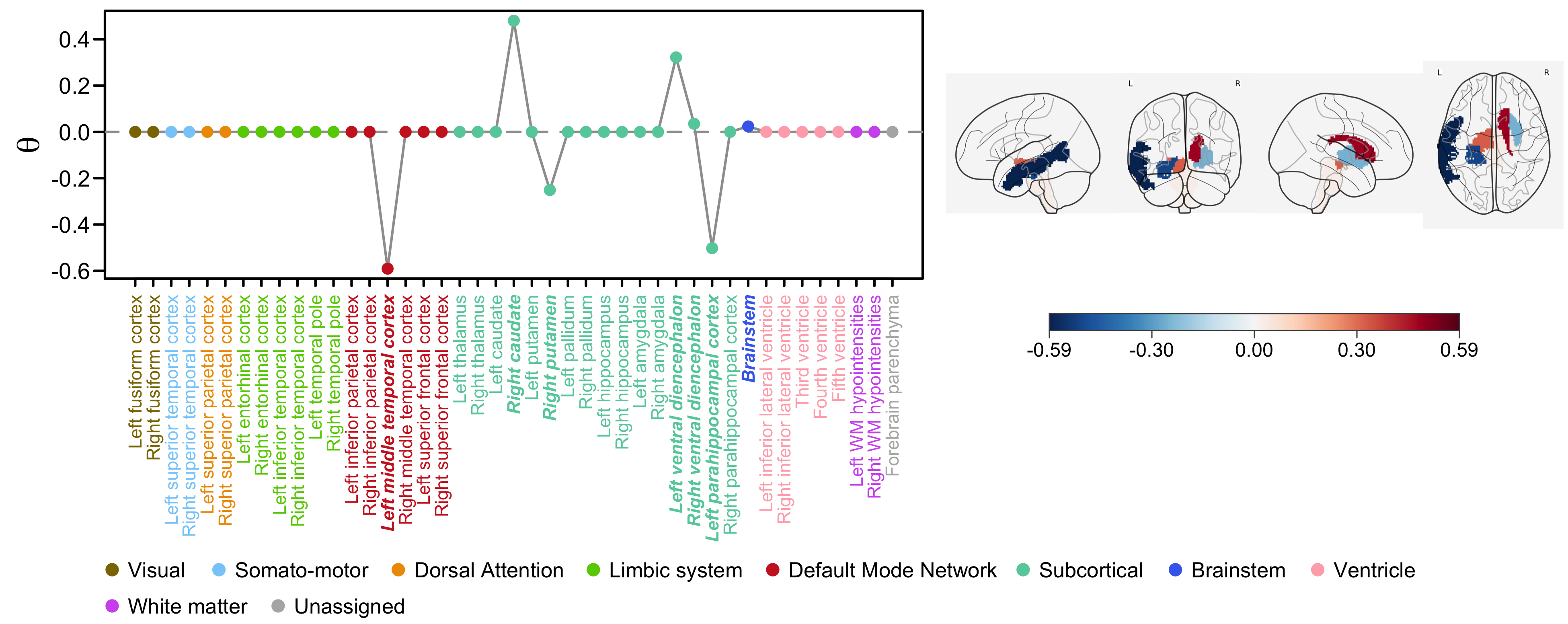}%
}
\caption{Index-weight profiles and brain maps estimated from SIME-I in the A4 study. Profile colors identify the anatomical and functional groups in the legend. Map colors show signed weights on subgroup-specific scales. L: left hemisphere. R: right hemisphere. WM: white matter.}
\label{fig:a4_index_profiles}
\end{figure}

Figure~\ref{fig:a4_index_profiles} pairs the signed index weights for all 43 MRI measures with the corresponding brain maps. The largest weights in magnitude are negative for right hippocampal volume change among carriers and left middle temporal volume change among non-carriers. The right caudate has a positive weight in both subgroups.

\begin{table}
\centering
\setlength{\abovecaptionskip}{0pt}
\setlength{\belowcaptionskip}{3pt}
\caption{Training-fold index and correction diagnostics for APOE-$\varepsilon4$ carriers ($n=487$). EDF: effective degrees of freedom. Cert.: excluded-gradient certificate ratio.}
\label{tab:a4_folds}
{\renewcommand{\baselinestretch}{1}\normalsize
\renewcommand{\arraystretch}{1}
\begin{tabular}{r r r r r r r r r r}
\hline
 & & & & & & \multicolumn{2}{c}{Link EDF} & & \\
\cline{7-8}
\multicolumn{1}{c}{\multirow{-2}{*}{Fold}} & \multicolumn{1}{c}{\multirow{-2}{*}{Train}} & \multicolumn{1}{c}{\multirow{-2}{*}{Test}} & \multicolumn{1}{c}{\multirow{-2}{*}{Screened}} & \multicolumn{1}{c}{\multirow{-2}{*}{Active}} & \multicolumn{1}{c}{\multirow{-2}{*}{Added}} & \multicolumn{1}{c}{Original} & \multicolumn{1}{c}{Refit} & \multicolumn{1}{c}{\multirow{-2}{*}{Cert.}} & \multicolumn{1}{c}{\multirow{-2}{*}{Max $\log\hat r$}} \\
\hline
1 & 389 & 98 & 19 & 13 & 0 & 3.841 & 2.018 & 1.104 & 0.265 \\ 
2 & 389 & 98 & 16 & 12 & 0 & 7.611 & 2.159 & 1.901 & 0.128 \\ 
3 & 390 & 97 & 11 & 7 & 0 & 4.056 & 1.853 & 1.431 & 0.094 \\ 
4 & 390 & 97 & 18 & 9 & 1 & 3.664 & 2.569 & 0.760 & 0.129 \\ 
5 & 390 & 97 & 13 & 9 & 0 & 8.365 & 6.929 & 1.759 & 0.333 \\ 
\hline
\end{tabular}
}
\end{table}

\begin{table}
\centering
\setlength{\abovecaptionskip}{0pt}
\setlength{\belowcaptionskip}{3pt}
\caption{Training-fold index and correction diagnostics for APOE-$\varepsilon4$ non-carriers ($n=304$). EDF: effective degrees of freedom. Cert.: excluded-gradient certificate ratio.}
\label{tab:a4_folds_noncarrier}
{\renewcommand{\baselinestretch}{1}\normalsize
\renewcommand{\arraystretch}{1}
\begin{tabular}{r r r r r r r r r r}
\hline
 & & & & & & \multicolumn{2}{c}{Link EDF} & & \\
\cline{7-8}
\multicolumn{1}{c}{\multirow{-2}{*}{Fold}} & \multicolumn{1}{c}{\multirow{-2}{*}{Train}} & \multicolumn{1}{c}{\multirow{-2}{*}{Test}} & \multicolumn{1}{c}{\multirow{-2}{*}{Screened}} & \multicolumn{1}{c}{\multirow{-2}{*}{Active}} & \multicolumn{1}{c}{\multirow{-2}{*}{Added}} & \multicolumn{1}{c}{Original} & \multicolumn{1}{c}{Refit} & \multicolumn{1}{c}{\multirow{-2}{*}{Cert.}} & \multicolumn{1}{c}{\multirow{-2}{*}{Max $\log\hat r$}} \\
\hline
1 & 243 & 61 & 9 & 7 & 0 & 7.723 & 5.285 & 1.792 & 0.151 \\ 
2 & 243 & 61 & 32 & 14 & 0 & 7.877 & 7.296 & 0.715 & 0.125 \\ 
3 & 243 & 61 & 11 & 6 & 0 & 7.423 & 4.466 & 1.633 & 0.023 \\ 
4 & 243 & 61 & 18 & 9 & 0 & 6.889 & 2.641 & 0.891 & 0.020 \\ 
5 & 244 & 60 & 35 & 16 & 0 & 8.873 & 8.228 & 0.762 & 0.101 \\ 
\hline
\end{tabular}
}
\end{table}

Table~\ref{tab:a4_split} compares 20 inner-fold seeds at fixed Stage~1 fits and evaluation folds with 20 outer fitting seeds that redraw the folds and refit Stage~1. In the outer checks, the inner inference seed is the fitting seed plus two. With the same analysis samples, variation is greater across outer seeds, particularly among non-carriers. All diagnostic intervals contain zero. The reported estimates and standard errors use the single split in Appendix~\ref{sub:app_a4_analysis}.

\begin{table}
\centering
\setlength{\abovecaptionskip}{0pt}
\setlength{\belowcaptionskip}{3pt}
\caption{Estimated SIME-I effects and sensitivity to inner-fold and cross-fitting randomization in the A4 study. SD: standard deviation. SE: standard error. CI: confidence interval.}
\label{tab:a4_split}
{\renewcommand{\baselinestretch}{1}\normalsize
\renewcommand{\arraystretch}{1}
\begin{tabular}{l l r r r r r r}
\hline
 & & & & \multicolumn{2}{c}{SD across 20 seeds} & \multicolumn{2}{c}{CIs excluding zero} \\
\cline{5-6}\cline{7-8}
\multicolumn{1}{c}{\multirow{-2}{*}{Subgroup}} & \multicolumn{1}{c}{\multirow{-2}{*}{Quantity}} & \multicolumn{1}{c}{\multirow{-2}{*}{Reported}} & \multicolumn{1}{c}{\multirow{-2}{*}{SE}} & \multicolumn{1}{c}{Inner} & \multicolumn{1}{c}{Outer} & \multicolumn{1}{c}{Inner} & \multicolumn{1}{c}{Outer} \\
\hline
 & NDE & -0.033 & 0.071 & 0.010 & 0.024 & 0 & 0 \\ 
\multirow{-2}{*}{Carrier} & NIE & -0.053 & 0.069 & 0.010 & 0.025 & 0 & 0 \\ 
\hline
 & NDE & 0.068 & 0.114 & 0.017 & 0.111 & 0 & 0 \\ 
\multirow{-2}{*}{Non-carrier} & NIE & 0.045 & 0.098 & 0.018 & 0.109 & 0 & 0 \\ 
\hline
\end{tabular}
}
\end{table}

Including the influential non-carrier increases the estimated NDE and reverses the sign of the estimated NIE. The intervals contain zero both with and without the exclusion.


\clearpage
\bibliographystyle{apalike}
\bibliography{Bibliography}

\begin{thebibliography}{}

\bibitem[Alquier and Biau, 2013]{alquier2013sparse}
Alquier, P. and Biau, G. (2013).
\newblock Sparse single-index model.
\newblock {\em Journal of Machine Learning Research}, 14(8):243--280.

\bibitem[Alves et~al., 2023]{alves2023accelerated}
Alves, F., Kalinowski, P., and Ayton, S. (2023).
\newblock Accelerated brain volume loss caused by anti-$\beta$-amyloid drugs: a systematic review and meta-analysis.
\newblock {\em Neurology}, 100(20):e2114--e2124.

\bibitem[Andrews and Didelez, 2021]{andrews2021insights}
Andrews, R.~M. and Didelez, V. (2021).
\newblock Insights into the cross-world independence assumption of causal mediation analysis.
\newblock {\em Epidemiology}, 32(2):209--219.

\bibitem[Bang and Robins, 2005]{bang2005doubly}
Bang, H. and Robins, J.~M. (2005).
\newblock Doubly robust estimation in missing data and causal inference models.
\newblock {\em Biometrics}, 61(4):962--973.

\bibitem[Baron and Kenny, 1986]{baron1986moderator}
Baron, R.~M. and Kenny, D.~A. (1986).
\newblock The moderator--mediator variable distinction in social psychological research: Conceptual, strategic, and statistical considerations.
\newblock {\em Journal of Personality and Social Psychology}, 51(6):1173--1182.

\bibitem[Cheng and Li, 2026]{cheng2026identification}
Cheng, C. and Li, F. (2026).
\newblock Identification and multiply robust estimation in causal mediation analysis across principal strata.
\newblock {\em Journal of the Royal Statistical Society Series B: Statistical Methodology}, 88(1):1--22.

\bibitem[Chernozhukov et~al., 2018]{chernozhukov2018double}
Chernozhukov, V., Chetverikov, D., Demirer, M., Duflo, E., Hansen, C., Newey, W., and Robins, J. (2018).
\newblock Double/debiased machine learning for treatment and structural parameters.
\newblock {\em The Econometrics Journal}, 21(1):C1--C68.

\bibitem[Chernozhukov et~al., 2022]{chernozhukov2022automatic}
Chernozhukov, V., Newey, W.~K., and Singh, R. (2022).
\newblock Automatic debiased machine learning of causal and structural effects.
\newblock {\em Econometrica}, 90(3):967--1027.

\bibitem[Cui et~al., 2025]{cui2025estimation}
Cui, S., Guo, X., and Zhang, Z. (2025).
\newblock Estimation and inference in ultrahigh-dimensional partially linear single-index models.
\newblock {\em Science China Mathematics}, 68(8):1807--1840.

\bibitem[Davatzikos et~al., 2009]{davatzikos2009longitudinal}
Davatzikos, C., Xu, F., An, Y., Fan, Y., and Resnick, S.~M. (2009).
\newblock Longitudinal progression of {Alzheimer's}-like patterns of atrophy in normal older adults: the {SPARE-AD} index.
\newblock {\em Brain}, 132(8):2026--2035.

\bibitem[Derkach et~al., 2019]{derkach2019high}
Derkach, A., Pfeiffer, R.~M., Chen, T.-H., and Sampson, J.~N. (2019).
\newblock High dimensional mediation analysis with latent variables.
\newblock {\em Biometrics}, 75(3):745--756.

\bibitem[Dickerson et~al., 2009]{dickerson2009cortical}
Dickerson, B.~C., Bakkour, A., Salat, D.~H., Feczko, E., Pacheco, J., Greve, D.~N., et~al. (2009).
\newblock The cortical signature of {Alzheimer's} disease: regionally specific cortical thinning relates to symptom severity in very mild to mild {AD} dementia and is detectable in asymptomatic amyloid-positive individuals.
\newblock {\em Cerebral Cortex}, 19(3):497--510.

\bibitem[Donohue et~al., 2014]{donohue2014preclinical}
Donohue, M.~C., Sperling, R.~A., Salmon, D.~P., Rentz, D.~M., Raman, R., Thomas, R.~G., Weiner, M., and Aisen, P.~S. (2014).
\newblock The preclinical {Alzheimer} cognitive composite: measuring amyloid-related decline.
\newblock {\em JAMA Neurology}, 71(8):961--970.

\bibitem[Eftekhari et~al., 2021]{eftekhari2021inference}
Eftekhari, H., Banerjee, M., and Ritov, Y. (2021).
\newblock Inference in high-dimensional single-index models under symmetric designs.
\newblock {\em Journal of Machine Learning Research}, 22(27):1--63.

\bibitem[Fan and Gijbels, 1996]{fan1996local}
Fan, J. and Gijbels, I. (1996).
\newblock {\em Local Polynomial Modelling and Its Applications}.
\newblock Chapman \& Hall, London.

\bibitem[Fan and Li, 2001]{fan2001variable}
Fan, J. and Li, R. (2001).
\newblock Variable selection via nonconcave penalized likelihood and its oracle properties.
\newblock {\em Journal of the American statistical Association}, 96(456):1348--1360.

\bibitem[Farbmacher et~al., 2022]{farbmacher2022causal}
Farbmacher, H., Huber, M., Laff{\'e}rs, L., Langen, H., and Spindler, M. (2022).
\newblock Causal mediation analysis with double machine learning.
\newblock {\em The Econometrics Journal}, 25(2):277--300.

\bibitem[H{\"a}rdle et~al., 1993]{hardle1993optimalbandwidth}
H{\"a}rdle, W., Hall, P., and Ichimura, H. (1993).
\newblock Optimal smoothing in single-index models.
\newblock {\em The Annals of Statistics}, 21(1):157--178.

\bibitem[Huang and Pan, 2016]{huang2016hypothesis}
Huang, Y.-T. and Pan, W.-C. (2016).
\newblock Hypothesis test of mediation effect in causal mediation model with high-dimensional continuous mediators.
\newblock {\em Biometrics}, 72(2):402--413.

\bibitem[Ichimura, 1993]{ichimura1993semiparametric}
Ichimura, H. (1993).
\newblock Semiparametric least squares ({SLS}) and weighted {SLS} estimation of single-index models.
\newblock {\em Journal of Econometrics}, 58(1--2):71--120.

\bibitem[Imai et~al., 2010]{imai2010identification}
Imai, K., Keele, L., and Yamamoto, T. (2010).
\newblock Identification, inference and sensitivity analysis for causal mediation effects.
\newblock {\em Statistical Science}, 25(1):51--71.

\bibitem[Imai and Yamamoto, 2013]{imai2013identification}
Imai, K. and Yamamoto, T. (2013).
\newblock Identification and sensitivity analysis for multiple causal mechanisms: {Revisiting} evidence from framing experiments.
\newblock {\em Political Analysis}, 21(2):141--171.

\bibitem[Jack et~al., 2015]{jack2015different}
Jack, Jr., C.~R., Wiste, H.~J., Weigand, S.~D., Knopman, D.~S., Mielke, M.~M., Vemuri, P., et~al. (2015).
\newblock Different definitions of neurodegeneration produce similar amyloid/neurodegeneration biomarker group findings.
\newblock {\em Brain}, 138(12):3747--3759.

\bibitem[Jack~Jr et~al., 2018]{jack2018nia}
Jack~Jr, C.~R., Bennett, D.~A., Blennow, K., Carrillo, M.~C., Dunn, B., Haeberlein, S.~B., Holtzman, D.~M., Jagust, W., Jessen, F., Karlawish, J., et~al. (2018).
\newblock {NIA-AA} research framework: toward a biological definition of {Alzheimer's} disease.
\newblock {\em Alzheimer's \& Dementia}, 14(4):535--562.

\bibitem[Jones et~al., 2025]{jones2025causal}
Jones, J., Ertefaie, A., and Strawderman, R.~L. (2025).
\newblock Causal mediation analysis: selection with asymptotically valid inference.
\newblock {\em Journal of the Royal Statistical Society Series B: Statistical Methodology}, 87(3):678--700.

\bibitem[Lancaster et~al., 2025]{lancaster2025reductions}
Lancaster, T.~M., Murphy, K., and Chandler, H. (2025).
\newblock Reductions in brainstem volume as a key macrostructural indicator in at-risk populations for {Alzheimer's} disease.
\newblock {\em Alzheimer's Research \& Therapy}, 17(1):177.

\bibitem[Ledoit and Wolf, 2004]{ledoit2004well}
Ledoit, O. and Wolf, M. (2004).
\newblock A well-conditioned estimator for large-dimensional covariance matrices.
\newblock {\em Journal of Multivariate Analysis}, 88(2):365--411.

\bibitem[Lee et~al., 2015]{lee2015model}
Lee, J.~D., Sun, Y., and Taylor, J.~E. (2015).
\newblock On model selection consistency of regularized {M}-estimators.
\newblock {\em Electronic Journal of Statistics}, 9(1):608--642.

\bibitem[Lee et~al., 2026]{lee2026atrophy}
Lee, S., Baratono, S.~R., Ha, J., Burt, G.~T., Palm, S.~T., Drew, W.~J., Zide, B.~S., Chiulli, N.~M., Lariviere, S., Zhang, S., et~al. (2026).
\newblock Atrophy in preclinical {Alzheimer's} disease maps to a network that predicts longitudinal decline.
\newblock {\em Molecular Psychiatry}.
\newblock Advance online publication.

\bibitem[Li, 1991]{li1991sliced}
Li, K.-C. (1991).
\newblock Sliced inverse regression for dimension reduction.
\newblock {\em Journal of the American Statistical Association}, 86(414):316--327.

\bibitem[Liu et~al., 2024]{liu2024general}
Liu, R., Williams, N.~T., Rudolph, K.~E., and D{\'\i}az, I. (2024).
\newblock General targeted machine learning for modern causal mediation analysis.
\newblock {\em arXiv preprint arXiv:2408.14620}.

\bibitem[Loh et~al., 2022]{loh2022nonlinear}
Loh, W.~W., Moerkerke, B., Loeys, T., and Vansteelandt, S. (2022).
\newblock Nonlinear mediation analysis with high-dimensional mediators whose causal structure is unknown.
\newblock {\em Biometrics}, 78(1):46--59.

\bibitem[Menze et~al., 2026]{menze2026joint}
Menze, I., Bernal, J., Kievit, R.~A., Tripathi, K.~P., Kaleck, T., Yakupov, R., Altenstein, S., Bartels, C., Buerger, K., Butryn, M., et~al. (2026).
\newblock Joint trajectories of brain atrophy, white matter hyperintensities and cognition quantify brain maintenance.
\newblock {\em Nature Communications}, 17:5846.

\bibitem[Na et~al., 2024]{na2024cortex}
Na, S., Kim, T., Song, I.-U., Hong, Y.~J., and Kim, S.-H. (2024).
\newblock Cortex-to-caudate volume ratio as a predictor of cognitive decline in {Alzheimer's} disease and mild cognitive impairment.
\newblock {\em Journal of the Neurological Sciences}, 462:123113.

\bibitem[Neykov et~al., 2016]{neykov2016l1}
Neykov, M., Liu, J.~S., and Cai, T. (2016).
\newblock L1-regularized least squares for support recovery of high dimensional single index models with {G}aussian designs.
\newblock {\em Journal of Machine Learning Research}, 17(87):1--37.

\bibitem[Pearl, 2001]{pearl2001direct}
Pearl, J. (2001).
\newblock Direct and indirect effects.
\newblock In {\em Proceedings of the seventeenth conference on uncertainty in artificial intelligence}, pages 411--420. Morgan Kaufmann Publishers Inc.

\bibitem[Radchenko, 2015]{radchenko2015high}
Radchenko, P. (2015).
\newblock High dimensional single index models.
\newblock {\em Journal of Multivariate Analysis}, 139:266--282.

\bibitem[Robins, 2003]{robins2003semantics}
Robins, J.~M. (2003).
\newblock Semantics of causal dag models and the identification of direct and indirect effects.
\newblock {\em Highly structured stochastic systems}, pages 70--81.

\bibitem[Robins and Greenland, 1992]{robins1992identifiability}
Robins, J.~M. and Greenland, S. (1992).
\newblock Identifiability and exchangeability for direct and indirect effects.
\newblock {\em Epidemiology}, pages 143--155.

\bibitem[Rubin, 1974]{rubin1974estimating}
Rubin, D.~B. (1974).
\newblock Estimating causal effects of treatments in randomized and nonrandomized studies.
\newblock {\em Journal of educational Psychology}, 66(5):688.

\bibitem[Song et~al., 2020]{song2020bayesian}
Song, Y., Zhou, X., Zhang, M., Zhao, W., Liu, Y., Kardia, S.~L., Diez~Roux, A.~V., Needham, B.~L., Smith, J.~A., and Mukherjee, B. (2020).
\newblock {Bayesian} shrinkage estimation of high dimensional causal mediation effects in omics studies.
\newblock {\em Biometrics}, 76(3):700--710.

\bibitem[Sperling et~al., 2023]{sperling2023trial}
Sperling, R.~A., Donohue, M.~C., Raman, R., Rafii, M.~S., Johnson, K., Masters, C.~L., van Dyck, C.~H., Iwatsubo, T., Marshall, G.~A., Yaari, R., et~al. (2023).
\newblock Trial of solanezumab in preclinical {Alzheimer's} disease.
\newblock {\em New England Journal of Medicine}, 389(12):1096--1107.

\bibitem[Sperling et~al., 2020]{sperling2020association}
Sperling, R.~A., Donohue, M.~C., Raman, R., Sun, C.-K., Yaari, R., Holdridge, K., Siemers, E., Johnson, K.~A., Aisen, P.~S., and {A4 Study Team} (2020).
\newblock Association of factors with elevated amyloid burden in clinically normal older individuals.
\newblock {\em JAMA Neurology}, 77(6):735--745.

\bibitem[Sperling et~al., 2014]{sperling2014a4}
Sperling, R.~A., Rentz, D.~M., Johnson, K.~A., Karlawish, J., Donohue, M., Salmon, D.~P., and Aisen, P. (2014).
\newblock The {A4} study: Stopping {AD} before symptoms begin?
\newblock {\em Science Translational Medicine}, 6(228):228fs13.

\bibitem[Tchetgen~Tchetgen and Shpitser, 2012]{tchetgen2012semiparametric}
Tchetgen~Tchetgen, E.~J. and Shpitser, I. (2012).
\newblock Semiparametric theory for causal mediation analysis: Efficiency bounds, multiple robustness and sensitivity analysis.
\newblock {\em The Annals of Statistics}, 40(3):1816--1845.

\bibitem[VanderWeele, 2015]{vanderweele2015explanation}
VanderWeele, T.~J. (2015).
\newblock {\em Explanation in Causal Inference: Methods for Mediation and Interaction}.
\newblock Oxford University Press.

\bibitem[VanderWeele and Vansteelandt, 2009]{vanderweele2009conceptual}
VanderWeele, T.~J. and Vansteelandt, S. (2009).
\newblock Conceptual issues concerning mediation, interventions and composition.
\newblock {\em Statistics and its Interface}, 2:457--468.

\bibitem[VanderWeele and Vansteelandt, 2014]{vanderweele2014mediation}
VanderWeele, T.~J. and Vansteelandt, S. (2014).
\newblock Mediation analysis with multiple mediators.
\newblock {\em Epidemiologic Methods}, 2(1):95--115.

\bibitem[Vansteelandt and Daniel, 2017]{vansteelandt2017interventional}
Vansteelandt, S. and Daniel, R.~M. (2017).
\newblock Interventional effects for mediation analysis with multiple mediators.
\newblock {\em Epidemiology (Cambridge, Mass.)}, 28(2):258.

\bibitem[Wearn et~al., 2026]{wearn2026locus}
Wearn, A., Onuska, K.~M., Shanks, H.~R., Gaurav, R., Leh{\'e}ricy, S., Baracchini, G., Hughes, C., Tremblay-Mercier, J., Narayanan, S., Breitner, J., et~al. (2026).
\newblock Locus coeruleus degeneration is associated with cortical tau deposition and cognitive decline in older adults at familial risk of {Alzheimer's} disease.
\newblock {\em Alzheimer's \& Dementia}, 22(4):e71427.

\bibitem[Wu et~al., 2023]{wu2023model}
Wu, Y., Wang, L., and Fu, H. (2023).
\newblock Model-assisted uniformly honest inference for optimal treatment regimes in high dimension.
\newblock {\em Journal of the American Statistical Association}, 118(541):305--314.

\bibitem[Xie et~al., 2025]{xie2025multi}
Xie, L., Das, S.~R., Li, Y., Wisse, L.~E., McGrew, E., Lyu, X., DiCalogero, M., Shah, U., Ilesanmi, A., Denning, A.~E., et~al. (2025).
\newblock A multi-cohort study of longitudinal and cross-sectional {Alzheimer's} disease biomarkers in cognitively unimpaired older adults.
\newblock {\em Alzheimer's \& Dementia}, 21(2):e14492.

\bibitem[Zhang, 2010]{zhang2010nearly}
Zhang, C.-H. (2010).
\newblock Nearly unbiased variable selection under minimax concave penalty.
\newblock {\em The Annals of Statistics}, pages 894--942.

\bibitem[Zhang et~al., 2016]{zhang2016estimating}
Zhang, H., Zheng, Y., Zhang, Z., Gao, T., Joyce, B., Yoon, G., Zhang, W., Schwartz, J., Just, A., Colicino, E., Vokonas, P., Zhao, L., Lv, J., Baccarelli, A., Hou, L., and Liu, L. (2016).
\newblock Estimating and testing high-dimensional mediation effects in epigenetic studies.
\newblock {\em Bioinformatics}, 32(20):3150--3154.

\bibitem[Zhao and Yu, 2006]{zhao2006model}
Zhao, P. and Yu, B. (2006).
\newblock On model selection consistency of lasso.
\newblock {\em The Journal of Machine Learning Research}, 7:2541--2563.

\bibitem[Zhao, 2024]{zhao2024mediation}
Zhao, Y. (2024).
\newblock Mediation analysis with multiple exposures and multiple mediators.
\newblock {\em Statistics in medicine}, 43(25):4887--4898.

\bibitem[Zhao et~al., 2020]{zhao2020sparse}
Zhao, Y., Lindquist, M.~A., and Caffo, B.~S. (2020).
\newblock Sparse principal component based high-dimensional mediation analysis.
\newblock {\em Computational Statistics \& Data Analysis}, 142:106835.

\bibitem[Zhao and Luo, 2022]{zhao2022pathway}
Zhao, Y. and Luo, X. (2022).
\newblock Pathway lasso: Pathway estimation and selection with high dimensional mediators.
\newblock {\em Statistics and Its Interface}, 15(1):39--50.

\bibitem[Zhou et~al., 2020]{zhou2020estimation}
Zhou, R.~R., Wang, L., and Zhao, S.~D. (2020).
\newblock Estimation and inference for the indirect effect in high-dimensional linear mediation models.
\newblock {\em Biometrika}, 107(3):573--589.

\end{thebibliography}

\end{document}